\documentclass[12pt]{iopart}

\input{packages.tex}
\newif\ifrefereecoloring
\refereecoloringfalse

\newcommand{\refA}[1]{\ifrefereecoloring \textcolor{orange}{#1}\else #1\fi}
\newcommand{\refB}[1]{\ifrefereecoloring \textcolor{black!30!green}{#1}\else #1\fi}
\newcommand{\refC}[1]{\ifrefereecoloring {\color{brown}#1}\else #1\fi}

\newcommand{\cO}{\mathcal{O}}

\newcommand{\cM}{\mathcal{M}}
\newcommand{\cN}{\mathcal{N}}

\newcommand{\rR}{\mathbb{R}}

\newcommand{\bK}{\mathbf{K}}

\newcommand{\vA}{{\vec{A}}}
\newcommand{\vtheta}{{\vec{\theta}}}
\newcommand{\veps}{{\vec{\epsilon}}}
\newcommand{\vsig}{{\vec{\sigma}}}
\newcommand{\dvsig}{{\delta\vec{\sigma}}}
\newcommand{\Dvsig}{{\Delta\vec{\sigma}}}

\def\exOne{11.54835442}
\def\exTwo{11.62359272}
\def\exThree{11.72316039}
\def\exFour{11.82807116}
\def\ionE{15.7596119}

\newcommand{\drawIntervals}[3]
{
   \addplot+ [
        line width=1.5,
        mark=none,
        solid, smooth,
        color=#2,
    ]
    table [x expr=\thisrowno{0} - #3, y index=1]{#1};
    
   \addplot+ [
       name path=sig3p,
        line width=0.5,
        mark=none,
        dashed, smooth,
        color=#2,
        opacity=0.2,
    ]
    table [x expr=\thisrowno{0} - #3, y index=6]{#1};
   \addplot+ [
       name path=sig3m,
        line width=0.5,
        mark=none,
        dashed, smooth,
        color=#2,
        opacity=0.2,
    ]
    table [x expr=\thisrowno{0} - #3, y index=7]{#1};
    \addplot+ [
        fill=#2,
        opacity=0.2,
    ]
    fill between[of = sig3p and sig3m,];

   \addplot+ [
       name path=sig2p,
        line width=1.0,
        mark=none,
        dashed, smooth,
        color=#2,
        opacity=0.3,
    ]
    table [x expr=\thisrowno{0} - #3, y index=4]{#1};
   \addplot+ [
       name path=sig2m,
        line width=1.0,
        mark=none,
        dashed, smooth,
        color=#2,
        opacity=0.3,
    ]
    table [x expr=\thisrowno{0} - #3, y index=5]{#1};
    \addplot+ [
        fill=#2,
        opacity=0.3,
    ]
    fill between[of = sig2p and sig2m,];

   \addplot+ [
       name path=sig1p,
        line width=1.0,
        mark=none,
        dashed, smooth,
        color=#2,
        opacity=0.4,
    ]
    table [x expr=\thisrowno{0} - #3, y index=2]{#1};
   \addplot+ [
       name path=sig1m,
        line width=1.0,
        mark=none,
        dashed, smooth,
        color=#2,
        opacity=0.4,
    ]
    table [x expr=\thisrowno{0} - #3, y index=3]{#1};
    \addplot+ [
        fill=#2,
        opacity=0.4,
    ]
    fill between[of = sig1p and sig1m,];
}

\refereecoloringfalse

\begin{document}

\title[]{Characterization of uncertainties in electron-argon collision cross sections}

\author{Seung Whan Chung$^1$,
Todd A. Oliver$^1$,
Laxminarayan L. Raja$^2$,
Robert D. Moser$^{1,3}$
}

\address{$^1$ Oden Institute for Computational Engineering and Sciences, University of Texas at Austin, 201 E. 24th Street, Austin, TX 78712}
\address{$^2$ Department of Aerospace Engineering \& Engineering Mechanics, University of Texas at Austin, 2617 Wichita Street, Austin, TX 78712}
\address{$^3$ Walker Department of Mechanical Engineering, University of Texas at Austin, 204 E. Dean Keeton Street, Austin, TX 78712}
\ead{chung28@llnl.gov}
\vspace{10pt}
\begin{indented}
\item[]June 2023
\end{indented}

\begin{abstract}
The predictive capability of a plasma discharge model depends on accurate representations of electron-impact collision cross sections,
which determine the corresponding reaction rates and electron transport properties.
The values of cross sections can be known only approximately either through experiments or simulations
and are thus subject to uncertainties.
Quantifying the uncertainties in plasma simulations allows us
to assess the reliability of simulations and to provide a basis
for interpreting discrepancies between simulations and experiments.
For such uncertainty quantification of plasma simulations,
it is essential to quantify the uncertainties of the underlying cross sections.
Although much effort has been committed to calibrate the cross section values,
their uncertainties are not well investigated.
We characterize uncertainties in electron-argon atom collision
cross sections using a Bayesian framework.  Six collision
processes---elastic momentum transfer, ionization, and four
excitations---are characterized with semi-empirical models, which
effectively capture the features important to
the macroscopic properties of the plasma.
A probability model for the uncertain parameters of these semi-empirical models is developed.
Specifically,
a Gaussian-process likelihood model is proposed to capture discrepancies among data sets,
as well as the model-form inadequacies of the semi-empirical models.
Two other likelihood models are compared with the proposed Gaussian-process model,
to illustrate the importance of the choice of the likelihood model.
The cross section models are calibrated using the electron-beam experiments and \textit{ab-inito} quantum simulations.
The resulting calibrated uncertainties capture well the scattering among the data sets.
The calibrated cross section models are further
validated against swarm-parameter experiments and zero-dimensional Boltzmann equation simulations
of widely used cross section datasets.
%
%
%
%
\end{abstract}
\submitto{\PSST}

%
%
%
%
%

\section{Introduction}

The cross sections of atomic and molecular collision processes are important inputs in plasma discharge model predictions~\cite{Gargioni2008}.
Most macroscopic observables of plasma, such as species compositions and temperatures,
depend on the energy distribution functions of the particles that comprise the plasma~\cite{Ferziger1973}.
The evolution of these energy distribution functions is governed by the species Boltzmann equation, in which the cross sections appear in the collision term
that represents the effect of particle collisional interactions.
Chemical reactions and transport processes are the macroscopic manifestation of these collision processes~\cite{Vlcek1989, Bultel2002, Kapper2011}, and
are determined by both the cross sections themselves and the energy distributions.
In this context, electron-impact collisions are particularly important for modeling non-equilibrium plasmas~\cite{Pitchford2013, Pitchford2017}, where the energy distribution functions cannot be determined from equilibrium considerations.
\par
Because of these dependencies, particle collision cross sections are necessary for a plasma discharge model.
They are clearly essential for simulations based on the Boltzmann equations, where they appear explicitly.
In particle-based Monte-Carlo simulations of the Boltzmann equation,
these cross sections are used to statistically determine the collision rates~\cite{Oran1998, Bird1994, Birdsall1991, Vahedi1993, Birdsall2004}.
In grid-based Boltzmann equation simulations, on the other hand,
the collision rates are evaluated deterministically as expectations over the evolving distributions\refC{~\cite{Hagelaar2005, petrov1997multi, winkler1986new, segur1983application}.}
In either case, the cross sections---represented as functions of electron energy,
either by analytical models or tabulated data---are direct inputs to the model.
Further, even in fluid-based models where the cross sections do not
appear explicitly in the model PDEs, the input data---specifically,
reaction rates and transport properties---are dependent on the
underlying cross sections~\cite{Hagelaar2005}.
\par
Because of this importance, the characterization of collision
cross sections has been a focus of experimental measurements for many
decades and, more recently, \textit{ab initio} quantum mechanical
simulations.  This work has resulted in numerous databases of
cross sections, e.g.,~\cite{Bederson1971, Alves2005, Gargioni2008, Pitchford2013, Biagi-data, BSR-data, Hayashi-data, IST-Lisbon-data, Puech-data}.
These datasets are mainly obtained from three sources:
electron beam experiments~\cite{Filippelli1994, Tachibana1986, Chilton1998},
\emph{ab initio} quantum calculations~\cite{Zatsarinny2013, Mceachran2014, Gangwar2012, Djuissi2022},
and swarm experiments~\cite{Pitchford2013, Milloy1977, Haddad1982, Nakamura1987, Biagi1989}.
In Section~\ref{sec:curation} we curated these cross section datasets
for the electron-argon atom collision processes as a target example of this study.
\par
\todo[inline]{$\downarrow$ here we introduce the uncertainty as a result of experiment/simulation errors.
Also we describe that our knowledge is not perfect.}
While these datasets provide credible tabulated values of 
cross sections that can be used for accurate plasma modeling,
the cross section values are
subject to uncertainties. These uncertainties arise from experimental
or numerical errors in measuring the cross sections, and from
uncharacterized systematic errors intrinsic to these studies~\cite{Gargioni2008}.
The sources of the uncertainties in cross section measurements and simulations
are described in details in Section~\ref{sec:curation}.
\refC{
As a result of uncertainties involved at multiple levels,
there are discrepancies among the datasets,
often beyond their reported measurement errors.
In this sense, it should be noted that
these cross section databases do not represent ``true'' collision cross sections,
rather the ``most credible'' values based on the specific experiment or simulation data.
This is, of course, well recognized throughout low-temperature plasma community~\cite{alves2018foundations}.
}
\par
The cross section uncertainties result in
uncertainties in the predicted plasma properties. To quantify
uncertainties in the predictions, it is necessary to quantify the
uncertainties in the cross sections. Such uncertainty quantification
(UQ) allows the reliability of simulation predictions to be assessed
and provides a basis for interpreting discrepancies between
simulations and experiments~\cite{Stuart2010, Smith2013}. A representation of collision
cross sections for use in a plasma simulation should therefore include
uncertainties.
\par
Nonetheless,
information on uncertainties in the existing cross section datasets is limited.
Swarm-derived datasets only provide ``least-square-fitted'' values against the swarm experiments,
but not their associated uncertainties and correlations.
Furthermore, the different cross sections derived from swarm data cannot be considered independent evaluations of these cross sections because of their dependence on the distribution function itself,
resulting in further complexity in the cross section uncertainties.
Some electron-beam experiments reported their measurement noises;
however, discrepancies among the datasets clearly indicate that
there are other types of uncertainties,
which are difficult to characterize and thus remain unevaluated.
\refC{
The uncertainties of the datasets from \textit{ab-initio} calculation are likewise uncharacterized.
To our knowledge,
there are only qualitative comparisons among the existing collision cross section datasets~\cite{Gargioni2008, Pitchford2013, Pitchford2017}.
Most recent UQ efforts in the low-temperature plasma community have focused on
forward propagation of chemical rate constant uncertainties to plasma prediction~\cite{alves2018foundations,
turner2015uncertainty,turner2016uncertainty,turner2017computer,koelman2019uncertainty,berthelot2017modeling}.
However,
the rate constant uncertainties are assigned
based on the recommended values from the authors of the datasets.
For electron-impact processes,
despite their strong dependencies on cross sections,
the uncertainties of cross sections are not considered
and assigned similarly as for heavy particle impact collision processes~\cite{turner2015uncertainty}.
A more detailed study is needed to accurately characterize
the uncertainties of electron-impact collision cross sections,
for the uncertainty analysis of the overall plasma prediction~\cite{turner2015uncertainty}.
}
\par
Here we introduce a Bayesian probabilistic model for the cross sections,
which is intended to represent our incomplete knowledge regarding their true values~\cite{Kennedy2001, Rasmussen2006, Foreman-Mackey2013}.
The probabilistic model is calibrated via a Bayesian update using data from multiple sources.
Unlike past efforts in the plasma community, which focus exclusively on determining the most credible values of the cross-sections,
this approach aims to provide a much richer characterization of the cross sections by representing their uncertainties as well.
\refC{Thus, the goals of this study are two-fold:
first, to provide a general, systematic framework to characterize uncertainties in collision cross sections given data;
and, second, to demonstrate this method for six important electron-argon collisions.}
\par
The application of Bayesian inference to the cross section uncertainty quantification
can be challenging in several aspects.
First, for the most part, the cross sections vary smoothly across the electron energy space,
implying that their uncertainties are also correlated along the electron energy space.
Fortunately, many past studies on cross sections 
provide semi-empirical models that accounts for large-scale variations
with a few tunable parameters~\cite{Gargioni2008,Haddad1982,Bretagne1986,Kim1994}.
We exploit these semi-empirical models, with an expectation that
these parametric representations will be sufficient to represent the essential features of the cross sections
to determine reaction rates and transport parameters in collisional plasma simulations.
For other applications where one is interested in the cross section at specific energies,
rather than integrated quantities, a richer representation may be necessary.
\par
Second, in Bayesian inference,
just as for the cross sections, the uncertainty itself must be represented with a proper model.
This makes the application of Bayesian inference to the cross sections challenging,
because most uncertainties in the determination of cross sections arise from uncharacterized systematic errors
in the experiments or the simulations.
These uncertainties manifest as large discrepancies among the datasets beyond their reported measurement errors.
While the cross section itself can be characterized with some semi-empirical models,
there is no knowledge or theory to quantify these systematic errors.
These include any errors introduced by the semi-empirical model form itself as well.
Nonetheless, this systematic error must be accounted in evaluating the uncertainty.
In our study, we `model' the systematic error with a Gaussian process representation~\cite{Kennedy2001}.
A Gaussian process model is a probabilistic
representation of a function that is characterized by a number of hyperparameters.
This is, of course, not the `true' representation of the systematic error.
However, Gaussian processes provide a general, flexible way to account for errors
about which we do not have sufficient information to pose a parametric model.
\refB{
In Section~\ref{sec:bayes}, as references,
we also present two `improper' models for the systematic error,
which are based on overly specific assumptions.
It will be shown in Section~\ref{sec:result}
how these improper assumptions result in poor representation of the uncertainties in cross sections.
}
\par
\refB{
Lastly, the data from which we infer cross sections can pose a challenge for Bayesian inference.
This is particularly the case for the swarm parameter experiments and the associated swarm-derived cross section datasets.
Electron swarm experiments measure electron fluid transport properties,
which depend on the collision cross sections~\cite{Bederson1971, Gargioni2008, Pitchford2013}.
Obtaining the cross sections from the transport properties determined in the swarm experiments
requires an inference procedure involving the (usually simplified) electron
Boltzmann equation, in which many collision processes are active
simultaneously~\cite{Milloy1977, Haddad1982, Nakamura1987, Biagi1989, Alves2005}.
The collective impacts of many collisions on swarm parameters result in
the inferred cross sections being correlated with each other,
which cannot be used as an individual cross section.
The swarm-derived cross section datasets likewise have--unreported--correlations among the cross sections,
and it is recommended to use the entire dataset as a whole~\cite{Alves2005,Pitchford2013,Pitchford2017}.
This limits the use of cross sections to applications that requires different set of collisions
with different plasma compositions than the calibrated dataset.
Also, when the uncertainties of cross sections are forward propagated with such correlations,
it is difficult to interpret the results of plasma simulations.
}
\par
\refC{
In this study, we infer cross sections and their uncertainties only from
the electron-beam measurements and the \textit{ab initio} calculations,
thereby providing independent uncertainties for the individual cross sections.
In fact, it turns out that these direct measurements of cross sections
are sufficient to infer the cross sections and their uncertainties
that can explain the swarm parameter experiment data, as demonstrated in Section~\ref{sec:result}.
}
\par
\refB{
In Section~\ref{sec:curation},
we curate the cross section data for Bayesian inference and validation.
The semi-empirical cross section models are introduced (Section~\ref{sec:model}).
The cross section uncertainty description for Bayesian calibration
is described in Section~\ref{sec:bayes}:
to establish the importance of choice of likelihood,
we first present the two inadequate probability models in Section~\ref{subsec:wrong-bayes-1}~and~\ref{subsec:wrong-bayes-2},
and lastly in Section~\ref{subsec:gp-bayes}
we present the Gaussian-process-based model.
The results of the Bayesian calibration and validation for three
different formulations of the probabilistic error model are presented
in Section~\ref{sec:result}.
The inferred probabilistic cross section models are further validated
against the swarm experiments in Section~\ref{sec:result}.
Finally, concluding remarks are provided
in Section~\ref{sec:conclusion}.
}

\section{Cross section measurement datasets}\label{sec:curation}
The available data are split into two subsets: one consisting of
electron-beam experiments and \textit{ab initio} simulation results,
and the second consisting of swarm experiment measurements.
The first subset is either measurements or calculations of individual collision cross sections,
which are used in this study for calibration of uncertain cross sections.
The second subset is then used for validating the calibrated cross sections
collectively for given experiment conditions.
The specific data sources as well as relevant features of the data are
described for each of these subsets in
Sections~\ref{subsec:electron-beam} and~\ref{subsec:swarm-data}.

\subsection{Electron-beam experiments and \textit{ab initio} calculations}\label{subsec:electron-beam}

\subsubsection{Data characterization}
\begin{table}[tbp]
\centering
\begin{tabular}{|c|c|c|}
\hline
Collision process & Author & Energy range $\epsilon$ ($eV$) \\
\hline
\multirow{5}{8em}{Elastic momentum-transfer, $\sigma_{el}(\epsilon)$} & Zatsarinny \& Bartschat (BSR), 2013~\cite{Pitchford2013, Zatsarinny2013} & $10^{-3}$--$300$\\\cline{2-3}
                                                                                                                & Srivastava \textit{et al.}, 1981~\cite{Srivastava1981} & 3--100\\\cline{2-3}
                                                                                                                & Gibson \textit{et al.}, 1996~\cite{Gibson1996} & 1--10\\\cline{2-3}
                                                                                                                & Panajotovic \textit{et al.}, 1997~\cite{Panajotovic1997} & 10--100 \\\cline{2-3}
                                                                                                                & Mielewska \textit{et al.}, 2004~\cite{Mielewska2004} & 5--10 \\
\hline
\multirow{3}{8em}{Direct ionization ($Ar^+$), $\sigma_{ion}(\epsilon)$} & Rapp \& Englander-Golden, 1965~\cite{Rapp1965} & Threshold--$10^3$ \\\cline{2-3}
								 & Wetzel \textit{et al.}, 1987~\cite{Wetzel1987} & Threshold--200 \\\cline{2-3}
								 & Straub \textit{et al.}, 1995~\cite{Straub1995} & Threshold--$10^3$ \\
\hline
\multirow{6}{8em}{Direct excitation, $\sigma_{ex}(\epsilon)$} & Zatsarinny \& Bartschat (BSR), 2013~\cite{Pitchford2013, Zatsarinny2013} & Threshold--$300$\\\cline{2-3}
                                                                    & Chutjian \& Cartwright, 1981~\cite{Chutjian1981} ($1s_5$--$1s_2$) & Threshold--100 \\\cline{2-3}
    								 & Li \textit{et al.}~\cite{Li1988} ($1s_4$, $1s_2$) & 400, 500 \\\cline{2-3}
    								 & Schappe \textit{et al.}~\cite{Schappe1994} ($1s_5$, $1s_3$) & Threshold--100 \\\cline{2-3}
    								 & Filipovic \textit{et al.}~\cite{Filipovic2000a} ($1s_2$) & 16--80 \\\cline{2-3}
    								 & Filipovic \textit{et al.}~\cite{Filipovic2000b} ($1s_5$--$1s_3$) & 20--80 \\\cline{2-3}
    								 & Khakoo \textit{et al.}~\cite{Khakoo2004} ($1s_5$--$1s_2$) & 14--100 \\
\hline
\end{tabular}
\caption{Summary of the curated electron-beam experiments and \textit{ab initio} calculation.}
\label{tab:crs-curation}
\end{table}
\begin{figure}
\input{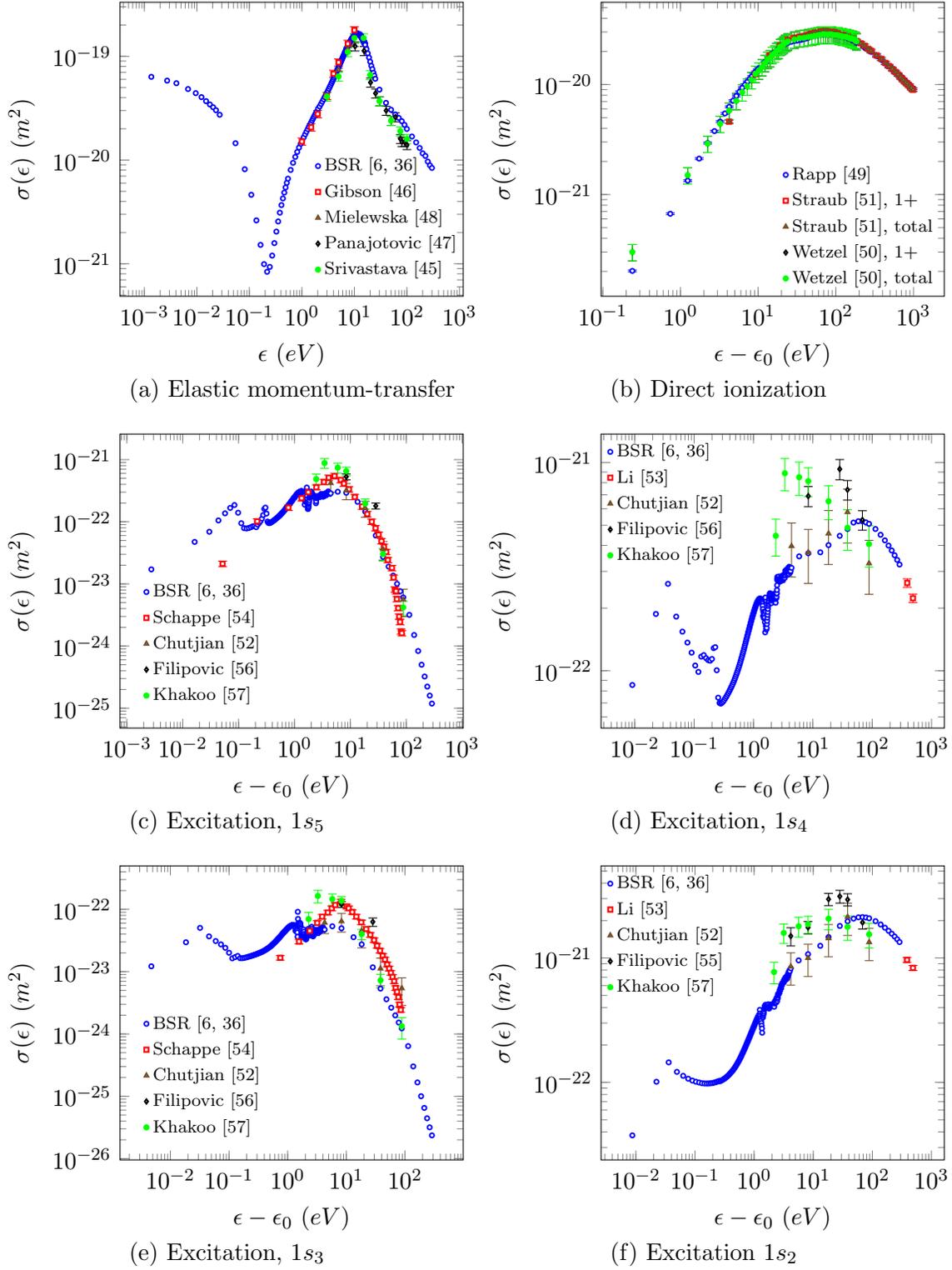}
\caption{
Electron-impact argon cross section data from the curated electron-beam experiments and \textit{ab initio} calculation.
The reported standard-deviation error, if available from the reference, is indicated as an error bar.
}
\label{fig:crs-curation}
\end{figure}
Electron-beam experiments measure the scattering of a mono-energetic
electron beam, to determine differential cross sections at specific
energies and angles \cite{Bederson1971, Gargioni2008}.  Combined with optical techniques, they
can also measure excitation cross sections \cite{Filippelli1994, Tachibana1986, Chilton1998}. These are direct
measurements of the cross sections,
which at least notionally report their measurement accuracy~\cite{Gargioni2008,Pitchford2013}.
Table~\ref{tab:crs-curation} and Figure~\ref{fig:crs-curation} summarize the curated cross section data.
Almost all experimental data are limited to energy$>1eV$,
and very low-energy data is scarcely available.
\par
Aside from experiments,
\emph{ab initio} quantum calculations also provide cross section data~\cite{Pitchford2013, Pitchford2017}.
These calculations involve the wavefunctions of a colliding electron and atom solved
through approximate forms of Schr\"{o}dinger's equation.
The cross section is computed as a dependent variable~\cite{Zatsarinny2013, Mceachran2014, Gangwar2012, Djuissi2022}.
Among various \emph{ab initio} datasets, the BSR data provides a comprehensive set of
electron-argon collision cross sections~\cite{Zatsarinny2004, Zatsarinny2013, Zatsarinny2014, Allan2006, Khakoo2011}.
The resulting cross section data for noble gases agree well with experimental data.
Though its accuracy is not reported,
the data is considered credible based on its accurate prediction of atomic differential cross section,
as well as its low dependency on other (uncertain) sources~\cite{Zatsarinny2013, Allan2006, Khakoo2011}.
The BSR dataset by Zatsarinny and Bartschat~\cite{Pitchford2013, Zatsarinny2013, BSR-data},
which provides highly resolved data over a wide energy range,
is thus included in this study.

\subsubsection{Discrepancies between datasets}

An observation from Figure~\ref{fig:crs-curation} is the large discrepancies among datasets.
Particularly for direct excitations in the $1-100eV$ range,
the datasets are scattered far outside the confidence intervals of each other,
indicating that their reported measurement errors cannot fully explain the discrepancy.
\par
This discrepancy is likely due to systematic errors that vary from experiment to experiment~\cite{Gargioni2008}.
There can be several factors contributing to such errors.
First, the detailed design of the electron gun varies between experiments as the energy range of interest usually spans a few $eV$ to several thousands of $eV$~\cite{Gargioni2008}.
Also, the cross-beam technique, which is widely used for this type of experiment, is usually limited to relative measurements of the cross section,
so the measurement is scaled with a reference value.
The choice of this reference value can differ by experiment.
For the elastic collision, the widely-used relative flow techniques often use the cross section of helium~\cite{Srivastava1981, Gibson1996, Mielewska2004},
or the same cross section from other measurements~\cite{Panajotovic1997};
reflecting this point, Zatsarinny and Bartschat~\cite{Zatsarinny2004} also rescaled the experimental data from Buckman~\textit{et al.}~\cite{Buckman1983} to compare with BSR dataset.
Similarly for ionization, the measurement is sometimes normalized with respect to the cross section of hydrogen~\cite{Rapp1965};
or the absolute value is measured at only one energy, and the measurements at other energy points are respectively scaled~\cite{Straub1995}.
Likewise, the excitation cross sections are also normalized with previous measurements~\cite{Chutjian1981, Filipovic2000a, Filipovic2000b, Khakoo2004}.
Furthermore, many of these measurements are obtained by integrating the corresponding differential cross section over angle using a numerical quadrature,
where low-angle values are either extrapolated/adopted from other datasets~\cite{Srivastava1981, Panajotovic1997, Mielewska2004, Chutjian1981, Li1988, Filipovic2000a, Filipovic2000b, Khakoo2004}.
While Gibson~\textit{et al.}~\cite{Gibson1996} utilized a phase-shift analysis to infer from the differential cross section,
this involves a model-fitting procedure which likewise induces another systematic error.
The optical emission measurements also involve multiple excitation and cascade processes,
so measuring one of them requires approximation/estimation of the others~\cite{Schappe1994}.
\refC{There can be many more factors that contribute to the systematic error, such as the degree of the purity of the gas.}
\par
There is not sufficient prior knowledge available to separate and
independently quantify all these factors. However, simply neglecting
this discrepancy leads to a significantly misleading conclusion for
the uncertainty in the cross sections.  Instead, a probabilistic model
is introduced to account for the uncertainty implied by these
discrepancies.  The details of this model and its implications are
discussed further in Section~\ref{sec:bayes}.

\subsection{Electron swarm experiments}\label{subsec:swarm-data}

\todo[inline]{
- describe experiments first, then swarm-derived datasets\\
- incorporate the description from old intro.
}

\subsubsection{Data chracterization}\label{subsubsec:characterization}
\begin{figure}
\input{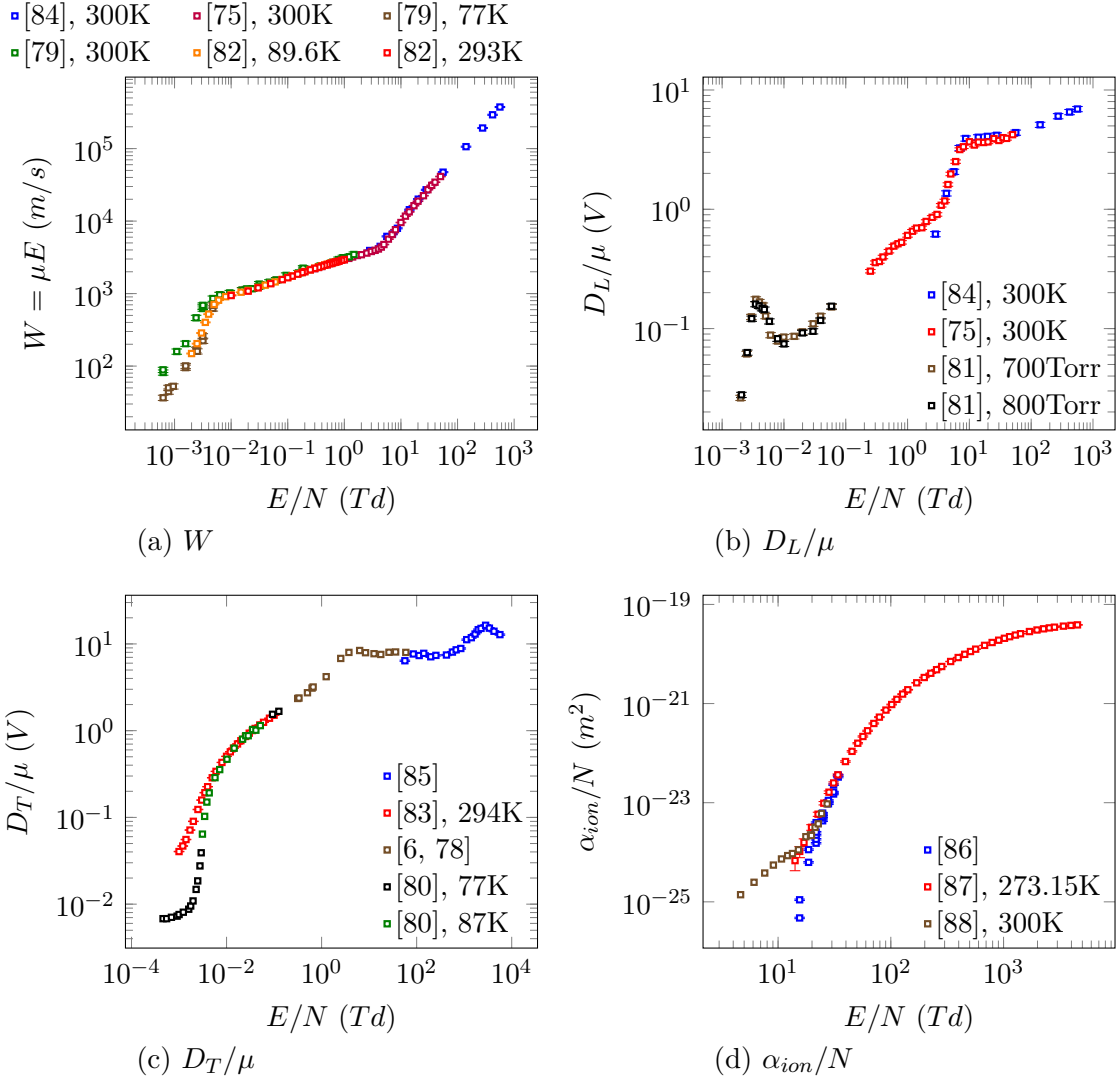}
\caption{
Swarm parameter experiment data used for validation of the calibrated cross sections:
(a) drift velocity,
(b) the ratio of longitudinal diffusivity to mobility,
(c) the ratio of transverse diffusivity to mobility, and
(d) the reduced ionization coefficient.
The error bar, if available, indicates the reported standard deviation of the measurement. 
}
\label{fig:swarm-parameter}
\end{figure}

\begin{figure}
\input{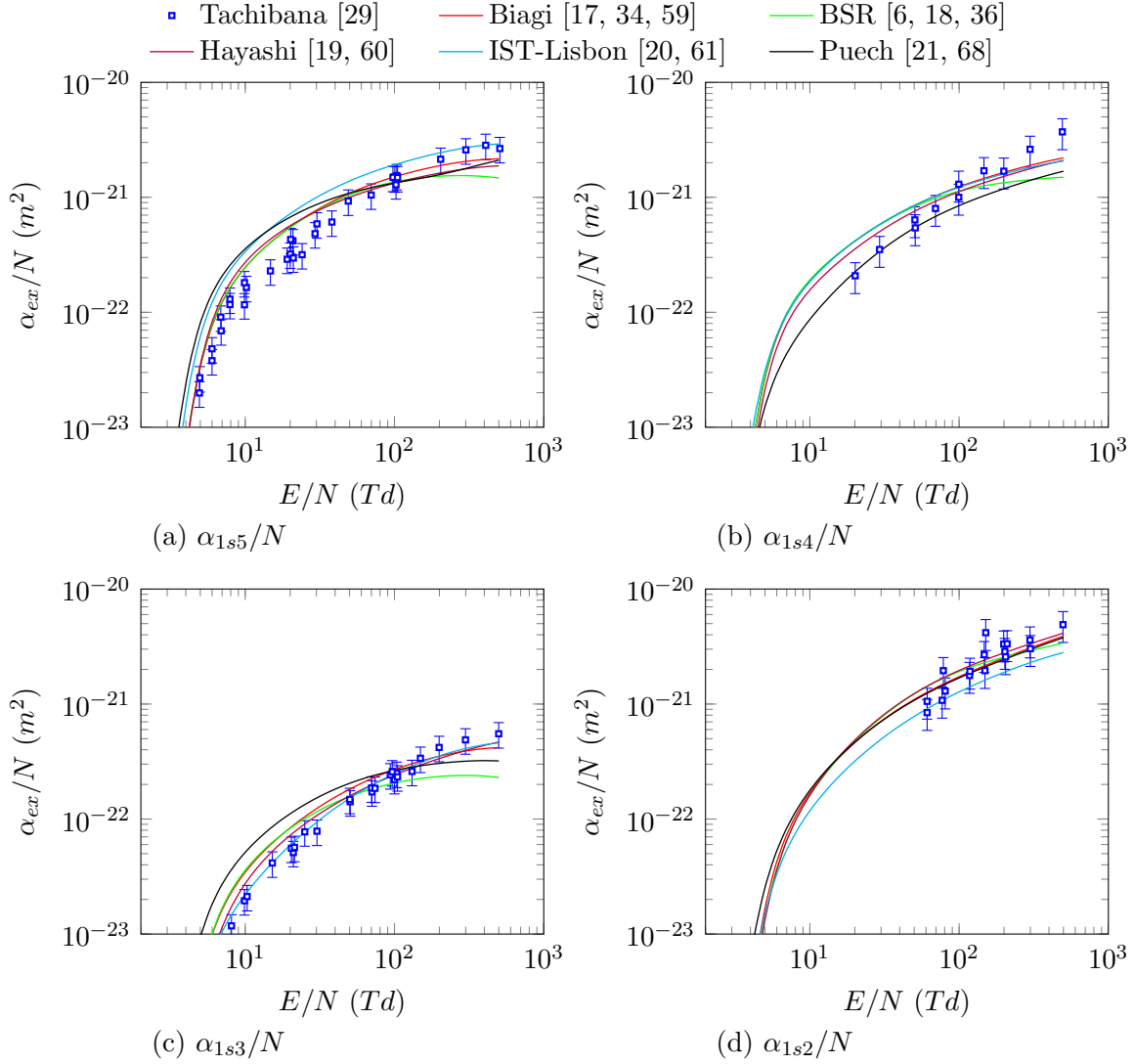}
\caption{
Swarm parameter data used for validation of the calibrated cross sections:
the reduced excitation coefficient to the level (a) $1s_5$, (b) $1s_4$, (c) $1s_3$, and (d) $1s_2$.
The error bar, if available, indicates the reported standard deviation of the measurement. 
}
\label{fig:swarm-parameter-ex}
\end{figure}

Electron swarm experiments measure electron transport parameters in
a population of electrons, which depend on the collision cross sections
\cite{Bederson1971, Gargioni2008, Pitchford2013}.
Table~\ref{tab:swarm-parameter} summarizes the swarm parameter measurements used in this work.  Further, these data
are shown in Figure~\ref{fig:swarm-parameter} and Figure~\ref{fig:swarm-parameter-ex}.
\begin{table}[tbp]
\centering
\begin{tabular}{|c|c|c|c|}
\hline
Type & Reference & Swarm parameters & Condition \\
\hline
\multirow{9}{*}{Transport} & Townsend \& Bailey\refB{*}~\cite{Townsend1922, Pitchford2013} & $D_T / \mu$ & \\\cline{2-4}
                                            & Pack \& Phelps\refB{*}~\cite{Pack1961} & $W = \mu E$ & $77K$, $300K$ \\\cline{2-4}
                                            & Warren \& Parker\refB{*}~\cite{Warren1962} & $D_T / \mu$ & $77K$, $88K$ \\\cline{2-4}
                                            & Robertson \& Rees\refB{*}~\cite{Robertson1972} & $D_L / \mu$ & $90K$, $700Torr$/$800Torr$ \\\cline{2-4}
                                            & Robertson~\cite{Robertson1977} & $W$ & $89.6K$, $293K$ \\\cline{2-4}
                                            & Milloy \& Crompton~\cite{Milloy1977ratio} & $D_T/\mu$ & $294K$ \\\cline{2-4}
                                            & Kucukarpaci \& Lucas\refB{*}~\cite{Kucukarpaci1981} & $W$, $D_L\mu$ & $300K$ \\\cline{2-4}
                                            & Al-Amin \& Lucas~\cite{AlAmin1987} & $D_T/\mu$ & \\\cline{2-4}
                                            & Nakamura \& Kurachi~\cite{Nakamura1988} & $W$, $D_L/N$ & $300K$ \\
\hline
\multirow{3}{*}{Ionization} & Golden \& Fisher\refB{*}~\cite{Golden1961} & \multirow{3}{*}{$\alpha_{ion}$} & \\\cline{2-2}\cline{4-4}
                                        & Kruithof~\cite{Kruithof1940} &  & $273.15K$ \\\cline{2-2}\cline{4-4}
                                        & Specht \textit{et al.}\refB{*}~\cite{Specht1980} &  & $300K$ \\
\hline
\multirow{6}{*}{Excitation} & Tachibana\refB{*}~\cite{Tachibana1986} & \multirow{6}{*}{$\alpha_{ex}$ ($1s_5$--$1s_2$)} & \\\cline{2-2}\cline{4-4}
                                         & Biagi\refB{$\dagger$}~\cite{Biagi1988, Biagi1989, Biagi-data} & & \multirow{5}{*}{$300K$} \\\cline{2-2}
                                         & BSR\refB{$\dagger$}~\cite{Pitchford2013, Zatsarinny2013,BSR-data} & & \\\cline{2-2}
                                         & Hayashi\refB{$\dagger$}~\cite{Hayashi2003, Hayashi-data} & &  \\\cline{2-2}
                                         & IST-Lisbon\refB{$\dagger$}~\cite{Alves2014, IST-Lisbon-data} & &  \\\cline{2-2}
                                         & Puech\refB{$\dagger$}~\cite{Puech1986, Puech-data} & & \\\cline{2-2}
\hline
\end{tabular}
\caption{Summary of swarm parameter experiment data.
\refB{* indicates that the data is digitized from the figures of the cited references.
$\dagger$ indicates that the data is obtained using \texttt{BOLSIG}, using the specified LXCat database.}
}
\label{tab:swarm-parameter}
\end{table}
The figures show that the scatter among datasets for the same conditions is small compared to that of the electron-beam experiments.
While the ionization Townsend coefficient from Specht \textit{et al.}~\cite{Specht1980} deviates from other experiments at low energy,
the authors recognized that in this regime their measurements include contributions from photoelectric emission by the metastables.
\par
The experimental data for the direct excitation rate coefficients is scarce,
since it is difficult to isolate these collision processes from others, especially radiative de-excitations.
While we use the experimental data from Tachibana~\cite{Tachibana1986} who estimated these radiative de-excitation rates from the emission spectroscopy,
we also adopt the swarm-analysis data from the LXCat community~\cite{Pitchford2013, Biagi-data, BSR-data, Hayashi-data, IST-Lisbon-data, Puech-data},
as shown in Figure~\ref{fig:swarm-parameter-ex}.
Here the excitation rate coefficients are obtained using a zero-dimensional Boltzmann solver (\texttt{BOLSIG+}~\cite{Hagelaar2005}),
with the nominal cross section datasets~\cite{Biagi-data, BSR-data, Hayashi-data, IST-Lisbon-data, Puech-data}.
The direct excitation rates to $2p$-levels are also computed from the same datasets,
which are then assumed to be radiatively de-excited to $1s$-levels based on radiative transition probabilities~\cite{Wiese1969}.
This approximation has been used for comparison with the experiment~\cite{Puech1986, Alves2005, Pitchford2013}.

\subsubsection{Swarm-derived cross section datasets}
Another type of widely used cross section data is based on swarm-analysis,
using data from the swarm experiments mentioned above~\cite{Gargioni2008, Pitchford2013}.
Obtaining the cross sections from the transport properties determined in the swarm experiments
requires an inference procedure involving the (usually simplified)
Boltzmann equation, in which many collision processes are active
simultaneously~\cite{Milloy1977, Haddad1982, Nakamura1987, Biagi1989, Alves2005}.
Some widely used cross section datasets from swarm analysis are listed in Table~\ref{tab:lxcat-adoption}.
In this analysis, it is often necessary to employ
prior information on cross sections from other sources, either for a
certain energy range or for a particular collision process
\cite{Pitchford2013}. Uncertainties in the inferred cross sections
arise from the error in measuring the electron transport properties,
errors arising from the simplifications in the analysis and from the
uncertainties in the cross sections from other sources that are adopted
in the analysis. The latter two uncertainties generally go
uncharacterized.
\begin{table}
\centering
\begin{tabular}{|c|c|c|c|}
\hline
Databases & \makecell{Adopted (adjusted)\\ cross sections} & Reference\\
\hline
\multirow{3}{*}{\refB{Biagi}~\cite{Biagi1988, Biagi1989, Biagi-data}} & $\sigma_{el}$ ($<1eV$) & Haddad \& O'Malley~\cite{Haddad1982} \\
                                                                & $\sigma_{ex}$ & BSR~\cite{BSR-data, Zatsarinny2013}\\
                                                                & $\sigma_{ion}$ & Rapp \& Englander-Golden~\cite{Rapp1965}, \cite{Straub1995} \\
\hline
Hayashi~\cite{Hayashi2003, Hayashi-data} & \multicolumn{2}{|c|}{Compilation of 1960 literatures} \\
\hline
\multirow{3}{*}{IST-Lisbon~\cite{Alves2014, IST-Lisbon-data}} & $\sigma_{el}$ & Phelps~\cite{Yamabe1983} \\
                                           & $\sigma_{ex}$ & Hayashi~\cite{Hayashi2003}, \cite{Khakoo2004,Chilton1998,Weber2003,Drawin1967,Lee1973,Bretagne1986,Bretagne1982}\\
                                           & $\sigma_{ion}$ & Rapp \& Englander-Golden~\cite{Rapp1965}\\
\hline
\multirow{1}{*}{Morgan~\cite{Morgan-data}} & not specified. & Phelps and Hayashi$^*$~\cite{Pitchford2017}\\
\hline
\multirow{3}{*}{Puech~\cite{Puech1986, Puech-data}} & $\sigma_{el}$ ($0.014 \le \epsilon \le 4eV$) & Milloy \textit{et al.}~\cite{Milloy1977}\\
                                    & $\sigma_{el}$(otherwise) & Phelps~\cite{Frost1964}\\
                                    & $\sigma_{ex}$, $\sigma_{ion}$ & \cite{Bretagne1986}\\
\hline
\multirow{4}{*}{Phelps \cite{Frost1964,Tachibana1981, Phelps-data}} & \refB{$\sigma_{m,eff}$ ($\sigma_{el}$, $\epsilon<4eV$)} & Milloy \textit{et al.}~\cite{Milloy1977}\\
                                                                    & \refB{$\sigma_{m,eff}$ ($\sigma_{el}$, $\epsilon>8eV$)} & \cite{Fletcher1972}\\
                                                                    & $\sigma_{ex, total}$ & \cite{Schaper1969} \\
                                                                    & $\sigma_{ion}$ & \cite{Smith1930} \\
\hline
\multirow{4}{*}{Nakamura~\cite{Nakamura1988}} & $\sigma_{el}$ ($\epsilon\le2.5eV$) & Milloy \textit{et al.}~\cite{Milloy1977} \\
                                                                                & $\sigma_{el}$ ($\epsilon\ge15eV$) & \cite{Fon1983} \\
                                                                                & $\sigma_{ex, total}$ & \cite{Ferreira1983} \\
                                                                                & $\sigma_{ion}$ & Rapp \& Englander-Golden~\cite{Rapp1965} \\
\hline
\multirow{1}{*}{Haddad (1982) \cite{Haddad1982}}  & $\sigma_{el}$ ($\epsilon>1eV$) & Milloy \textit{et al.}~\cite{Milloy1977} \\
\hline
\end{tabular}
\caption{References of adopted cross sections in swarm-analysis datasets from LXCat community.
Authors of frequently adopted datasets are explicitly specified.
\refB{$\sigma_{m,eff}$ corresponds to the effective momentum-transfer cross section.}}
\label{tab:lxcat-adoption}
\end{table}
\par
\refB{
Due to such collective impacts of many collisions on swarm parameters,
the cross sections inferred from the swarm experiment data are by nature
correlated with each other, and they cannot be used as an individual cross section.
Many of these datasets are developed with particular plasma models in mind
and the individual cross sections are not intended to be used outside of that context~\cite{carbone2021data}.
Thus in the low-temperature plasma community it is strongly recommended
to use the dataset as a whole~\cite{Alves2005,Pitchford2013,Pitchford2017,carbone2021data}.
}

\subsubsection{Implicit, inter-dependent swarm-derived cross sections}

The swarm-derived cross section datasets are valuable assets
for many plasma application/predictions and thus a consideration must be given in our study.
However, for quantifying uncertainty in the cross sections,
they are not readily usable due to many confounding factors in characterizing their uncertainties.
First of all, the cross sections are calibrated in a least-square sense,
and the uncertainties in the experiments are not backward-propagated toward the cross sections.
Thus, most sources only provide the maximum-likelihood estimate of the cross section
and do not attempt to assess its uncertainty.
To our knowledge,
the only swarm-analysis dataset which does provide its uncertainty
is the elastic momentum-transfer cross section at low energy by Milloy \textit{et al.}~\cite{Milloy1977},
where the uncertainty is estimated via a sensitivity analysis.
\par
\begin{figure}
\begin{tikzpicture}[font=\small,]
    \begin{groupplot}[
        group style={
            group name = my plots,
            group size= 2 by 1,
            xlabels at =edge bottom,
            horizontal sep=2.5cm,
            vertical sep=3.5cm,
        },
        name=chung,
        height = 0.4\textwidth,
        width = 0.45\textwidth,
    ]    
\pgfplotsset{set layers=standard}%

        \nextgroupplot[
            ylabel={$\sigma_{el}(\epsilon)$ ($m^2$)},
            xlabel={$\epsilon$ ($eV$)},
            tick scale binop ={\times},
            xmode=log, ymode=log,
            legend style={
                font=\small,
                draw=none, fill=none,
                at={(rel axis cs: 0., 1.0)},
                anchor=south west,
                nodes={scale=1.0},
                legend cell align={left},
                legend columns=4,
                /tikz/every even column/.append style={column sep=0.5cm},
            },
            legend image post style={mark options={scale=1.0, fill=white, line width=1.0}},
        ]
        
           \addplot+ [
                line width=1.0,
                smooth, solid,
                mark=none,
                color=green!50!black,
            ]
            table [x index=0, y index=1]{./data/figure2/Milloy.elastic.txt};
           \addplot+ [
                line width=0.5,
                smooth,
                mark=*,
                only marks,
                mark options={fill=white,solid,scale=0.3,},
            ]
            table [x index=0, y index=1]{./data/figure2/Biagi.raw.elastic.txt};
           \addplot+ [
                line width=0.5,
                smooth,
                mark=*,
                only marks,
                mark options={fill=white,solid,scale=0.3,},
            ]
            table [x index=0, y index=1]{./data/figure2/Hayashi.raw.elastic.txt};
           \addplot+ [
                line width=0.5,
                smooth,
                mark=*,
                only marks,
                mark options={fill=white,solid,scale=0.3,},
            ]
            table [x index=0, y index=1]{./data/figure2/IST-Lisbon.raw.elastic.txt};
           \addplot+ [
                line width=0.5,
                smooth,
                mark=*,
                only marks,
                mark options={fill=white,solid,scale=0.3,},
            ]
            table [x index=0, y index=1]{./data/figure2/Morgan.raw.elastic.txt};
           \addplot+ [
                line width=0.5,
                smooth, solid,
                mark=*,
                only marks,
                mark options={fill=white,solid,scale=0.3,},
                orange,
            ]
            table [x index=0, y index=1]{./data/figure2/Phelps.raw.elastic.txt};
           \addplot+ [
                line width=0.5,
                smooth, solid,
                mark=*,
                only marks,
                mark options={fill=white,solid,scale=0.3,},
                color=burntorange,
            ]
            table [x index=0, y index=1]{./data/figure2/Puech.raw.elastic.txt};
            
            \legend{Milloy~\cite{Milloy1977}, Biagi~\cite{Biagi-data}, Hayashi~\cite{Hayashi-data}, IST-Lisbon~\cite{IST-Lisbon-data}, Morgan~\cite{Morgan-data}, Phelps~\cite{Phelps-data}, Puech~\cite{Puech-data}}
            
            \draw[
                black,
                line width=0.2,
                solid,
            ] (axis cs: 0.014, 1e-22) -- (axis cs: 0.014, 1e-18)
            (axis cs: 4.0, 1e-22) -- (axis cs: 4.0, 1e-18);
            \draw[
                black,
                ->,
                line width=0.5,
                solid,
            ] (rel axis cs: 0.2, 0.3) -- (rel axis cs: 0.3, 0.3);
            \draw[
                black,
                ->,
                line width=0.5,
                solid,
            ]  (rel axis cs: 0.67, 0.3) -- (rel axis cs: 0.57, 0.3);
            
        \nextgroupplot[
            ylabel={$\sigma_{el}(\epsilon)$ ($m^2$)},
            xlabel={$\epsilon$ ($eV$)},
            tick scale binop ={\times},
            xmode=log, ymode=log,
            xmin=0.014, xmax=4.0,
            legend style={
                font=\small,
                draw=none, fill=none,
                at={(rel axis cs: 0.1, 0.1)},
                anchor=south west,
                nodes={scale=1.0},
                legend cell align={left},
                legend columns=1,
                /tikz/every even column/.append style={column sep=0.5cm},
            },
        ]
        
           \addplot+ [
                line width=1.0,
                smooth, solid,
                mark=none,
                color=green!50!black,
            ]
            table [x index=0, y index=1]{./data/figure2/Milloy.elastic.txt};
           \addplot+ [
                line width=0.5,
                smooth,
                mark=*,
                only marks,
                mark options={fill=white,solid,scale=0.3,},
            ]
            table [x index=0, y index=1]{./data/figure2/Biagi.raw.elastic.txt};
           \addplot+ [
                line width=0.5,
                smooth,
                mark=*,
                only marks,
                mark options={fill=white,solid,scale=0.3,},
            ]
            table [x index=0, y index=1]{./data/figure2/Hayashi.raw.elastic.txt};
           \addplot+ [
                line width=0.5,
                smooth,
                mark=*,
                only marks,
                mark options={fill=white,solid,scale=0.3,},
            ]
            table [x index=0, y index=1]{./data/figure2/IST-Lisbon.raw.elastic.txt};
           \addplot+ [
                line width=0.5,
                smooth,
                mark=*,
                only marks,
                mark options={fill=white,solid,scale=0.3,},
            ]
            table [x index=0, y index=1]{./data/figure2/Morgan.raw.elastic.txt};
           \addplot+ [
                line width=0.5,
                smooth, solid,
                mark=*,
                only marks,
                mark options={fill=white,solid,scale=0.3,},
                orange,
            ]
            table [x index=0, y index=1]{./data/figure2/Phelps.raw.elastic.txt};
           \addplot+ [
                line width=0.5,
                smooth, solid,
                mark=*,
                only marks,
                mark options={fill=white,solid,scale=0.3,},
                color=burntorange,
            ]
            table [x index=0, y index=1]{./data/figure2/Puech.raw.elastic.txt};
            
 
  \end{groupplot}
 \node[below = 1.5cm of my plots c1r1.south west,
            anchor=west,
        ] {(a)};
\node[below = 1.5cm of my plots c2r1.south west,
            anchor=west,
        ] {(b) $0.014eV$ -- $4eV$};
\end{tikzpicture}
\caption{
Elastic momentum-transfer cross sections from swarm-analysis datasets:
(a) on the overall energy range, and
(b) on $\epsilon\in[0.014, 4]$ ($eV$),
to show the agreement with the data from Milloy \textit{et al.}~\cite{Milloy1977}.
}
\label{fig:lxcat-elastic}
\end{figure}
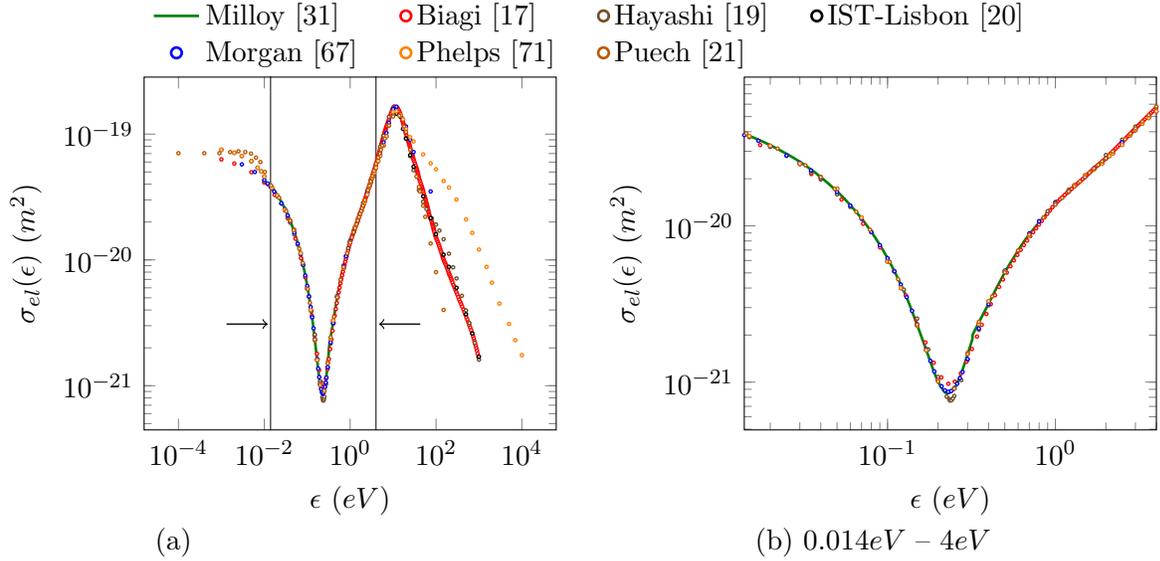
A more fundamental challenge in using the swarm-derived datasets for uncertainty quantification lies in their inter-dependency.
The swarm parameters depend on the electron energy distribution function, which is determined by the entire set of collision processes.
The set of collision processes that can be inferred from the transport/chemistry data is not unique,
unless the experimental condition reduces the number of possible collision processes~\cite{Crompton1994, Gargioni2008}.
This is possible only for the elastic momentum-transfer processes in the low-energy range for noble gases~\cite{Crompton1994, Gargioni2008, Milloy1977}.
A common practice for other collision processes is to calibrate only a target collision process, while the other necessary cross sections are taken from the literature, sometimes with mild adjustments.
As a result, the cross sections from swarm analysis datasets often show small discrepancies among each other;
\refC{however,} this result does not necessarily imply that the actual uncertainty is small\refC{,
let alone that the correlation with other collision processes must be considered.}
The elastic momentum-transfer cross section shown in Figure~\ref{fig:lxcat-elastic} is one stark example.
While these datasets are mildly scattered either for $\epsilon < 0.014 eV$ or for $\epsilon > 4 eV$,
all of them are strikingly well aligned within $\epsilon\in[0.014, 4]$ interval, except a small variation at the Ramsauer minimum.
While it may be tempting to say that the cross section in this range has an extremely low uncertainty,
it is rather the result of everyone adopting and mildly adjusting the dataset from Milloy \textit{et al}~\cite{Milloy1977}.
Table~\ref{tab:lxcat-adoption} shows these swarm-analysis datasets which are known to have adopted Milloy \textit{et al.}~\cite{Milloy1977} with mild adjustments.
Employing all these datasets independently will induce a strong, artificial bias to that from Milloy~\textit{et al}~\cite{Milloy1977}.
While the elastic collision may be considered accurate in this range,
almost all other cross sections---which are not deemed as accurate as the momentum-transfer---are also adopted from the literature as shown in Table~\ref{tab:lxcat-adoption},
thus having an inter-dependency with each other.
Such bias is much harder to characterize than the systematic errors
we observe from the electron-beam experiements in Figure~\ref{fig:crs-curation}.
\par
For these reasons,
\refC{
the swarm-derived cross section datasets are not readily usable for uncertainty quantification,
despite their usefulness in plasma modeling and prediction.
}
The swarm-derived cross section datasets are thus not used for calibration in this work.
Instead, as mentioned in Section~\ref{subsubsec:characterization},
they are used to obtain the excitation rate coefficients as part of
validation datasets~\cite{Biagi-data, BSR-data, Hayashi-data, IST-Lisbon-data, Puech-data}.

\subsubsection{Swarm-parameter experiments as a validation case}
In principle,
the swarm parameter experiment data in Table~\ref{tab:swarm-parameter}
can be also used for Bayesian inference.
This requires an inverse problem of finding the semi-empirical cross section models,
whose outputs through the Boltzmann equation match the swarm parameter measurements.
This is in essence a reproduction of the conventional swarm-analysis, but in a more comprehensive way:
the entire set of cross sections would be calibrated together;
and the associated uncertainty would be evaluated,
as well as their most credible values.
\par
\refB{
However, despite the small discrepancies among the measurements,
including the swarm experiments in the calibration
can bring many complications and limitations in the resulting uncertain cross sections.
First, calibrating uncertainties in cross sections as well as their most credible values}
involves much larger statistical sampling than simply fitting the models,
even with a simplified Boltzmann model such as \texttt{BOLSIG+}~\cite{Hagelaar2005}.
\refB{
Second,
just as for the swarm-analysis,
the uncertainty is given for the entire cross section dataset as a whole,
being correlated among individual cross sections.
This limits the use of the calibrated uncertainties
where some transitions or states of a species are replaced with a lumped transition or an effective state,
respectively.
Furthermore,
the cross sections (and their uncertainties) are calibrated under a zero-dimensional Boltzmann equation model,
which are deemed accurate only under certain experimental conditions.
Thus the calibrated uncertainties either can be used only in such context,
or the model form error of the Boltzmann model must be accounted in terms of uncertainty,
which involves additional uncertainty modeling and calibration procedures.
}
\par
In order to avoid these potential complications,
it is decided to use only the direct cross section measurements in Table~\ref{fig:crs-curation}
for the Bayesian inference procedure.
The swarm parameter experiment data in Table~\ref{tab:swarm-parameter}
are instead used for validation of the calibrated models resulting from the Bayesian inference.
It will be shown in Section~\ref{sec:result}
that the uncertainty calibrated only from the electron beam experiments and the \textit{ab initio} calculations
is sufficient to explain the scatter in the swarm parameter experiments.
Furthermore, the resulting swarm paramter predictions match very well overall with the experiments.

\section{Semi-empirical cross section models}\label{sec:model}

We observe from Figure~\ref{fig:crs-curation} that, for the most part,
the cross sections vary smoothly in the energy ($\epsilon$) space.
This implies that cross sections values are correlated in a neighborhood
of $\epsilon$-space---the closer the more strongly, presumably---while
their reported measurement errors might be assumed independent.
Further, it suggests that the cross section
behavior can be adequately described with a low-dimensional parameter
set.
\par
To take advantage of this behavior,
we represent the cross sections $\sigma(\epsilon)$ with semi-empirical models $f(\epsilon;\vec{A})$, where $\vec{A}$ denotes the parameters of the model.
The parametric uncertainty for $\vec{A}$ represents the uncertainty of the cross section.
The semi-empirical models are based on theoretical arguments combined with a few empirical factors~\cite{Gargioni2008,Haddad1982,Bretagne1986,Kim1994}.
These model forms are intended to capture the essential features of the cross section using a
a relatively low-dimensional set of uncertain parameters.
\par
While some cross sections---excitations in particular---exhibit rich
fine-scale structures~\cite{Buckman1983, Zatsarinny2013}, most
datasets do not have the resolution necessary to support inference of
these features.  Further, the plasma transport and reaction properties
that are the primary quantities of interest in this work are integrals
over energy, and are therefore insensitive to these fine features in
the cross sections.  This insensitivity is demonstrated via
sensitivity analysis in \ref{app:sensitivity}. Thus, even when some
small-scale features are present---and a richer model would be
necessary to capture these features---we rely upon low-dimensional,
semi-empirical models.


\subsection{Elastic momentum-transfer}\label{subsec:model-elastic}
For low electron-impact energy ($<1eV$),
the modified effective range theory (MERT) model has been widely used for the elastic momentum-transfer cross section~\cite{OMalley1963, Milloy1977, Haddad1982}.
The cross section is expressed as partial wave expansions,
\begin{subequations}\label{eq:mert}
\begin{equation}
f_{MERT}(\epsilon; \vec{A}_{MERT}) = \frac{4\pi}{k^2} \sum_{L=0}^{\infty} \refC{(L+1)} \sin^2(\eta_L - \eta_{L+1}),
\end{equation}
where $k^2 = (2m_e/\hbar^2)\epsilon$ with the electron mass $m_e$ and the reduced Planck constant $\hbar$.
The MERT formula determines the electron-atom scattering phase shift $\eta_L$ with 4 empirical factors $\vec{A}_{MERT} = \{A, D, F, \epsilon_1\}$,
\begin{equation}
k^{-1}\tan\eta_0 = -A\left[ 1 + \frac{4\alpha}{3a_0}k^2\log(ka_0) \right]
- \frac{\pi\alpha}{3a_0}k + D k^\refC{2} + F k^\refC{3},
\end{equation}
\begin{equation}
k^{-1}\tan\eta_1 = \frac{\pi}{15}\alpha k\left(1 - \sqrt{\frac{\epsilon}{\epsilon_1}}\right)
\end{equation}
\begin{equation}
k^{-1}\tan\eta_L = \frac{\pi\alpha k}{(2L+3)(2L+1)(2L-1)a_0}
\qquad \text{for } L>1,
\end{equation}
\end{subequations}
with $a_0=5.29177\times10^{-11}m$ the Bohr radius and $\alpha = 11.08a_0^3$ the static polarizability of argon atom.
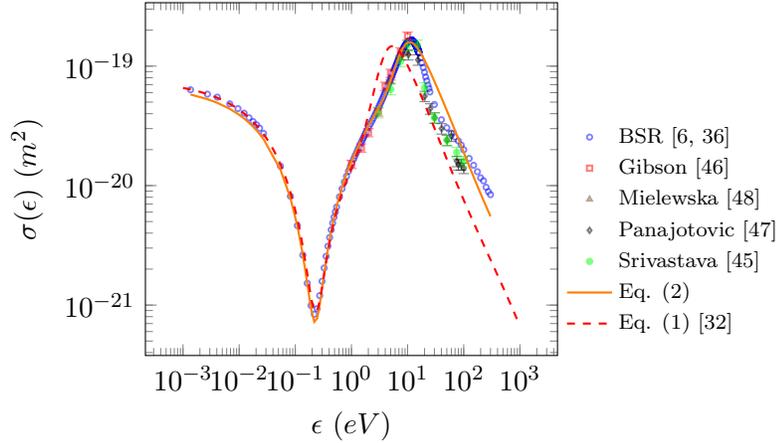
\begin{figure}
\centering
\begin{tikzpicture}[font=\small,]
    \begin{groupplot}[
        group style={
            group name = my plots,
            group size= 1 by 1,
            xlabels at =edge bottom,
            horizontal sep=2.2cm,
            vertical sep=2.2cm,
        },
        name=chung,
        height = 0.4\textwidth,
        width = 0.45\textwidth,
    ]    
\pgfplotsset{set layers=standard}%

        \nextgroupplot[
            enlarge x limits={true, abs value = 5mm},
            ylabel={$\sigma(\epsilon)$ ($m^2$)},
            xlabel={$\epsilon$ ($eV$)},
            tick scale binop ={\times},
            xmode=log, ymode=log,
            legend style={
                font=\scriptsize,
                draw=none, fill=none,
                at={(rel axis cs: 1.0, 0.03)},
                anchor=south west,
                nodes={scale=1.0},
                legend cell align={left},
                legend columns=1,
                /tikz/every even column/.append style={column sep=0.5cm},
            },
            clip mode=individual,
        ]
        
           \addplot+ [
               opacity=0.5,
                line width=0.7,
                mark=*,
                mark options={fill=white,solid,scale=0.5,},
                only marks,
            ]
            table [x index=0, y index=1]{./data/figure1/momentum/BSR.momentum.txt};
           \addplot+ [
               opacity=0.5,
                line width=0.8,
                mark=square*,
                mark options={fill=white,solid,scale=0.5,},
                only marks,
                /pgfplots/error bars/.cd,
                    x dir=none,
                    y dir=both,
                    y explicit,
                    error mark={-}
            ]
            table [x index, y index=1,y error expr=\thisrowno{2}]{./data/figure1/momentum/Gibson1996.momentum.txt};
           \addplot+ [
               opacity=0.5,
                line width=0.8,
                mark=triangle*,
                mark options={fill=white,solid,scale=0.5,},
                only marks,
                /pgfplots/error bars/.cd,
                    x dir=none,
                    y dir=both,
                    y explicit,
                    error mark={-}
            ]
            table [x index=0, y index=1,y error expr=\thisrowno{2}]{./data/figure1/momentum/Mielewska2004.momentum.txt};
           \addplot+ [
               opacity=0.5,
                line width=0.8,
                mark=diamond*,
                mark options={fill=white,solid,scale=0.5,},
                only marks,
                /pgfplots/error bars/.cd,
                    x dir=none,
                    y dir=both,
                    y explicit,
                    error mark={-}
            ]
            table [x index=0, y index=1,y error expr=\thisrowno{2}]{./data/figure1/momentum/Panajotovic1997.momentum.txt};
           \addplot+ [
               opacity=0.5,
                line width=0.8,
                mark=*,
                mark options={fill=green,solid,scale=0.5,},
                color=green,
                only marks,
                /pgfplots/error bars/.cd,
                    x dir=none,
                    y dir=both,
                    y explicit,
                    error mark={-}
            ]
            table [x index=0, y index=1,y error expr=\thisrowno{2}]{./data/figure1/momentum/Srivastava1981.momentum.txt};

           \addplot+ [
                line width=0.8,
                mark=none,
                color=orange,
                solid,
            ]
            table [x index=0, y index=1]{./data/figure4/Elastic.MERT_shifted.txt};            
           \addplot+ [
                line width=0.8,
                mark=none,
                color=red,
                dashed,
            ]
            table [x index=0, y index=1]{./data/figure4/Elastic.MERT.txt};
            
            \legend{BSR~\cite{Pitchford2013, Zatsarinny2013}, Gibson~\cite{Gibson1996}, Mielewska~\cite{Mielewska2004}, Panajotovic~\cite{Panajotovic1997}, Srivastava~\cite{Srivastava1981}, {Eq. (\ref{eq:mert-shift})}, {Eq. (\ref{eq:mert})~\cite{Haddad1982}}}
         
  \end{groupplot}
\end{tikzpicture}
%
\caption{
The semi-empirical model (\ref{eq:mert-shift}) for the elastic momentum-transfer cross section,
together with the electron-beam experiment datasets.
The parameters are calibrated in the least-square sense:
$A=-1.4175$, $D=62.5779$, $F=-85.2362$, $\epsilon_1=0.7364$, $t_1=0.4463$, $t_2=1.6514$, and $t_3=1.9992$.
The MERT model (\ref{eq:mert}) by Haddad and O'Malley~\cite{Haddad1982} is also included for a reference.
}
\label{fig:elastic-model}
\end{figure}
Figure~\ref{fig:elastic-model} shows the MERT model calibrated by Haddad and O'Malley~\cite{Haddad1982},
which agrees well with BSR cross section data up to $1 eV$ and deviates from the electron-beam experiment data for the larger energy range.
\par
We note that, while the MERT model deviates for $>1eV$,
it exhibits a qualitatively similar trend with the electron-beam data,
and that it only requires a minor shift along $\epsilon$-axis.
Thus we designed an extension to the MERT model,
\begin{subequations}\label{eq:mert-shift}
\begin{equation}
f_{el}(\epsilon; \vec{A}_{el}) = f_{MERT}(\tilde{\epsilon}; \vec{A}_{MERT}),
\end{equation}
where $\vec{A}_{el}=\{\vec{A}_{MERT}, t_1, t_2, t_3\}$ includes three additional parameters,
which are used to determine the shift factor applied to $\epsilon$,
\begin{equation}\label{eq:mert-shift-factor}
\tilde{\epsilon} = \epsilon\left[ \frac{1 + t_1}{2} - \frac{1 - t_1}{2} \tanh\left(\frac{\epsilon - t_2}{t_3}\right) \right].
\end{equation}
\end{subequations}
In essence, (\ref{eq:mert-shift-factor}) amplifies $\epsilon$ by a factor of $t_1$ if $\epsilon > t_2$.
Figure~\ref{fig:elastic-model} shows the model with $\vec{A}_{el}$ least-square-fitted against the BSR dataset,
which seems flexible enough to capture the overall trend of the momentum-transfer cross section.
This model approximates the high-energy ($>10eV$) range with only $\cO(\epsilon^{-1})$ asymptote,
neglecting smaller-scale variations.
However, a sensitivity analysis using a zero-dimensional Boltzmann solver (\texttt{BOLSIG+}~\cite{Hagelaar2005})
indicates that the resulting swarm parameters are insensitive to these variations (see \ref{app:sensitivity}).

\subsection{Ionization}
Kim and Rudd~\cite{Kim1994, Gargioni2008} suggested a simple empirical form for the total ionization cross section
based on the binary-encounter-dipole model,
\begin{equation}\label{eq:ion}
f_i(\epsilon; \vec{A}_i) = \frac{4\pi a_0^2}{\epsilon / \epsilon_i}\left[a\log\frac{\epsilon}{\epsilon_i} + b\left(1 - \frac{\epsilon_i}{\epsilon}\right) + c\frac{\log(\epsilon/\epsilon_i)}{\epsilon/\epsilon_i + 1}\right],
\end{equation}
with $\epsilon_i = 15.760eV$ the first ionization energy of argon and $\vec{A}_i = \{a, b, c\}$ the model parameters for $f_i$.
\begin{figure}
\centering
\begin{tikzpicture}[font=\small,]
    \begin{groupplot}[
        group style={
            group name = my plots,
            group size= 1 by 1,
            xlabels at =edge bottom,
            horizontal sep=2.2cm,
            vertical sep=2.2cm,
        },
        name=chung,
        height = 0.4\textwidth,
        width = 0.45\textwidth,
    ]    
\pgfplotsset{set layers=standard}%

        \nextgroupplot[
            enlarge x limits={true, abs value = 5mm},
            ylabel={$\sigma(\epsilon)$ ($m^2$)},
            xlabel={$\epsilon - \epsilon_0$ ($eV$)},
            tick scale binop ={\times},
            xmode=log, ymode=log,
            ymin=1e-22,
            legend style={
                font=\scriptsize,
                draw=none, fill=none,
                at={(rel axis cs: 1.0, 0.)},
                anchor=south east,
                nodes={scale=1.0},
                legend cell align={left},
                legend columns=1,
                /tikz/every even column/.append style={column sep=0.5cm},
            },
            clip mode=individual,
        ]
        
           \addplot+ [
               opacity=0.5,
                line width=0.7,
                mark=*,
                mark options={fill=white,solid,scale=0.5,},
                only marks,
                /pgfplots/error bars/.cd,
                    x dir=none,
                    y dir=both,
                    y explicit,
                    error mark={-}
            ]
            table [x expr=\thisrowno{0}-\ionE, y index=1, y error expr=\thisrowno{2}]{./data/figure1/ion/Rapp1965.ion.total.txt};
           \addplot+ [
               opacity=0.5,
                line width=0.8,
                mark=square*,
                mark options={fill=white,solid,scale=0.5,},
                only marks,
                /pgfplots/error bars/.cd,
                    x dir=none,
                    y dir=both,
                    y explicit,
                    error mark={-}
            ]
            table [x expr=\thisrowno{0}-\ionE, y index=1,y error expr=\thisrowno{2}]{./data/figure1/ion/Straub1995.ion.1+.txt};
           \addplot+ [
               opacity=0.5,
                line width=0.8,
                mark=triangle*,
                mark options={fill=white,solid,scale=0.5,},
                only marks,
                /pgfplots/error bars/.cd,
                    x dir=none,
                    y dir=both,
                    y explicit,
                    error mark={-}
            ]
            table [x expr=\thisrowno{0}-\ionE, y index=1,y error expr=\thisrowno{2}]{./data/figure1/ion/Straub1995.ion.total.txt};
           \addplot+ [
               opacity=0.5,
                line width=0.8,
                mark=diamond*,
                mark options={fill=white,solid,scale=0.5,},
                only marks,
                /pgfplots/error bars/.cd,
                    x dir=none,
                    y dir=both,
                    y explicit,
                    error mark={-}
            ]
            table [x expr=\thisrowno{0}-\ionE, y index=1,y error expr=\thisrowno{2}]{./data/figure1/ion/Wetzel1987.ion.1+.txt};
           \addplot+ [
               opacity=0.5,
                line width=0.8,
                mark=*,
                mark options={fill=green,solid,scale=0.5,},
                color=green,
                only marks,
                /pgfplots/error bars/.cd,
                    x dir=none,
                    y dir=both,
                    y explicit,
                    error mark={-}
            ]
            table [x expr=\thisrowno{0}-\ionE, y index=1,y error expr=\thisrowno{2}]{./data/figure1/ion/Wetzel1987.ion.total.txt};

           \addplot+ [
                line width=0.8,
                mark=none,
                color=orange,
                solid,
            ]
            table [x expr=\thisrowno{0}-\ionE, y index=1]{./data/figure5/Ionization.BED.txt};    
            
            \legend{Rapp~\cite{Rapp1965}, {Straub~\cite{Straub1995}, $1+$}, {Straub~\cite{Straub1995}, total}, {Wetzel~\cite{Wetzel1987}, $1+$}, {Wetzel~\cite{Wetzel1987}, total}, {Eq. (\ref{eq:ion})}}
         
  \end{groupplot}
\end{tikzpicture}
%
\caption{
The semi-empirical model (\ref{eq:ion}) for the total ionization cross section,
together with the electron-beam experiment datasets.
The parameters are calibrated in the least-square sense:
$a=4.4765$, $b=-2.3637$, and $c=-3.4470$.
}
\label{fig:ion-model}
\end{figure}
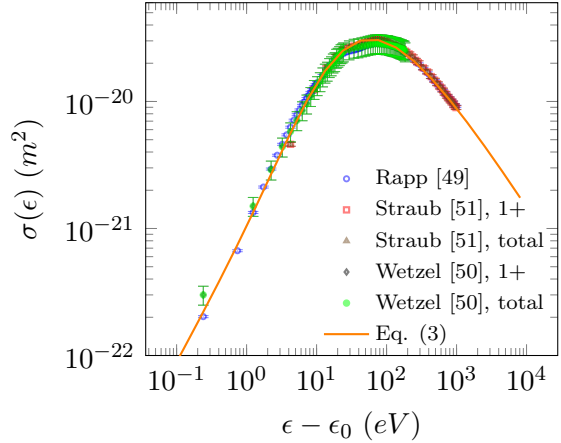
Figure~\ref{fig:ion-model} shows the empirical model evaluated using least-squares-fitted parameters, compared with the electron-beam datasets.

\subsection{Excitation---metastables}
For $1s_5$, $1s_3$ excitation cross sections,
we employ a semi-empirical model suggested by Bretagne \textit{et al.}~\cite{Bretagne1986},
\begin{equation}\label{eq:metastable}
f_{me}(\epsilon; \vec{A}_{me}) = \frac{4\pi a_0^2R}{\epsilon}b_{me}\left(1 - \frac{\epsilon_{me}}{\epsilon}\right)\left(2\epsilon\right)^{-\gamma_{me}},
\quad
me = 1s_5, 1s_3,
\end{equation}
with $R = 13.606eV$ the Rydberg energy,
$\epsilon_{1s5} = 11.548eV$, $\epsilon_{1s3} = 11.723eV$ the excitation energies for $1s_5$ and $1s_3$ levels, respectively,
and the model parameters $\vec{A}_{me} = \{b_{me}, \gamma_{me}\}$.
The first term in (\ref{eq:metastable}) corresponds to the large-energy asymptotic limit from Bethe-Born approximation,
while the last two terms are the empirical factors for the low-energy range.
\begin{figure}
\centering
\begin{tikzpicture}[font=\small,]
    \begin{groupplot}[
        group style={
            group name = my plots,
            group size= 2 by 1,
            xlabels at =edge bottom,
            horizontal sep=2.4cm,
            vertical sep=2.2cm,
        },
        name=chung,
        height = 0.4\textwidth,
        width = 0.45\textwidth,
    ]    
\pgfplotsset{set layers=standard}%

        \nextgroupplot[
            enlarge x limits={true, abs value = 5mm},
            ylabel={$\sigma(\epsilon)$ ($m^2$)},
            xlabel={$\epsilon - \epsilon_0$ ($eV$)},
            tick scale binop ={\times},
            xmode=log, ymode=log,
            legend style={
                font=\scriptsize,
                draw=none, fill=none,
                at={(rel axis cs: 0.0, 1.0)},
                anchor=south west,
                nodes={scale=1.0},
                legend cell align={left},
                legend columns=3,
                /tikz/every even column/.append style={column sep=0.5cm},
            },
        ]
        
           \addplot+ [
                line width=0.7,
                mark=*,
                mark options={fill=white,solid,scale=0.5,},
                only marks,
            ]
            table [x expr=\thisrowno{0}-\exOne, y index=1]{./data/figure1/1s5/BSR.1s5.txt};
           \addplot+ [
                line width=0.8,
                mark=square*,
                mark options={fill=white,solid,scale=0.5,},
                only marks,
                /pgfplots/error bars/.cd,
                    x dir=none,
                    y dir=both,
                    y explicit,
                    error mark={-}
            ]
            table [x expr=\thisrowno{0}-\exOne, y index=1,y error expr=\thisrowno{2}]{./data/figure1/1s5/Schappe1994.1s5.txt};
           \addplot+ [
                line width=0.8,
                mark=triangle*,
                mark options={fill=white,solid,scale=0.5,},
                only marks,
                /pgfplots/error bars/.cd,
                    x dir=none,
                    y dir=both,
                    y explicit,
                    error mark={-}
            ]
            table [x expr=\thisrowno{0}-\exOne, y index=1,y error expr=\thisrowno{2}]{./data/figure1/1s5/Chutjian1981.1s5.txt};
           \addplot+ [
                line width=0.8,
                mark=diamond*,
                mark options={fill=white,solid,scale=0.5,},
                only marks,
                /pgfplots/error bars/.cd,
                    x dir=none,
                    y dir=both,
                    y explicit,
                    error mark={-}
            ]
            table [x expr=\thisrowno{0}-\exOne, y index=1,y error expr=\thisrowno{2}]{./data/figure1/1s5/Filipovic2000b.1s5.txt};
           \addplot+ [
                line width=0.8,
                mark=*,
                mark options={fill=green,solid,scale=0.5,},
                color=green,
                only marks,
                /pgfplots/error bars/.cd,
                    x dir=none,
                    y dir=both,
                    y explicit,
                    error mark={-}
            ]
            table [x expr=\thisrowno{0}-\exOne, y index=1,y error expr=\thisrowno{2}]{./data/figure1/1s5/Khakoo2004.1s5.txt};
            
           \addplot+ [
                line width=1.0,
                mark=none,
                color=burntorange,
                smooth, solid,
            ]
            table [x expr=\thisrowno{0}-\exOne, y index=1,]{./data/figure6/1s5.excitation.model.ref.txt};
   
            \legend{BSR~\cite{Pitchford2013, Zatsarinny2013}, Schappe~\cite{Schappe1994}, Chutjian~\cite{Chutjian1981}, Filipovic~\cite{Filipovic2000b}, Khakoo~\cite{Khakoo2004}, Eq. (\ref{eq:metastable})}
            
        \nextgroupplot[
            enlarge x limits={true, abs value = 5mm},
            ylabel={$\sigma(\epsilon)$ ($m^2$)},
            xlabel={$\epsilon - \epsilon_0$ ($eV$)},
            tick scale binop ={\times},
            xmode=log, ymode=log,
            ymin=1e-26,
            legend style={
                font=\scriptsize,
                draw=none, fill=none,
                at={(rel axis cs: 0.0, 1.0)},
                anchor=south west,
                nodes={scale=1.0},
                legend cell align={left},
                legend columns=2,
                /tikz/every even column/.append style={column sep=0.5cm},
            },
        ]
        
           \addplot+ [
                line width=0.7,
                mark=*,
                mark options={fill=white,solid,scale=0.5,},
                only marks,
            ]
            table [x expr=\thisrowno{0}-\exThree, y index=1]{./data/figure1/1s3/BSR.1s3.txt};
           \addplot+ [
                line width=0.8,
                mark=square*,
                mark options={fill=white,solid,scale=0.5,},
                only marks,
                /pgfplots/error bars/.cd,
                    x dir=none,
                    y dir=both,
                    y explicit,
                    error mark={-}
            ]
            table [x expr=\thisrowno{0}-\exThree, y index=1,y error expr=\thisrowno{2}]{./data/figure1/1s3/Schappe1994.1s3.txt};
           \addplot+ [
                line width=0.8,
                mark=triangle*,
                mark options={fill=white,solid,scale=0.5,},
                only marks,
                /pgfplots/error bars/.cd,
                    x dir=none,
                    y dir=both,
                    y explicit,
                    error mark={-}
            ]
            table [x expr=\thisrowno{0}-\exThree, y index=1,y error expr=\thisrowno{2}]{./data/figure1/1s3/Chutjian1981.1s3.txt};
           \addplot+ [
                line width=0.8,
                mark=diamond*,
                mark options={fill=white,solid,scale=0.5,},
                only marks,
                /pgfplots/error bars/.cd,
                    x dir=none,
                    y dir=both,
                    y explicit,
                    error mark={-}
            ]
            table [x expr=\thisrowno{0}-\exThree, y index=1,y error expr=\thisrowno{2}]{./data/figure1/1s3/Filipovic2000b.1s3.txt};
           \addplot+ [
                line width=0.8,
                mark=*,
                mark options={fill=green,solid,scale=0.5,},
                color=green,
                only marks,
                /pgfplots/error bars/.cd,
                    x dir=none,
                    y dir=both,
                    y explicit,
                    error mark={-}
            ]
            table [x expr=\thisrowno{0}-\exThree, y index=1,y error expr=\thisrowno{2}]{./data/figure1/1s3/Khakoo2004.1s3.txt};
            
           \addplot+ [
                line width=1.0,
                mark=none,
                color=burntorange,
                smooth, solid,
            ]
            table [x expr=\thisrowno{0}-\exThree, y index=1,]{./data/figure6/Bretagne.1s3.model.txt};
            
         
  \end{groupplot}
 \node[below = 2cm of my plots c1r1.south west,
            anchor=west,
        ] {(a) Excitation, $1s_5$};
\node[below = 2cm of my plots c2r1.south west,
            anchor=west,
        ] {(b) Excitation, $1s_3$};
\end{tikzpicture}
%
\caption{
The semi-empirical model (\ref{eq:metastable}) for the (a) $1s_5$ and (b) $1s_3$ level excitation cross section,
together with the electron-beam experiment datasets.
}
\label{fig:metastable-model}
\end{figure}
Figure~\ref{fig:metastable-model} shows the models with the literature values from Bretagne \textit{et al.}~\cite{Bretagne1986}.
The trends agree well with the measurement data.
The model, of course, cannot capture the fine resonance structures~\cite{Buckman1983} and the small-scale behaviors for $\epsilon - \epsilon_0 < 10^{-1} eV$,
which are observed in the BSR dataset.
A sensitivity analysis via \texttt{BOLSIG+}~\cite{Hagelaar2005}, however,
revealed that the plasma chemistry and transport properties are insensitive to these fine features (see \ref{app:sensitivity}).

\subsection{Excitation---resonances}
For resonance level excitations $1s_4$ and $1s_2$,
we employ a similar model to that from Bretagne \textit{et al.}~\cite{Bretagne1986} which accounts for the relativistic effect for electrons,
\begin{equation}\label{eq:resonance}
f_{r}(\epsilon; \vec{A}_r) = \frac{4\pi a_0^2R^2}{\epsilon}\frac{F_{0,r}}{\epsilon_r}\left[\log\left(\frac{\epsilon}{(1 - \beta^2)\epsilon_r}\right) - \beta^2\right]\exp\left[-\frac{\gamma_r}{1 + \epsilon/\epsilon_r}\right],
\quad
r = 1s_4, 1s_2,
\end{equation}
where $\beta = \sqrt{2\epsilon/m_ec^2}$ the relativistic factor with the electron mass $m_e$ and the speed of light $c$,
and the model parameters $\vec{A}_r = \{F_{0,r}, \gamma_r\}$.
The excitation energies $\epsilon_{1s4} = 11.624eV$ and $\epsilon_{1s2} = 11.828eV$ are chosen from the NIST atomic spectra database~\cite{Kramida2022}.
The difference of (\ref{eq:resonance}) from the model by Bretagne \textit{et al.}~\cite{Bretagne1986} is the last empirical factor,
which adjusts the low-energy behavior.
This adjustment is similar to the low-energy empirical factor suggested by Paretzke~\cite{Gargioni2008}.
\begin{figure}
\centering
\begin{tikzpicture}[font=\small,]
    \begin{groupplot}[
        group style={
            group name = my plots,
            group size= 2 by 1,
            xlabels at =edge bottom,
            horizontal sep=2.4cm,
            vertical sep=2.2cm,
        },
        name=chung,
        height = 0.4\textwidth,
        width = 0.45\textwidth,
    ]    
\pgfplotsset{set layers=standard}%

        \nextgroupplot[
            enlarge x limits={true, abs value = 5mm},
            ylabel={$\sigma(\epsilon)$ ($m^2$)},
            xlabel={$\epsilon - \epsilon_0$ ($eV$)},
            tick scale binop ={\times},
            xmode=log, ymode=log,
            ymin=1e-23,
            legend style={
                font=\scriptsize,
                draw=none, fill=none,
                at={(rel axis cs: 0., 1.0)},
                anchor=south west,
                nodes={scale=1.0},
                legend cell align={left},
                legend columns=3,
                /tikz/every even column/.append style={column sep=0.5cm},
            },
        ]
        
           \addplot+ [
                line width=0.7,
                mark=*,
                mark options={fill=white,solid,scale=0.5,},
                only marks,
            ]
            table [x expr=\thisrowno{0}-\exTwo, y index=1]{./data/figure1/1s4/BSR.1s4.txt};
           \addplot+ [
                line width=0.8,
                mark=square*,
                mark options={fill=white,solid,scale=0.5,},
                only marks,
                /pgfplots/error bars/.cd,
                    x dir=none,
                    y dir=both,
                    y explicit,
                    error mark={-}
            ]
            table [x expr=\thisrowno{0}-\exTwo, y index=1,y error expr=\thisrowno{2}]{./data/figure1/1s4/Li1988.1s4.txt};
           \addplot+ [
                line width=0.8,
                mark=triangle*,
                mark options={fill=white,solid,scale=0.5,},
                only marks,
                /pgfplots/error bars/.cd,
                    x dir=none,
                    y dir=both,
                    y explicit,
                    error mark={-}
            ]
            table [x expr=\thisrowno{0}-\exTwo, y index=1,y error expr=\thisrowno{2}]{./data/figure1/1s4/Chutjian1981.1s4.txt};
           \addplot+ [
                line width=0.8,
                mark=diamond*,
                mark options={fill=white,solid,scale=0.5,},
                only marks,
                /pgfplots/error bars/.cd,
                    x dir=none,
                    y dir=both,
                    y explicit,
                    error mark={-}
            ]
            table [x expr=\thisrowno{0}-\exTwo, y index=1,y error expr=\thisrowno{2}]{./data/figure1/1s4/Filipovic2000b.1s4.txt};
           \addplot+ [
                line width=0.8,
                mark=*,
                mark options={fill=green,solid,scale=0.5,},
                color=green,
                only marks,
                /pgfplots/error bars/.cd,
                    x dir=none,
                    y dir=both,
                    y explicit,
                    error mark={-}
            ]
            table [x expr=\thisrowno{0}-\exTwo, y index=1,y error expr=\thisrowno{2}]{./data/figure1/1s4/Khakoo2004.1s4.txt};
            
           \addplot+ [
                line width=1.0,
                mark=none,
                color=burntorange,
                smooth, solid,
            ]
            table [x expr=\thisrowno{0}-\exTwo, y index=1,]{./data/figure7/1s4.excitation.model.ref.txt};
            
            \legend{BSR~\cite{Pitchford2013, Zatsarinny2013}, Li~\cite{Li1988}, Chutjian~\cite{Chutjian1981}, Filipovic~\cite{Filipovic2000b, Filipovic2000a}, Khakoo~\cite{Khakoo2004}, Eq. (\ref{eq:resonance})}
            
        \nextgroupplot[
            enlarge x limits={true, abs value = 5mm},
            ylabel={$\sigma(\epsilon)$ ($m^2$)},
            xlabel={$\epsilon - \epsilon_0$ ($eV$)},
            tick scale binop ={\times},
            xmode=log, ymode=log,
            ymin=1e-23,
            legend style={
                font=\scriptsize,
                draw=none, fill=none,
                at={(rel axis cs: 0., 1.)},
                anchor=north west,
                nodes={scale=1.0},
                legend cell align={left},
                legend columns=1,
                /tikz/every even column/.append style={column sep=0.5cm},
            },
        ]
        
           \addplot+ [
                line width=0.7,
                mark=*,
                mark options={fill=white,solid,scale=0.5,},
                only marks,
            ]
            table [x expr=\thisrowno{0}-\exFour, y index=1]{./data/figure1/1s2/BSR.1s2.txt};
           \addplot+ [
                line width=0.8,
                mark=square*,
                mark options={fill=white,solid,scale=0.5,},
                only marks,
                /pgfplots/error bars/.cd,
                    x dir=none,
                    y dir=both,
                    y explicit,
                    error mark={-}
            ]
            table [x expr=\thisrowno{0}-\exFour, y index=1,y error expr=\thisrowno{2}]{./data/figure1/1s2/Li1988.1s2.txt};
           \addplot+ [
                line width=0.8,
                mark=triangle*,
                mark options={fill=white,solid,scale=0.5,},
                only marks,
                /pgfplots/error bars/.cd,
                    x dir=none,
                    y dir=both,
                    y explicit,
                    error mark={-}
            ]
            table [x expr=\thisrowno{0}-\exFour, y index=1,y error expr=\thisrowno{2}]{./data/figure1/1s2/Chutjian1981.1s2.txt};
           \addplot+ [
                line width=0.8,
                mark=diamond*,
                mark options={fill=white,solid,scale=0.5,},
                only marks,
                /pgfplots/error bars/.cd,
                    x dir=none,
                    y dir=both,
                    y explicit,
                    error mark={-}
            ]
            table [x expr=\thisrowno{0}-\exFour, y index=1,y error expr=\thisrowno{2}]{./data/figure1/1s2/Filipovic2000a.1s2.txt};
           \addplot+ [
                line width=0.8,
                mark=*,
                mark options={fill=green,solid,scale=0.5,},
                color=green,
                only marks,
                /pgfplots/error bars/.cd,
                    x dir=none,
                    y dir=both,
                    y explicit,
                    error mark={-}
            ]
            table [x expr=\thisrowno{0}-\exFour, y index=1,y error expr=\thisrowno{2}]{./data/figure1/1s2/Khakoo2004.1s2.txt};
            
           \addplot+ [
                line width=1.0,
                mark=none,
                color=burntorange,
                smooth, solid,
            ]
            table [x expr=\thisrowno{0}-\exFour, y index=1,]{./data/figure7/1s2.excitation.model.ref.txt};
            
         
  \end{groupplot}
 \node[below = 2cm of my plots c1r1.south west,
            anchor=west,
        ] {(a) Excitation, $1s_4$};
\node[below = 2cm of my plots c2r1.south west,
            anchor=west,
        ] {(b) Excitation, $1s_2$};
\end{tikzpicture}
%
\caption{
The semi-empirical model (\ref{eq:metastable}) for the (a) $1s_4$ and (b) $1s_2$ level excitation cross section,
together with the electron-beam experiment datasets.
}
\label{fig:resonance-model}
\end{figure}
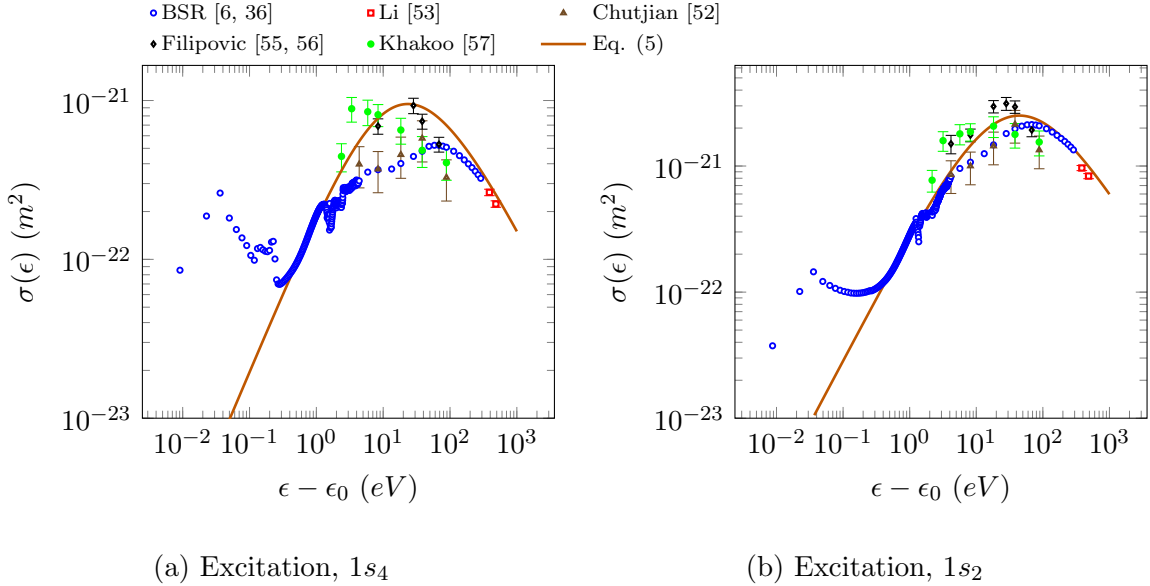
Figure~\ref{fig:resonance-model} shows results using reference parameter values.
The qualitative trends agree well with the measurement data.
\par

\section{Uncertainty quantification via Bayesian inference}\label{sec:bayes}

\subsection{Overall framework}
\refB{
We formulate an uncertainty representation of a general collision cross section $\sigma(\epsilon)$
given the corresponding data $\mathcal{M}$.
}
Each cross section has corresponding data $\mathcal{M}$ composed of multiple datasets
either from experiments or simulations,
\begin{subequations}
\begin{equation}
\cM = \{\cM_1, \cM_2, \ldots, \cM_M\}.
\end{equation}
\refB{
For each of the six collision processes we consider in this study,
$\cM$ corresponds to each subfigure in Figure~\ref{fig:crs-curation}.
The individual datasets $\cM_m$ ($m=1,\ldots,M$) for each of the collisional processes
are specified with different markers.
}
Each dataset $\cM_m$ consists of data points $(\veps_m, \vsig_m)$ and reported measurement uncertainties $\Dvsig_m$,
\begin{equation}
\cM_m = (\veps_m, \vsig_m, \Dvsig_m),
\quad
m = 1, \cdots, M,
\end{equation}
where $\veps_m, \vsig_m, \Dvsig_m \in \rR^{N_m}$, and $N_m$ denotes the number of data points in the dataset $\cM_m$.
\end{subequations}
\par
\refB{
The cross sections $\sigma(\epsilon)$ are represented by the semi-empirical models $f(\epsilon;\vec{A})$
introduced in Section~\ref{sec:model},
and the cross-section uncertainty is represented through the uncertainty of the parameters $\vec{A}$.
These parameters are calibrated given the data $\cM$ via a Bayesian inference process,
the output of which is $P[\vA\vert\cM]$
the probabilistic distribution for the model parameters conditioned on $\cM$.
}
\par
Specifically, according to the Bayes' law, $P[\vA\vert\cM]$ is
\begin{equation}\label{eq:bayes0}
\underbrace{P[\vA\vert\cM]}_{\text{posterior}} \propto \underbrace{P[\cM \vert \vA]}_{\text{likelihood}}\; \underbrace{P[\vA]}_{\text{prior}}.
\end{equation}
The prior distribution $P[\vA]$ is a probabilistic representation of any information about the parameters $\vA$ that is available separate from the data $\mathcal{M}$.
\refB{
For example, the recommended value from the authors of the semi-empirical model 
can provide us an order-of-magnitude estimate for $\vA$.
The prior distribution for each specific collision cross section 
will be discussed in Section~\ref{subsec:prior}.
}
The likelihood $P[\cM \vert \vA]$ is the probability distribution for the data conditioned on $\vA$.
\refB{
For given data $\mathcal{M}$,
an $\vA$ value that makes the data $\cM$ a more probable outcome of the model
would have a higher likelihood than $\vA$ values that make the observation less likely,
all other things being equal.
}
%
\par
The formulation of the right-hand side of (\ref{eq:bayes0}) is not unique.
That is, in most complicated problems at least, there are not clearly establishable,
true distributions for the prior and likelihood.
Instead, these distributions are mathematical models that represent a given state of information.
Different models could reflect, for instance, different interpretations of the data or different hypotheses
about the nature of the important uncertainties.
Thus, the formulation can differ depending on what uncertain factors are considered and how they are modeled.
\refB{
In this study,
there is not much information to consider for the prior,
and thus the prior formulation remains the same for any probabilistic model.
It is the likelihood that must be modeled under a certain interpretation or a hypothesis.
There can be many choices in the modeling, though not all of them are appropriate.
}
\par
To illustrate this fact and the implications of different choices, three likelihood formulations are introduced below.
In Section~\ref{subsec:wrong-bayes-1}, we introduce the simplest likelihood formulation that only considers the measurement uncertainty.
\refB{
This neglects the systematic error between the measurement datasets, though it is clearly observed in the data in Figure~\ref{fig:crs-curation}.
}
An alternative formulation is provided in Section~\ref{subsec:wrong-bayes-2},
where the systematic error is also considered as a type of parametric uncertainty.
Lastly in Section~\ref{subsec:gp-bayes},
we propose a likelihood formulation where the systematic error is not absorbed into the cross section model parameters.
It will be shown in Section~\ref{sec:result}
that this approach is most appropriate in the present setting,
while the first two can lead to misleading conclusions.
\par
\refB{
In order to distinguish between different likelihood models,
we denoted the probability terms $P_k$ with a subscript $k=1,2,3$
for three likelihood models in sections~\ref{subsec:wrong-bayes-1},~\ref{subsec:wrong-bayes-2},~and~\ref{subsec:gp-bayes}, respectively.
Any probability term that is used commonly for all likelihood models
is denoted with $P$ without a subscript.
}

\subsection{Prior formulations for cross section models}\label{subsec:prior}
\refB{
The prior $P[\vA]$ is imposed based on any prior knowledge about the collision cross section models.
For the direct ionization cross section, there is no reported value for $\vA_i$ in (\ref{eq:ion}),
and therefore the uniform prior $P[\vA_i] = 1$ is imposed.
For the elastic momentum-transfer collision,
a Gaussian probability is imposed on $\vA_{MERT}$ based on the literature value from Haddad and O'Malley~\cite{Haddad1982}
with $50\%$ relative standard deviation,
\begin{equation}\label{eq:prior-el}
P[\vA_{el}] = P[\vA_{MERT}] = \cN\left[\vA_{MERT, ref}, \mathrm{diag}\left((0.5\vA_{MERT, ref})^2\right) \right],
\end{equation}
with $\vA_{MERT, ref} = \{-1.488, 65.4, -84.3, 0.983\}$.
A smaller uncertainty is assigned to the elastic momentum-transfer collision,
considering the widely-accepted use of the MERT model.
For the metastable-level excitation cross sections,
a Gaussian probability is imposed on $\vA_{me}$ based on the literature values from Bretagne \textit{et al.}~\cite{Bretagne1986}
with $100\%$ relative standard deviation,
\begin{equation}
P[\vA_{me}] = \cN\left[ \vA_{me, ref}, \mathrm{diag}\left(\vA_{me, ref}^2\right) \right],
\end{equation}
with $\vA_{1s5} = \{51.2, 2.0\}$ and $\vA_{1s3} = \{10.4, 2.0\}$.
For the resonance-level excitation cross sections,
a Gaussian probability is imposed on $F_{0,r}$ based on the literature values from Bretagne \textit{et al.}~\cite{Bretagne1986}
with $100\%$ relative standard deviation,
\begin{equation}\label{eq:prior-resonance}
P[\vA_{r}] = P(F_{0,r}) = \cN\left[ F_{0,r,ref}, F_{0,r,ref}^2 \right],
\end{equation}
with $F_{0, 1s4, ref} = 0.061$ and $F_{0, 1s2, ref} = 0.254$.
}

\subsection{A likelihood formulation with the measurement error only}\label{subsec:wrong-bayes-1}
We first consider the reported measurement errors as the only uncertainty.
With the assumption that each dataset is obtained independently,
the resulting likelihood of the overall data $\cM$ is
\begin{equation}\label{eq:wrong-bayes1-total-likelihood1}
P_1\left[\cM \big| \vA \right] = \prod_{m=1}^M P_1\left[\cM_m \big| \vA \right].
\end{equation}
For the $n$-th data point $(\epsilon_{m,n}, \sigma_{m,n})$ in a dataset $\cM_m$,
we consider a \textit{measurement} model with a multiplicative measurement error $\delta\sigma_{m,n}$,
\begin{equation}\label{eq:wrong-bayes1-measurement-model}
\sigma_{m,n} = f(\epsilon_{m,n}; \vA) \times \delta\sigma_{m,n},
\end{equation}
so that
\begin{equation}\label{eq:wrong-bayes1-statement}
\log\sigma_{m,n} = \log f(\epsilon_{m,n}; \vA) + \log\delta\sigma_{m,n}.
\end{equation}
We assume the logarithm of the measurement error $\log\dvsig_m$ follows an independent Gaussian distribution based on the reported relative error,
\begin{equation}\label{eq:measurement-error}
P[\log\dvsig_m] = \cN\left[ 0, \mathrm{diag}\left(\log^2(1 + \frac{\Dvsig_m}{\vsig_m} )\right) \right],
\end{equation}
where all arithmetic operations between vectors are element-wise.
This logarithmic scale model is justified by the fact that
the majority of the measurement data report their accuracy in terms of the relative error $\Dvsig / \vsig$.
While the BSR dataset does not report the accuracy of the dataset,
we imposed a $10\%$ relative error based on its credibility as described in Section~\ref{subsec:electron-beam}.
With the probability model (\ref{eq:measurement-error}),
the likelihood of a dataset $\cM_m$ based on (\ref{eq:wrong-bayes1-statement}) is
\begin{equation}\label{eq:wrong-bayes1-total-likelihood2}
P_1\left[ \cM_m \big| \vA \right] = \cN\left[ \log f(\veps_m; \vA), \mathrm{diag}\left( \log^2 (1 + \frac{\Dvsig_m}{\vsig_m}) \right) \right].
\end{equation}
\par
To examine how the data impacts the uncertainty of $\vA$ under this formulation,
we consider two posterior formulations.  In the first, each data set is used individually, so that the posterior is given by
\begin{equation}\label{eq:wrong-bayes1-indiv}
P_1\left[ \vA \big| \cM_m \right] \propto P_1\left[ \cM_m \big| \vA \right] P[\vA],
\end{equation}
with the individual likelihood (\ref{eq:wrong-bayes1-total-likelihood2}) and the priors (\ref{eq:prior-el}--\ref{eq:prior-resonance}).
In the second, all the data are used simultaneously, so that
\begin{equation}\label{eq:wrong-bayes1-total}
P_1\left[ \vA \big| \cM \right] \propto \prod_{m=1}^M P_1\left[ \cM_m \big| \vA \right] P[\vA],
\end{equation}
with the overall likelihood (\ref{eq:wrong-bayes1-total-likelihood1}) and the same priors (\ref{eq:prior-el}--\ref{eq:prior-resonance}).

\subsection{All systematic error represented as the parametric uncertainty}\label{subsec:wrong-bayes-2}
In Section~\ref{sec:curation}, we note that the measurement data $\cM$ exhibits a large scattering among the datasets $\cM_m$
which is much larger than their reported observation errors $\Dvsig_m$.
This systematic error may not be negligible and thus should be accounted for in the probabilistic model using in the inference process.
One approach is to model the systematic error as an additional parametric uncertainty,
so each dataset $\cM_m$ may have a disturbance $\delta \vA_m$ in its parameter value,
\begin{equation}\label{eq:A-model1}
\vA_m = \vA + \delta \vA_m.
\end{equation}
\refB{
So then the discrepancies among the datasets $\cM_m$
are the result of the different realizations $f(\epsilon; \vA_m)$, induced by the uncertain disturbance $\delta\vA_m$.
}
\par
\refB{
There is no specific information for this uncertain disturbance.
In this case,}
this disturbance may be modeled to follow a joint Gaussian distribution over the parameter $\vA$-space
with a covariance,
\begin{equation}\label{eq:dA-model}
P_2\left[ \delta\vA_m \big| \vtheta \right] = \cN\left[ \vec{0}, \Sigma(\vtheta) \right],
\end{equation}
where, considering the symmetry of the covariance, $\vtheta$ corresponds to all the entries of the upper-triangular part of $\Sigma$.
Aside from the true $\vA$, several additional variables are considered uncertain in this viewpoint:
$\{\vA_m\}_{m=1}^M$ the realizations of $\vA$ for all datasets and the model parameter $\vtheta$.
The Bayes' law (\ref{eq:bayes0}) then should reflect all the uncertain variables,
having the resulting joint posterior,
\begin{equation}\label{eq:wrong-bayes2-joint-posterior}
\begin{split}
P_2\left[ \vA, \vtheta, \{\vA_m\}_{m=1}^M \big| \cM \right]
&\propto
P_2\left[ \cM, \{\vA_m\}_{m=1}^M \big| \vA, \vtheta \right]
\;
P[\vA]\; P_2[\vtheta]\\
&\propto
\prod_{m=1}^M P_2\left[ \cM_m, \vA_m \big| \vA, \vtheta \right]
\;
P[\vA]\; P_2[\vtheta],
\end{split}
\end{equation}
again assuming that each dataset is obtained independently.
\par
The same priors (\ref{eq:prior-el}--\ref{eq:prior-resonance}) are used for $P[\vA]$.
For the systematic error model parameter $\Sigma(\vtheta)$,
there is little information available to inform the probablilty for $\vtheta$,
and we can only expect that the corresponding variances would not be extremely small or large
compared to the parameter value itself.
Thus we assume the corresponding variances to be uniform on the logarithmic scale,
\begin{subequations}\label{eq:wrong-bayes2-hyper-prior}
\begin{equation}
P_2[\vtheta] = \prod_{d=1}^{\mathrm{dim}(\Sigma)}P_2\left[V_d(\vtheta)\right]
\end{equation}
\begin{equation}
P_2\left[V_d(\vtheta)\right] = 
\begin{cases}
\frac{1}{V_d(\log V_{d, max} - \log V_{d, min})}
& V_d \in [V_{d, min}, V_{d, max}]\\
0 & \text{otherwise,}
\end{cases}
\end{equation}
\end{subequations}
with $V_d(\vtheta)$ the variance along the $d$-th principal axis of $\Sigma(\vtheta)$
and $V_{d, min} = 10^{-2}$ and $V_{d, max} = 10^{1}$.
In actual evaluation, $V_d$ is determined as the $d$-th singular value of $\Sigma(\vtheta)$ given the parameter $\vtheta$.
\par
\refB{
The individual dataset posterior can be described
with the likelihood for $\cM_m$ and $\vA_m$,
\begin{equation}\label{eq:wrong-bayes2-indiv-posterior}
    \begin{split}
    P_2\left[ \cM_m, \vA_m \big| \vA, \vtheta \right]
    &\propto
    P_2\left[ \cM_m \big| \vA_m, \vA, \vtheta \right] \; P_2\left[ \vA_m \big| \vA, \vtheta \right].
    \end{split}
\end{equation}
With (\ref{eq:A-model1}) and (\ref{eq:dA-model}), the likelihood for $\vA_m$ is
\begin{equation}
P_2\left[ \vA_m \big| \vA, \vtheta \right] = \cN\left[ \vA, \Sigma(\vtheta) \right].
\end{equation}
When the parameter is realized as $\vA_m$ in the measurement $m$, the likelihood of the dataset $\cM_m$ is
\begin{equation}\label{eq:wrong-bayes2-likelihood}
P_2\left[ \cM_m \big| \vA_m, \vA, \vtheta \right] = \cN\left[ \log f(\veps_m; \vA_m), \mathrm{diag}\left( \log^2 (1 + \frac{\Dvsig_m}{\vsig_m}) \right) \right],
\end{equation}
which is similar to (\ref{eq:wrong-bayes1-total-likelihood2}) but the expected cross section is based on $\vA_m$,
reflecting the systematic error specific to the dataset $\cM_m$.
}
\par
For the cross section model parameter $\vA$ as our primary interest,
we consider all possibilities for $\{\vA_m\}_{m=1}^M$ and $\vtheta$ and thus marginalize $P_2\left[ \vA, \vtheta, \{\vA_m\}_{m=1}^M \big| \cM \right]$ over $\{\vA_m\}_{m=1}^M$ and $\vtheta$,
\begin{equation}\label{eq:wrong-bayes2}
P_2\left[ \vA \big| \cM \right]
= \int P_2\left[ \vA, \vtheta, \{\vA_m\}_{m=1}^M \big| \cM \right]\;d\vtheta d\vA_1 \cdots d\vA_M.
\end{equation}
The posterior (\ref{eq:wrong-bayes2}) is then evaluated with (\ref{eq:wrong-bayes2-joint-posterior}), (\ref{eq:wrong-bayes2-indiv-posterior}), (\ref{eq:prior-el}--\ref{eq:prior-resonance}) and (\ref{eq:wrong-bayes2-hyper-prior}).
\par
This description is predicated upon a strong assumption that the
model-form error is negligible and that systematic differences between
the experiments are due to each separate experiment realizing
different values of the semi-empirical model parameters.
It shall be seen in Section~\ref{sec:result} how this assumption impacts the results for $\vA$.

\subsection{A probabilistic description with a model for the systematic error}\label{subsec:gp-bayes}

%

While the measurement data $\cM$ exhibits a systematic error among the datasets $\cM_m$,
the employed semi-empirical models also do not necessarily capture all the features at all scales.
For example, the BSR excitation cross sections in Figure~\ref{fig:crs-curation}~(c--f) exhibits
small-scale variations near $\epsilon-\epsilon_0 \sim 1 eV$,
which cannot be represented at all by the model forms introduced in Section~\ref{sec:model}.
Thus, in addition to the systematic errors in the data, a model-form error is present.
The combined effect of the systematic data errors and the model inadequacies is denoted by $\delta f$ here.
Specifically, similar to (\ref{eq:wrong-bayes1-measurement-model}),
we consider a measurement model for the $n$-th data point $(\epsilon_{m,n}, \sigma_{m,n})$ in a dataset $\cM_m$,
where $\delta f$, which depends on additional parameters denoted $\vtheta$, appears as an additional multiplicative factor,
\begin{equation}
\sigma_{m,n} = f(\epsilon_{m,n}; \vA) \times \delta f(\epsilon_{m,n}; \vtheta) \times \delta\sigma_{m,n},
\end{equation}
which implies that
\begin{equation}\label{eq:log-sigma-model1}
\log\sigma_{m,n} = \log f(\epsilon_{m,n}; \vA) + \log\delta f(\epsilon_{m,n}; \vtheta) + \log\delta\sigma_{m,n}.
\end{equation}
\par
The resulting joint posterior analogous to (\ref{eq:wrong-bayes2-joint-posterior}) is
\begin{equation}\label{eq:gp-bayes-joint-post}
P_3\left[ \vA, \vtheta \big| \cM \right] \propto P_3\left[\cM \big| \vA, \vtheta \right]\;P\left[\vA \right]\;P_3\left[\vtheta \right],
\end{equation}
where the parameters $\vA$ and $\vtheta$ are taken to be independent in the prior.
Similar to (\ref{eq:wrong-bayes2}),
we consider all possibilities for $\vtheta$ and thus marginalize $P\left[ \vA, \vtheta \big| \cM \right]$ over $\vtheta$,
\begin{equation}\label{eq:gp-bayes-margin-post}
P_3\left[ \vA \big| \cM \right] \propto \int P_3\left[\cM \big| \vA, \vtheta \right]\;P\left[\vA \right]\;P_3\left[\vtheta \right]\; d\vtheta.
\end{equation}
\par
With the assumption that each dataset is obtained independently,
the resulting likelihood of the overall data $\cM$ is
\begin{equation}\label{eq:gp-bayes-total-likelihood}
P_3\left[\cM \big| \vA, \vtheta \right] = \prod_{m=1}^M P_3\left[\cM_m \big| \vA, \vtheta \right],
\end{equation}
where $P_3\left[\cM_m \big| \vA, \vtheta \right]$ will be determined from (\ref{eq:log-sigma-model1})
with the uncertainty models for $\log\dvsig_m$ and $\delta f(\vec{\epsilon}_m;\vA)$,
which are introduced below.
\par
For the measurement uncertainty $\log\dvsig_m$, the same model (\ref{eq:measurement-error}) is used.
On the other hand, modeling $\delta f(\epsilon; \vtheta)$ requires further development.
Unlike the cross section itself,
it is difficult to model the discrepancy as a simple analytical function as $f(\epsilon;\vA)$.
As described in Section~\ref{sec:curation},
there are many potential factors involved in the discrepancy among datasets, each of which is hard to quantify.
This is further aggravated by the fact
that we have no knowledge to separate the model inadequacy from the systematic error so that they may be modeled individually.
\par
In this situation,
Gaussian processes (GP) provide a flexible probabilistic framework for representing the observed combined effects of these errors.
\refB{In essence, GP only dictates how the error $\delta f(\epsilon)$ is correlated between two electron energy $\epsilon$ points,
without imposing a specific form of $\delta f(\epsilon)$ on electron energy space.}
Specifically, for a given $\vtheta$ and a dataset $\cM_m$, we model
$\log\delta f(\veps_m)$ as a Gassian random function with zero mean and covariance $\bK(\veps_m, \veps_m; \vtheta)\in\rR^{N_m\times N_m}$,
\begin{equation}\label{eq:gp-systematic-error}
P_3\left[ \log\delta f(\veps_m) \;\big|\; \vtheta \right] = \cN\left[ \vec{0}, \bK(\veps_m, \veps_m; \vtheta)\right].
\end{equation}
Each entry of the covariance matrix is determined by the covariance kernel $k$,
\begin{equation}
\bK(\veps_m, \veps_m; \vtheta)_{ij} = k( d_{m, ij} ; \vtheta),
\end{equation}
where $d_{m, ij}$ the distance between two data points $\epsilon_{m,i}$ and $\epsilon_{m,j}$ is determined in the logarithmic scale, based on the scaling behavior of each collision cross section.
For the elastic cross section,
\begin{subequations}
\begin{equation}
d_{m, ij} = \vert \log\epsilon_{m,i} - \log\epsilon_{m,j} \vert,
\end{equation}
and for the other collisions with threshold energy $\epsilon_0$,
\begin{equation}
d_{m, ij} = \vert \log(\epsilon_{m,i} - \epsilon_0) - \log(\epsilon_{m,j} - \epsilon_0) \vert.
\end{equation}
\end{subequations}
The widely used Mat\'{e}rn kernel is used for $k(d; \vtheta)$ with model parameters $\vtheta = \{V_{\theta}, d_{\theta}\}$,
\begin{equation}\label{eq:cov-kernel}
k(d; \vtheta) = V_{\theta}\frac{2^{1-\nu}}{\Gamma(\nu)}\left( \sqrt{2\nu}\frac{d}{d_{\theta}} \right)^{\nu}K_{\nu}\left( \sqrt{2\nu}\frac{d}{d_{\theta}} \right),
\end{equation}
where $\Gamma$ is the gamma function, $K_\nu$ is the modified Bessel function of the second kind, and $\nu = 1.5$.
$V_{\theta}$ and $d_{\theta}$ are the main parameters of the Mat\'{e}rn kernel that represent the variance and correlation scale, respectively.
\par
\todo{Kevin: Raja, I added a description for the covariance kernel in this paragraph $\downarrow$.
While it could be noteworthy, its detailed discussion is not so important for our conclusion,
since we consider all possible hyperparameters. This is discussed here.}
In GP, it is the covariance kernel (\ref{eq:cov-kernel}) that
dictates overall characteristics of a random function (here $\log\delta f(\veps)$).
In essence, for given two points in $\epsilon$-space,
the closer their distance $d$ is, the less their function values vary.
The variation amplitude is determined by $V_{\theta}$,
and the variation scale is determined by $d_{\theta}$.
In particular, functions chosen with long $d_{\theta}$
vary only smoothly along large scale of $\epsilon$-space,
while those with short $d_{\theta}$ exhibit fine-scale variations.
Typically these parameters are selected to maximize the marginal likelihood of the given dataset~\cite{Rasmussen2006}.
For our purpose of uncertainty quantification, however,
we cannot simply calibrate these parameters to the most credible values,
but their uncertainties also must be considered.
Thus, all possibilities of $\theta=\{V_\theta, d_{\theta}\}$ are considered via (\ref{eq:gp-bayes-margin-post}).
For other choices of these parameters, we refer readers to Williams and Rasmussen~\cite{Rasmussen2006}.
\par
With the probability models (\ref{eq:gp-systematic-error}) and (\ref{eq:measurement-error}),
the likelihood of a dataset $\cM_m$ based on (\ref{eq:log-sigma-model1}) is
\begin{equation}\label{eq:gp-bayes-indiv-likelihood}
\begin{split}
P_3\left[\cM_m \big| \vA, \vtheta \right] &= P_3\left[\log\vsig_m \big| \vA, \vtheta \right]\\
&= \cN\left[ \log\delta f(\veps_m;\vA), \bK(\veps_m, \veps_m; \vtheta) + \mathrm{diag}\left( \log^2 (1 + \frac{\Dvsig_m}{\vsig_m} ) \right) \right].
\end{split}
\end{equation}
\par
The same priors (\ref{eq:prior-el}--\ref{eq:prior-resonance}) are used for $P[\vA]$.
Analogous to (\ref{eq:wrong-bayes2-hyper-prior}), for the prior $P_3[\vtheta]$,
we assume $V_{\theta}$ to be uniform on the logarithmic scale,
\begin{equation}\label{eq:prior-gp-var}
P_3[\vtheta] = P_3[V_{\theta}] =
\begin{cases}
\frac{1}{V_{\theta}(\log V_{\theta, max} - \log V_{\theta, min})}
& V_{\theta} \in [V_{\theta, min}, V_{\theta, max}]\\
0 & \text{otherwise,}
\end{cases}
\end{equation}
with the range $V_{\theta, min} = 10^{-2}$ and $V_{\theta, max} = 10^{1}$.
\par
The likelihood (\ref{eq:gp-bayes-total-likelihood}), the priors (\ref{eq:prior-el}--\ref{eq:prior-resonance}) and (\ref{eq:prior-gp-var})
then constitutes the posteriors (\ref{eq:gp-bayes-joint-post}) and (\ref{eq:gp-bayes-margin-post}) for all 6 collision cross sections.

\par

\section{The inferred cross section uncertainty}\label{sec:result}

The probabilistic models from Section~\ref{sec:bayes}
are first compared for the $1s_3$ excitation cross section as an example.
Then the parametric uncertainties of all 6 cross sections are calibrated and presented.

\subsection{Comparison of the parametric posterior distributions}
The posterior distributions from Section~\ref{sec:bayes} are sampled using a standard Markov chain Monte Carlo (MCMC) algorithm implemented in the \texttt{emcee} package~\cite{Foreman-Mackey2013}.
For all posteriors, the total sample size exceeds at least 20 times the auto-correlation length, typically around 50 times,
in order to obtain enough effectively independent samples~\cite{Foreman-Mackey2013}.
\par
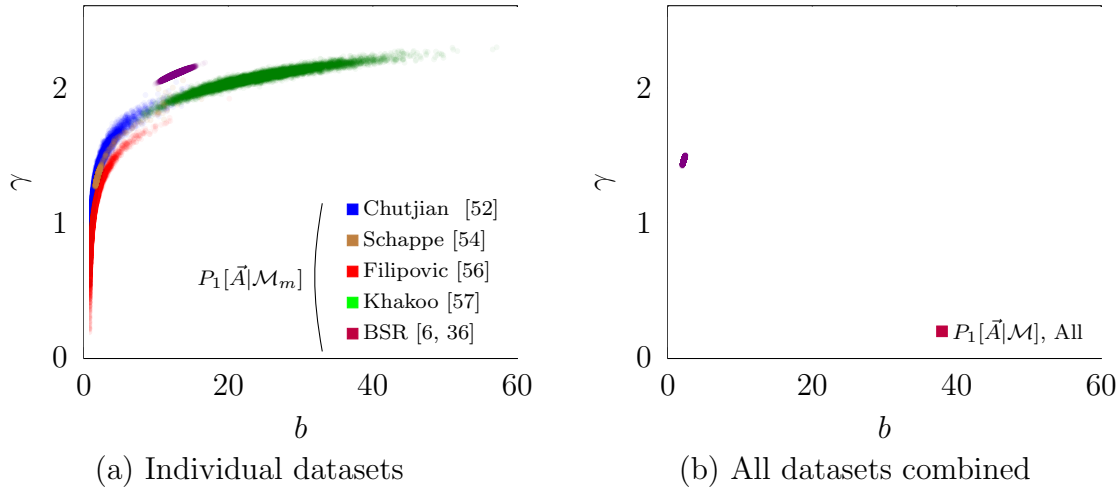
\begin{figure}
\centering
\begin{tikzpicture}[font=\normalsize,]
    \begin{groupplot}[
        group style={
            group name = my plots,
            group size= 2 by 1,
            xlabels at =edge bottom,
            horizontal sep=2cm,
            vertical sep=3.5cm,
        },
        name=chung,
        height = 0.4\textwidth,
        width = 0.5\textwidth,
    ]    
\pgfplotsset{set layers=standard}%

        \nextgroupplot[
            enlarge x limits={false, abs value = 5mm},
            xlabel={$b$},
            ylabel={$\gamma$},
            tick scale binop ={\times},
            ymin=-0.01, ymax=2.61,
            xmin=0.0, xmax=60.0,
            xticklabel style={yshift=-2pt},
            yticklabel style={xshift=-2pt},
            legend style={
                font=\scriptsize,
                draw=none, fill=none,
                at={(rel axis cs: 0.99, 0.01)},
                anchor=south east,
                nodes={scale=1.0},
                legend cell align={left},
                legend columns=1,
                /tikz/every even column/.append style={column sep=0.5cm},
            },
        ]
        
            \addlegendimage{only marks, blue, mark=square*, draw=none, mark options={color=blue}}
            \addlegendentry{Chutjian~~\cite{Chutjian1981}}
            \addlegendimage{only marks, magenta, mark=square*, mark options={color=brown}}
            \addlegendentry{Schappe~\cite{Schappe1994}}
            \addlegendimage{only marks, red, mark=square*, mark options={color=red}}
            \addlegendentry{Filipovic~\cite{Filipovic2000b}}
            \addlegendimage{only marks, green, mark=square*, mark options={color=green}}
            \addlegendentry{Khakoo~\cite{Khakoo2004}}
            \addlegendimage{only marks, purple, mark=square*, mark options={color=purple}}
            \addlegendentry{BSR~\cite{Pitchford2013, Zatsarinny2013}}
            
            \edef\imagepath{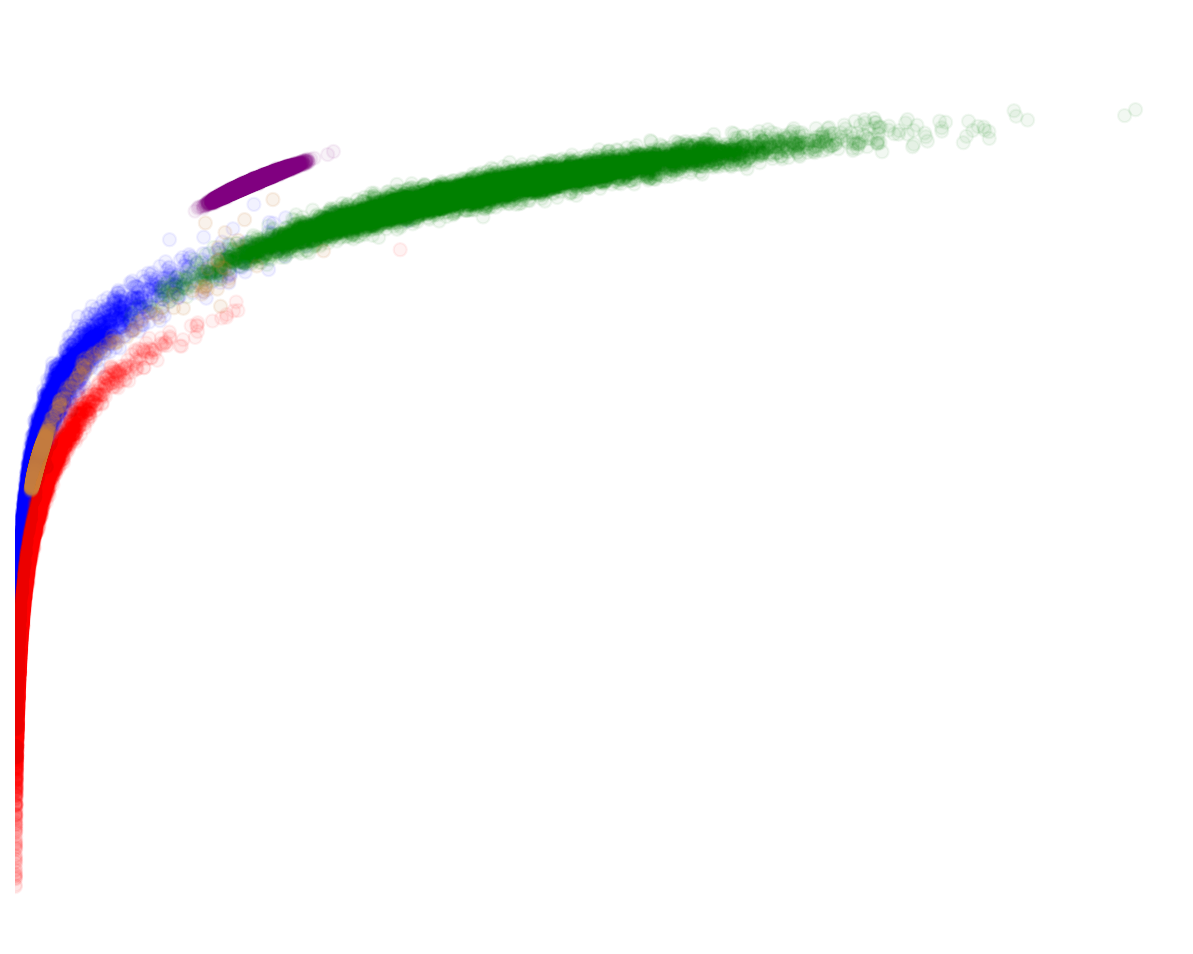}
            \addplot graphics[xmin=0.0,xmax=60.0,ymin=0.0,ymax=2.6]{\imagepath};
            
            \draw[
                black,
                solid,
            ] (rel axis cs: 0.55, 0.44) to[out=-100, in=100] node[anchor=east, font=\scriptsize] {$P_1[\vA \vert \cM_m]$} (rel axis cs: 0.55, 0.02);
            
        \nextgroupplot[
            enlarge x limits={false, abs value = 5mm},
            xlabel={$b$},
            ylabel={$\gamma$},
            tick scale binop ={\times},
            ymin=-0.01, ymax=2.61,
            xmin=0.0, xmax=60.0,
            xticklabel style={yshift=-2pt},
            yticklabel style={xshift=-2pt},
            legend style={
                font=\scriptsize,
                draw=none, fill=none,
                at={(rel axis cs: 0.99, 0.01)},
                anchor=south east,
                nodes={scale=1.0},
                legend cell align={left},
                legend columns=1,
                /tikz/every even column/.append style={column sep=0.5cm},
            },
        ]
        
            \addlegendimage{only marks, yellow, mark=square*, mark options={color=purple}}
            \addlegendentry{$P_1[\vA \vert \cM]$, All}
            
            \edef\imagepath{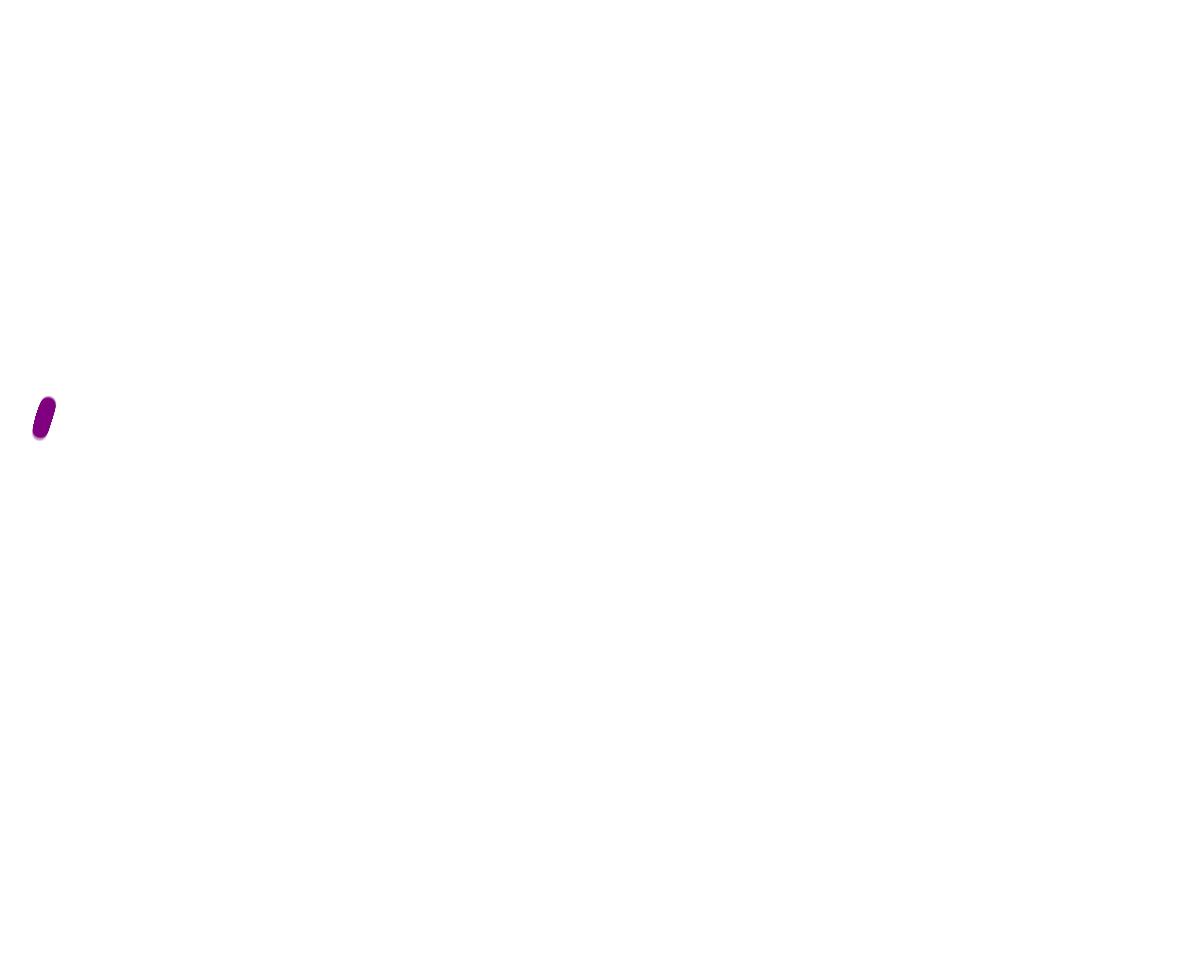}
            \addplot graphics[xmin=0.0,xmax=60.0,ymin=0.0,ymax=2.6]{\imagepath};

  \end{groupplot}
\node[below = 1.5cm of my plots c1r1.south west,
            anchor=west,
        ] {(a) Individual datasets};
\node[below = 1.5cm of my plots c2r1.south west,
            anchor=west,
        ] {(b) All datasets combined};
\end{tikzpicture}
\caption{
Sample scatter plots of the posterior distributions based
on the measurement-uncertainty-only representation (Section~\ref{subsec:wrong-bayes-1}) for $1s_3$ excitation cross section:
(a) with individual datasets ($P_1[\vA\vert\cM_m]$ in Eq.~\ref{eq:wrong-bayes1-indiv})
and (b) with all datasets combined ($P_1[\vA \vert \cM]$ in Eq.~\ref{eq:wrong-bayes1-total}).
}
\label{fig:wrong-bayes1}
\end{figure}
To illustrate the implications of neglecting the systematic error,
the two posteriors described in the first model in Section~\ref{subsec:wrong-bayes-1} are evaluated for the $1s_3$ excitation cross section.
Figure~\ref{fig:wrong-bayes1}~(a) shows samples from the posteriors $P_1[\vA \vert \cM_m]$ (\ref{eq:wrong-bayes1-indiv})
obtained from individual datasets $\cM_m$.
The individual posteriors are scattered in $\vA$-space,
and mostly do not overlap with each other.
This again reflects the scattering among the cross section datasets in Figure~\ref{fig:crs-curation}~(e),
implying that there is a large uncertainty in the cross section model parameters.
When evaluated using all the datasets simultaneously, however,
the probability description in Section~\ref{subsec:wrong-bayes-1} cannot capture this uncertainty.
In Figure~\ref{fig:wrong-bayes1}~(b),
the overall posterior $P_1[\vA \vert \cM]$ (\ref{eq:wrong-bayes1-total}) is highly concentrated in a narrow region,
indicating a very low uncertainty in $\vA$.
This is obviously a misleading conclusion given the scatter among the individual datasets.
\par
This does not mean that the Bayesian inference is wrong.
It is mathematically consistent given the likelihoods (\ref{eq:wrong-bayes1-total-likelihood2}) and (\ref{eq:wrong-bayes1-total-likelihood1}).
From (\ref{eq:wrong-bayes1-total-likelihood2}), each dataset $\cM_m$ claims a different best value of $\vA$ with its own measurement error $\delta \vsig_m$.
The overall likelihood (\ref{eq:wrong-bayes1-total-likelihood1}), which is simply the product of the likelihood for each dataset, naturally leads to a concentration of the posterior probability around a value somewhere in the middle.
This unrealistic result reflects the fact that the problem statement (\ref{eq:wrong-bayes1-statement}) does not properly account for all sources of uncertainty in the likelihood formulation.
\par
\begin{figure}
\centering
\begin{tikzpicture}[font=\small,]
    \begin{groupplot}[
        group style={
            group name = my plots,
            group size= 2 by 1,
            xlabels at =edge bottom,
            horizontal sep=1.5cm,
            vertical sep=3.5cm,
        },
        name=chung,
        height = 0.4\textwidth,
        width = 0.5\textwidth,
    ]    
\pgfplotsset{set layers=standard}%

        \nextgroupplot[
            enlarge x limits={false, abs value = 5mm},
            xlabel={$b$},
            ylabel={$\gamma$},
            tick scale binop ={\times},
            ymin=-0.01, ymax=2.61,
            xmin=0.0, xmax=60.0,
            xticklabel style={yshift=-2pt},
            yticklabel style={xshift=-2pt},
        ]
            \edef\imagepath{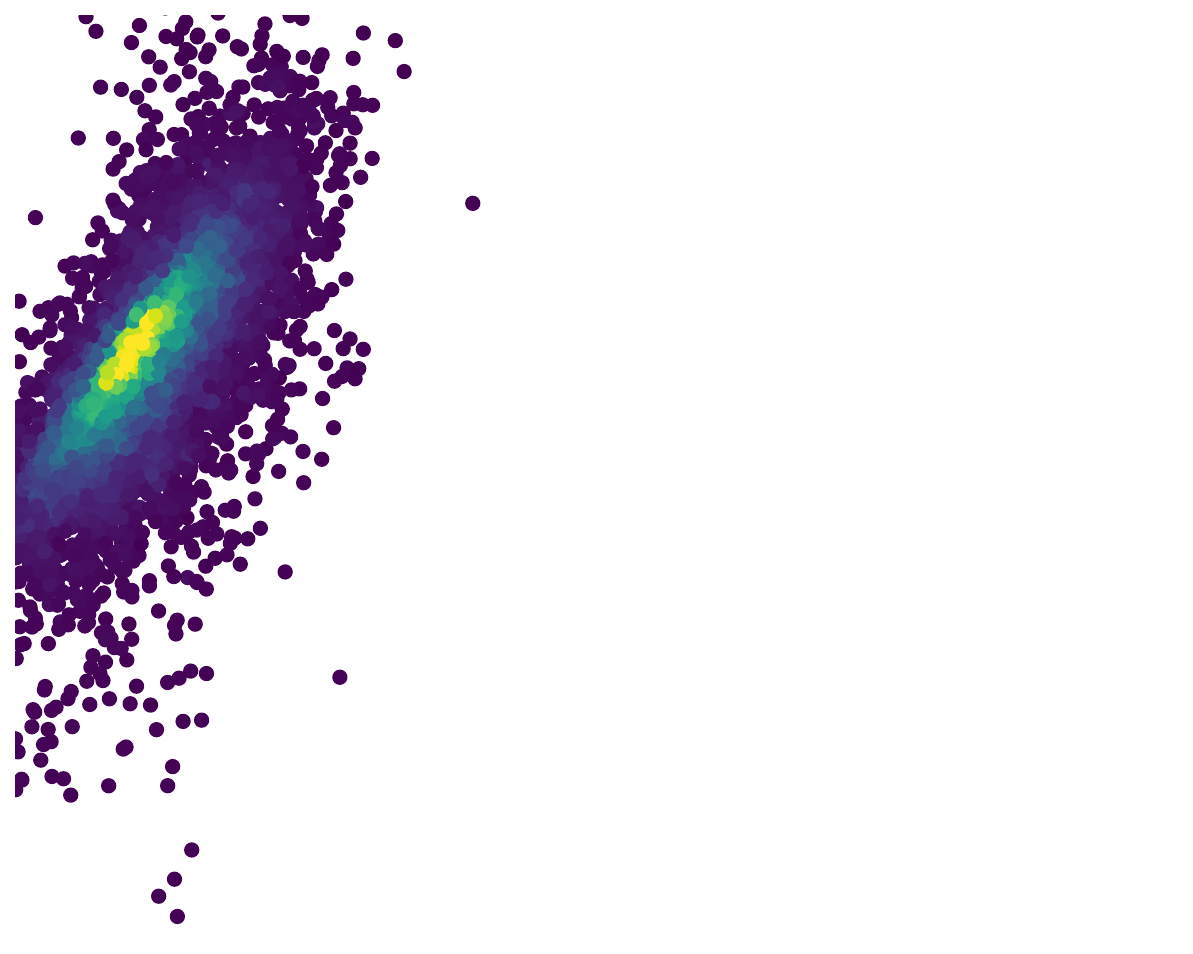}
            \addplot graphics[xmin=-1.0,xmax=60.0,ymin=0.0,ymax=2.7]{\imagepath};
            
        \nextgroupplot[
            enlarge x limits={false, abs value = 5mm},
            xlabel={$b$},
            ylabel={$\gamma$},
            tick scale binop ={\times},
            ymin=-0.01, ymax=2.61,
            xmin=0.0, xmax=60.0,
            xticklabel style={yshift=-2pt},
            yticklabel style={xshift=-2pt},
            point meta min=0.0, point meta max=6.7e-3,
            colorbar, colormap/viridis,
            colorbar style={
                font=\scriptsize,
                xticklabel pos=upper,
                scaled y ticks=false,
                ytick={0, 1e-3, 2e-3, 3e-3, 4e-3, 5e-3, 6e-3},
                yticklabels={$0$, $1$, $2$, $3$, $4$, $5$, {$6\times10^{-3}$}},
                /pgf/number format/precision=4,
                at={(rel axis cs: 2.25, 0.)}, anchor=south west,
                xlabel=$P(\vA \vert \cM)$,
            }
        ]
        
            \edef\imagepath{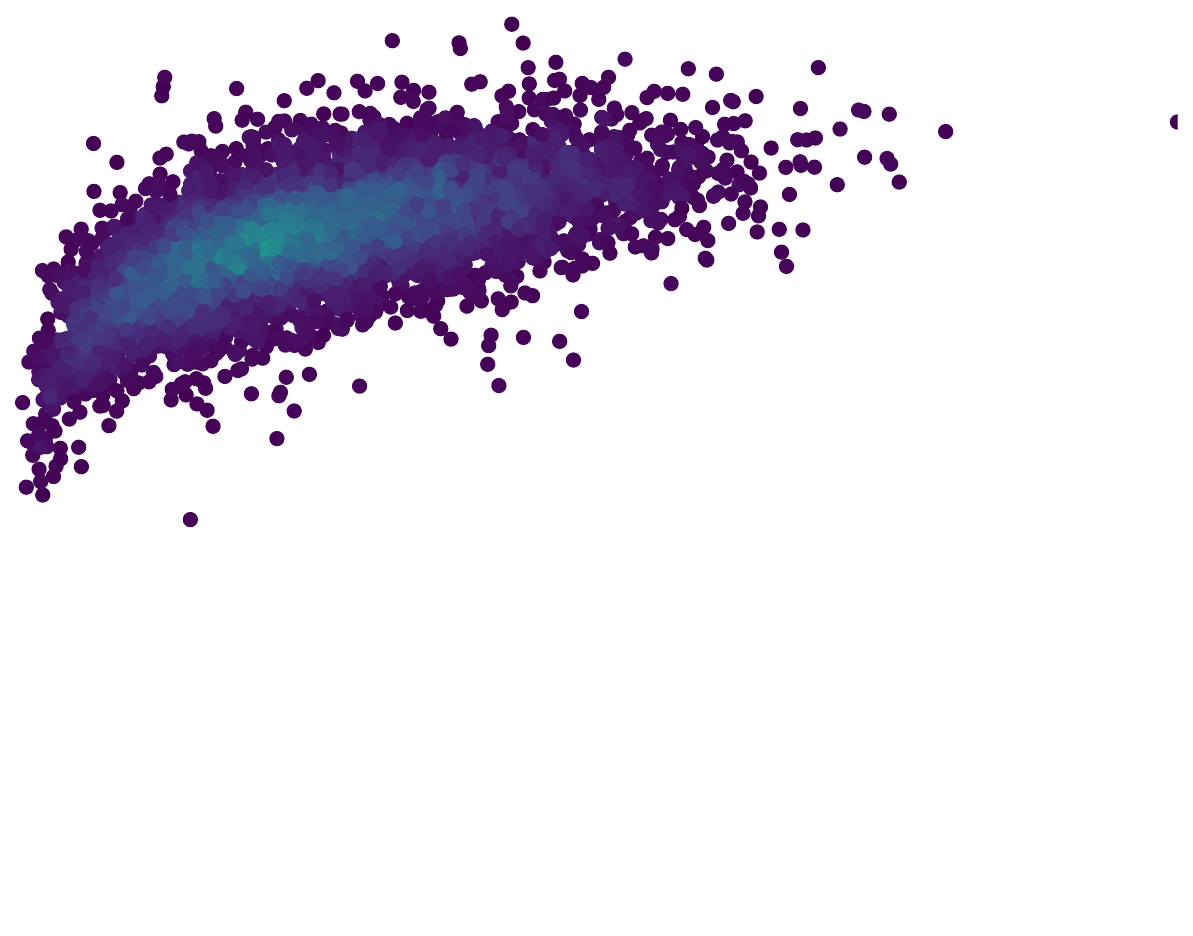}
            \addplot graphics[xmin=0.0,xmax=60.0,ymin=0.0,ymax=2.6]{\imagepath};

  \end{groupplot}
\node[below = 1.5cm of my plots c1r1.south west,
            anchor=west,
        ] {(a) $P_2[\vA\vert\cM]$ (\ref{eq:wrong-bayes2})};
\node[below = 1.5cm of my plots c2r1.south west,
            anchor=west,
        ] {(b) $P_3[\vA\vert\cM]$ (\ref{eq:gp-bayes-margin-post})};
\end{tikzpicture}
\caption{
Sample scatter plots of the posterior distributions of $1s_3$ excitation cross section:
(a) $P_2[\vA \vert \cM]$ (\ref{eq:wrong-bayes2}) based on the internal parametric uncertainty representation (Section~\ref{subsec:wrong-bayes-2})
and (b) $P_3[\vA \vert \cM]$ (\ref{eq:gp-bayes-margin-post}) based on the Gaussian process representation (Section~\ref{subsec:gp-bayes}).
In both figures, samples are colored according to their posterior probability.
}
\label{fig:wrong-bayes2}
\end{figure}
The uncertainty underlying the scattering among datasets
can be captured only when it is explicitly considered in the probabilistic description.
Figure~\ref{fig:wrong-bayes2} shows the posterior distributions evaluated with the systematic error models.
Whether as an additional parameter uncertainty or as a Gaussian process,
considering the possibility of the systematic error
significantly increases the uncertainty of $\vA$.
While both posteriors are peaked at similar parameter values,
their distributions have qualitatively different shapes.
In turn, each systematic error model leads to a different conclusion about the uncertainties of the cross section model parameters:
with $P_2$ (\ref{eq:wrong-bayes2}), we may conclude that the uncertainty in $(\gamma, b)$ follows a Gaussian distribution with the uncertainty in $\gamma$ more pronounced,
while $P_3$ (\ref{eq:gp-bayes-margin-post}) rather seems to conclude that $b$ is more uncertain.
\par
\begin{figure}
\centering
\input{figures/figure10.tex}
\caption{
Marginal posterior distribution $P[\sigma(\epsilon) \vert \cM]$ for $1s_3$ excitation cross section with
(a) $P_1[\vA \vert \cM]$ (\ref{eq:wrong-bayes1-total}) based on the measurement-uncertainty-only representation (Section~\ref{subsec:wrong-bayes-1});
(b) $P_2[\vA \vert \cM]$ (\ref{eq:wrong-bayes2}) based on the internal parametric uncertainty representation (Section~\ref{subsec:wrong-bayes-2});
and (c) $P_3[\vA \vert \cM]$ (\ref{eq:gp-bayes-margin-post}) based on the Gaussian process representation (Section~\ref{subsec:gp-bayes}).
Shaded areas indicate the $68\%$, $95\%$ and $99.7\%$ credibility intervals of the posterior distribution.
}
\label{fig:wrong-bayes2-marginal}
\end{figure}
The difference between the two systematic error models appears more drastically when the uncertainty of the cross section $\sigma(\epsilon)$ is considered.
Figure~\ref{fig:wrong-bayes2-marginal} shows the distribution of $\sigma(\epsilon)$ obtained by forward propagating each $P[\vA \vert \cM]$ through the parametric model compared to the calibration data.
Once again, the results from (\ref{eq:wrong-bayes1-total}) do not capture the systematic errors among datasets,
with the probability concentrated on almost a single line.
On the other hand,
while the Gaussian process model seems to properly capture the scattering among the datasets,
the resulting posterior from the internal parametric uncertainty model is overly uncertain about $\vA$,
with its credibility intervals exceeding far beyond the measurement datasets.
Moreover, considering the median value of the cross section in $\epsilon>10^2eV$,
the internal parametric uncertainty model deviates from the datasets compared to the Gaussian process model,
implying a slight miscalibration of the parameter value.
\par
\begin{figure}
\centering
\begin{tikzpicture}[font=\small,]
    \begin{groupplot}[
        group style={
            group name = my plots,
            group size= 2 by 1,
            xlabels at =edge bottom,
            horizontal sep=1.5cm,
            vertical sep=3.5cm,
        },
        name=chung,
        height = 0.4\textwidth,
        width = 0.5\textwidth,
    ]    
\pgfplotsset{set layers=standard}%

        \nextgroupplot[
            enlarge x limits={false, abs value = 5mm},
            xlabel={$b$},
            ylabel={$\gamma$},
            tick scale binop ={\times},
            ymin=-0.01, ymax=2.71,
            xmin=3e-2, xmax=1e2,
            xmode=log,
            xticklabel style={yshift=-2pt},
            yticklabel style={xshift=-2pt},
            legend style={
                font=\scriptsize,
                draw=none, fill=none,
                at={(rel axis cs: -0.1, 1.01)},
                anchor=south west,
                nodes={scale=1.0},
                legend cell align={left},
                legend columns=5,
                /tikz/every even column/.append style={column sep=0.5cm},
            },
        ]
            \addlegendimage{only marks, blue, mark=square*, draw=none, mark options={color=blue}}
            \addlegendentry{Chutjian~~\cite{Chutjian1981}}
            \addlegendimage{only marks, magenta, mark=square*, mark options={color=magenta}}
            \addlegendentry{Schappe~\cite{Schappe1994}}
            \addlegendimage{only marks, red, mark=square*, mark options={color=red}}
            \addlegendentry{Filipovic~\cite{Filipovic2000b}}
            \addlegendimage{only marks, green, mark=square*, mark options={color=green}}
            \addlegendentry{Khakoo~\cite{Khakoo2004}}
            \addlegendimage{only marks, orange, mark=square*, mark options={color=orange}}
            \addlegendentry{BSR~\cite{Pitchford2013, Zatsarinny2013}}
        
            \edef\imagepath{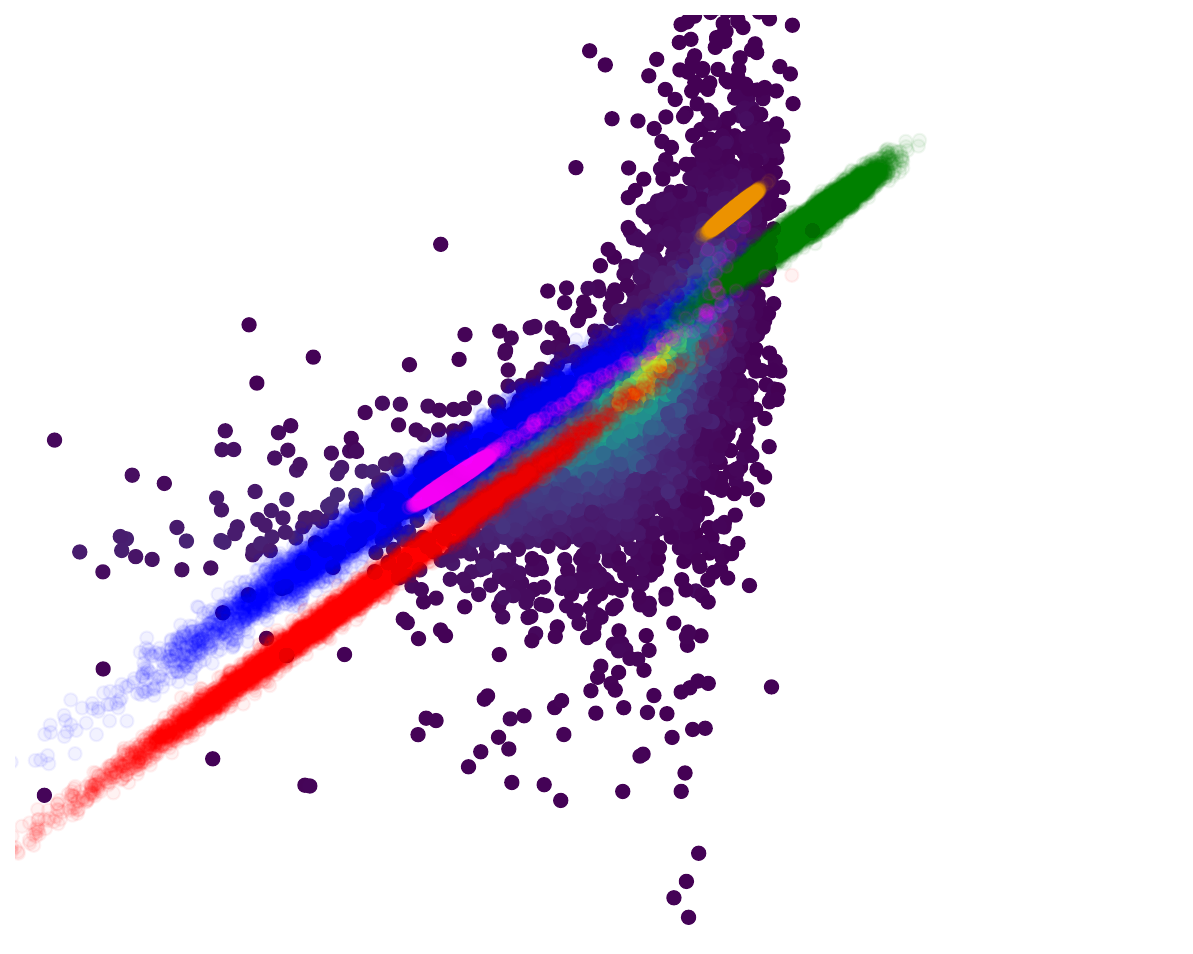}
            \addplot graphics[xmin=3e-2,xmax=5e2,ymin=0.0,ymax=2.7]{\imagepath};
            
        \nextgroupplot[
            enlarge x limits={false, abs value = 5mm},
            xlabel={$b$},
            ylabel={$\gamma$},
            tick scale binop ={\times},
            ymin=-0.01, ymax=2.71,
            xmin=3e-2, xmax=1e2,
            xmode=log,
            xticklabel style={yshift=-2pt},
            yticklabel style={xshift=-2pt},
            point meta min=0.0, point meta max=6.7e-3,
            colorbar, colormap/viridis,
            colorbar style={
                font=\scriptsize,
                xticklabel pos=upper,
                scaled y ticks=false,
                ytick={0, 1e-3, 2e-3, 3e-3, 4e-3, 5e-3, 6e-3},
                yticklabels={$0$, $1$, $2$, $3$, $4$, $5$, {$6\times10^{-3}$}},
                /pgf/number format/precision=4,
                at={(rel axis cs: 2.75, 0.)}, anchor=south west,
                xlabel=$P(\vA \vert \cM)$,
            }
        ]
        
            \edef\imagepath{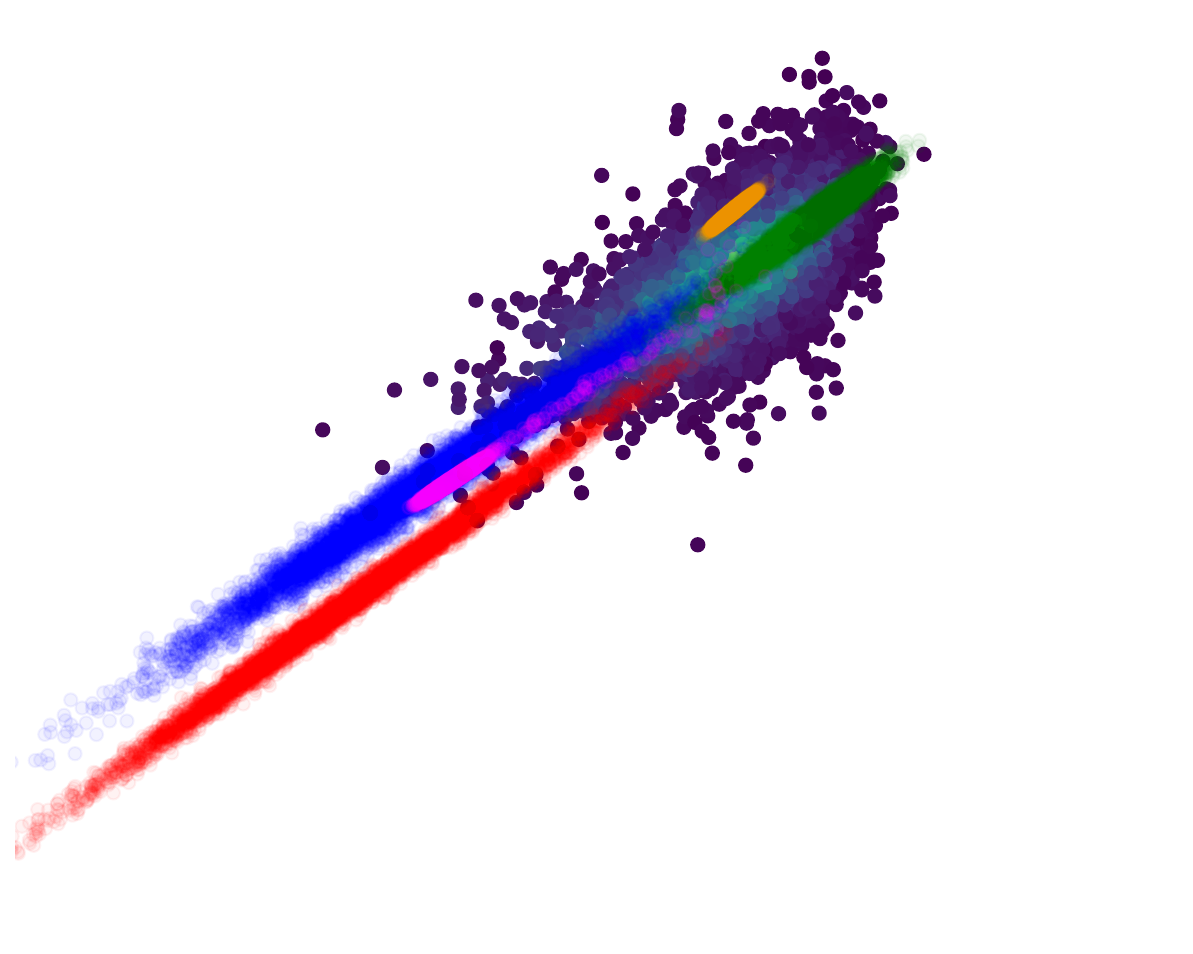}
            \addplot graphics[xmin=3e-2,xmax=5e2,ymin=0.0,ymax=2.7]{\imagepath};

  \end{groupplot}
\node[below = 1.5cm of my plots c1r1.south west,
            anchor=west,
        ] {(a) $P_2[\vA\vert\cM]$ (\ref{eq:wrong-bayes2})};
\node[below = 1.5cm of my plots c2r1.south west,
            anchor=west,
        ] {(b) $P_3[\vA\vert\cM]$ (\ref{eq:gp-bayes-margin-post})};
\end{tikzpicture}
\caption{
Sample scatter plots of the posterior distributions of $1s_3$ excitation cross section on log scale of $b$:
(a) the internal parametric uncertainty represenation $P_2[\vA \vert \cM]$ (Eq.~\ref{eq:wrong-bayes2} in Section~\ref{subsec:wrong-bayes-2})
and (b) the Gaussian process representation $P_3[\vA \vert \cM]$ (Eq.~\ref{eq:gp-bayes-margin-post} in Section~\ref{subsec:gp-bayes})
compared with the measurement-uncertainty-only representation of individual datasets
$P_1[\vA\vert\cM_m]$ (Eq.~\ref{eq:wrong-bayes1-indiv} in Section~\ref{subsec:wrong-bayes-1}).
Samples for both $P_2[\vA \vert \cM]$ and $P_3[\vA \vert \cM]$ are colored according to their posterior probability.
}
\label{fig:wrong-bayes2-log}
\end{figure}
A deeper analysis of the posterior distributions can expose
the flaw of the internal parametric model and explain the overstated uncertainty.
In Section~\ref{subsec:wrong-bayes-2},
the systematic error is modeled using a Gaussian on $\vA$ as shown in (\ref{eq:dA-model}),
without considering the actual shape of the cross section model $f(\epsilon; \vA)$.
The resulting posterior cannot exactly capture the correlation
which inherently exists between parameters given the cross section model and the datasets.
For the $1s_3$ excitation, due to the model definition (\ref{eq:metastable}),
the two parameters $\{b, \gamma\}$ should be nonlinearly correlated roughly along $\gamma \sim \log b$.
Figure~\ref{fig:wrong-bayes2-log} shows the individual posteriors in Figure~\ref{fig:wrong-bayes1} using a log scale for $b$,
which exhibits a clear correlation of $\gamma\sim\log b$.
$P_2[\vA \vert \cM]$ (\ref{eq:wrong-bayes2}),
which is a joint Gaussian over $(b, \gamma)$-space,
now looks completely distorted in Figure~\ref{fig:wrong-bayes2-log}~(a).
A significant portion of the distribution deviates from the parameter values that are inferred from the datasets,
which contributes to the biased and overstated uncertainty observed earlier.
\par
Of course, in this case, it may be possible to improve the uncertainty model (\ref{eq:dA-model}) by assuming it to be a joint Gaussian distribution over $(\log b, \gamma)$-space, instead of $(b, \gamma)$.
However, such an improvement is predicated upon prior knowledge of the cross section model and the appropriate resulting correlation between the model parameters.
Such prior information is often unavailable, particularly for more complex or higher-dimensional models where the implicit dependencies among the parameters are more difficult to understand and characterize.
\par
On the other hand,
the Gaussian-process model considers the systematic error on the measurement space $(\epsilon, \sigma)$ directly.
As a result, both the goodness-of-fit and the systematic error are evaluated together via (\ref{eq:gp-bayes-indiv-likelihood}), against the actual measurement data $\cM_m$.
The resulting posterior properly captures the intrinsic correlation between $b$ and $\gamma$ arising from the model,
without explicitly requiring prior information be built into the model form.
\par
\begin{figure}
\centering
\input{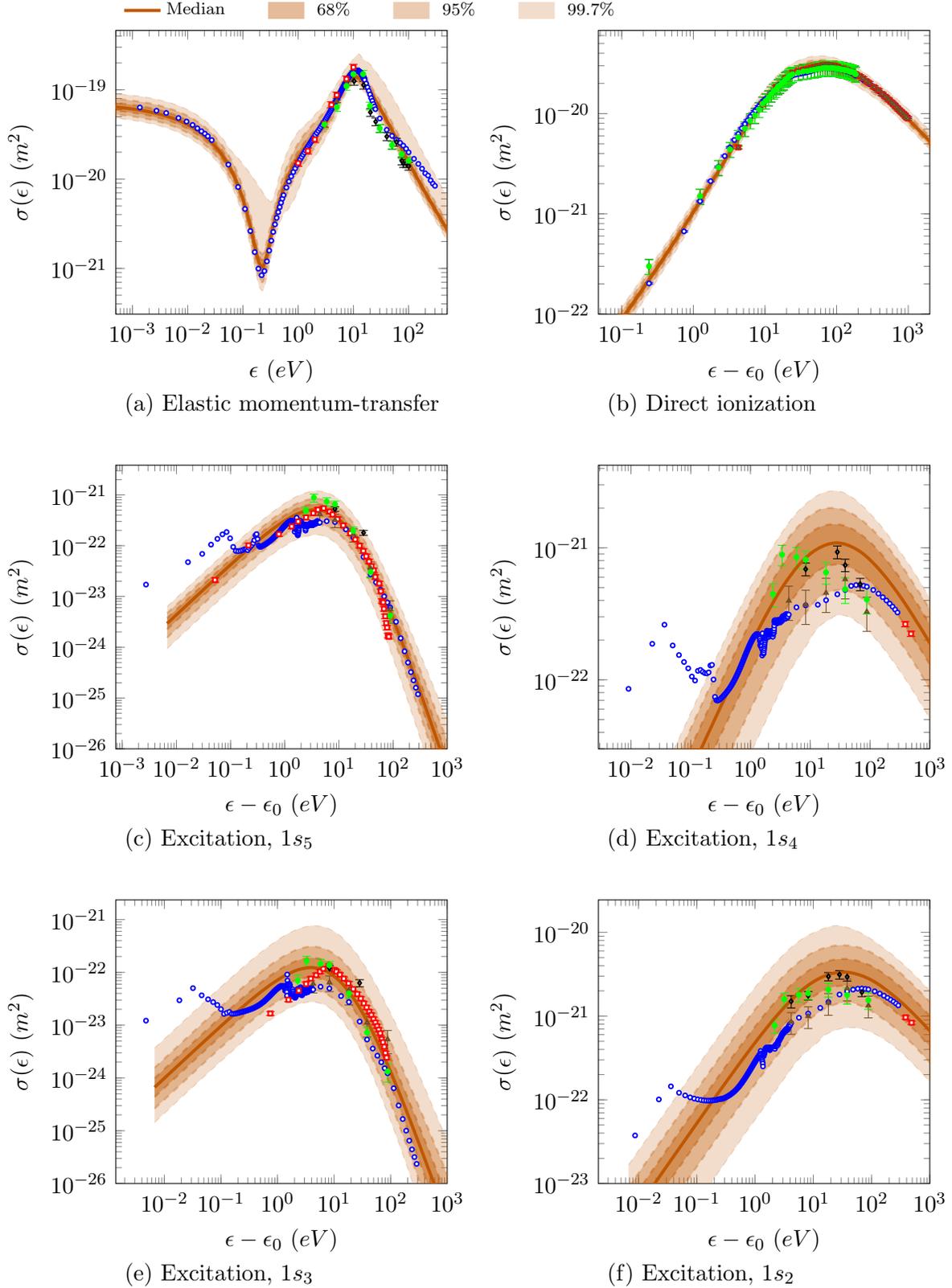}
\caption{
Marginal cross section uncertainty of 6 electron-impact argon collision cross sections,
with the $P_3$ Gaussian process representation (Eq.~\ref{eq:gp-bayes-margin-post} in Section~\ref{subsec:gp-bayes}).
Shaded areas indicate the $68\%$, $95\%$ and $99.7\%$ credibility intervals of the posterior distribution.
The measurement data are marked the same as in Figure~\ref{fig:crs-curation}.
}
\label{fig:gp-bayes-all}
\end{figure}
This comparison leads to a conclusion that, among three different perspectives,
the probabilistic description in Section~\ref{subsec:gp-bayes} with a Gaussian-process systematic error
most properly reflects the nature of the measurement data.
All the cross section uncertainties are thus calibrated with the posterior formulation (\ref{eq:gp-bayes-margin-post}) from Section~\ref{subsec:gp-bayes}.
Figure~\ref{fig:gp-bayes-all}
shows the marginal cross section posteriors for all 6 collision processes.
For all cross sections,
the scattering among the datasets is largely captured within the credibility intervals.
There are some data that lie outside the credibility intervals:
in Figure~\ref{fig:gp-bayes-all}~(a),
except some outliers between $95$--$99.7\%$ interval,
the elastic cross section samples lie closely around the median value
and do not capture the small variation in $\epsilon > 10 eV$;
and all the excitation cross sections in Figure~\ref{fig:gp-bayes-all}~(c--f)
do not capture the BSR data in $\epsilon - \epsilon_0 < 0.1 eV$.
These are, of course, expected results from the cross section models chosen for this uncertainty quantification because the models chosen in Section~\ref{sec:model} cannot represent these features.
Their parametric uncertainties, therefore, do not reflect any deviation which the models cannot capture,
rather ascribing it to the systematic error (including the model inadequacy).
However, the resulting parametric uncertainty is still able to encompass the uncertainty in the plasma chemistry and transport properties,
which will be shown subsequently.

\subsection{Validation with swarm-parameter experiments}\label{subsec:validation}

The calibrated parametric models for the 6 collision cross sections
are forward-propagated to the swarm parameters via \texttt{BOLSIG+}
and compared with the experimental data from Table~\ref{tab:swarm-parameter}.
For the forward propagation,
the swarm parameters are evaluated with 7200 samples from the posterior distributions of the model parameters.
For each sample,
the collision cross section models from Section~\ref{sec:model} are evaluated and then used as inputs in \texttt{BOLSIG+} simulations.
\texttt{BOLSIG+} simulations are run at both the default condition of $300K$~\cite{Pitchford2013},
and also at different background temperatures to match with the experiment conditions in Table~\ref{tab:swarm-parameter}.
\refA{Recombination and step-wise ionization processes are considered negligible at these experiment conditions
and thus not included in swarm parameter calculations.}
\par
\begin{figure}
\input{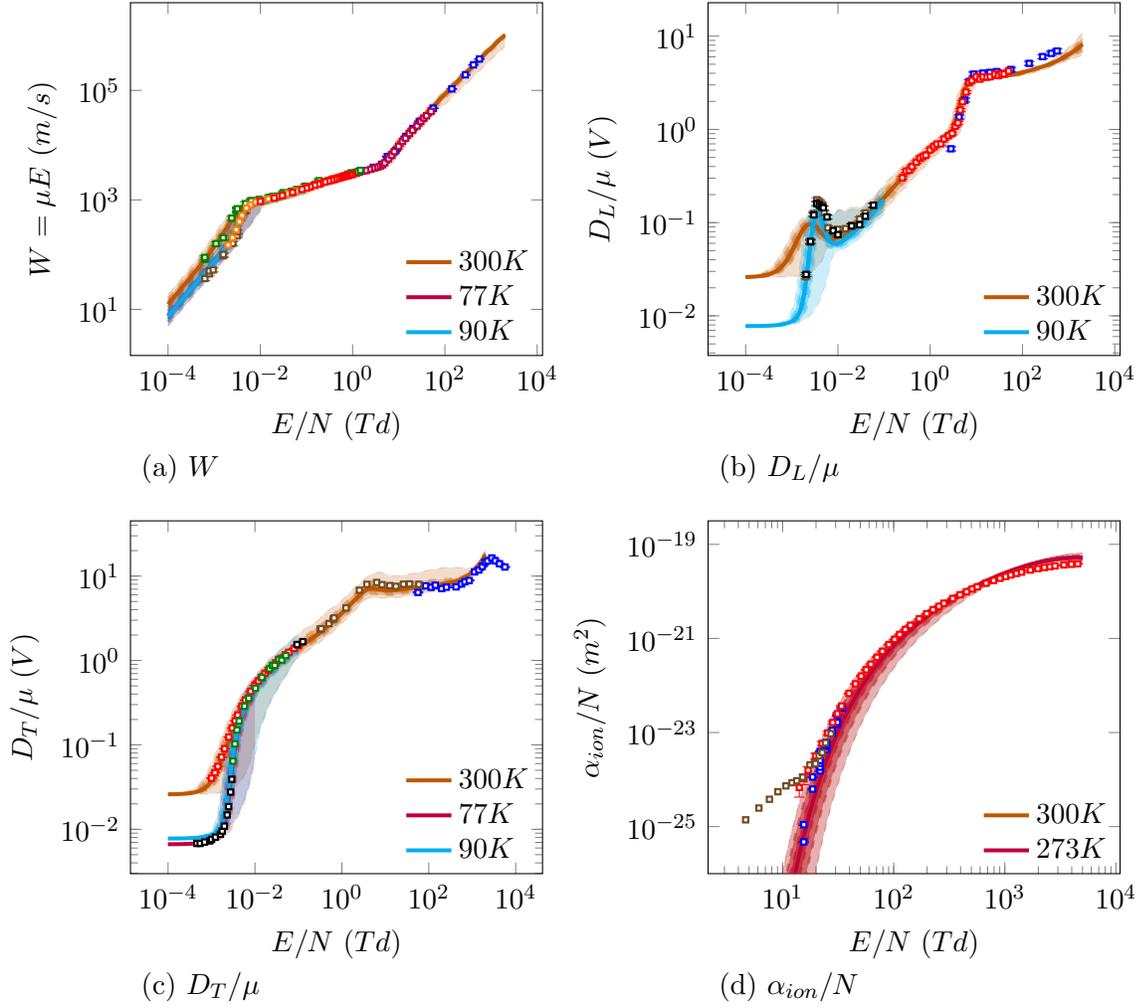}
\caption{
The cross section uncertainty from the model $P_3$ propagated unto the swarm parameters,
shown with 3 credibility intervals ($68\%$, $95\%$, and $99.7\%$):
(a) drift velocity,
(b) the ratio of longitudinal diffusivity to mobility,
(c) the ratio of transverse diffusivity to mobility, and
(d) the reduced ionization coefficient.
The measurement data are marked the same as in Figure~\ref{fig:swarm-parameter}.
}
\label{fig:uncertainty-swarm}
\end{figure}
\begin{figure}
\input{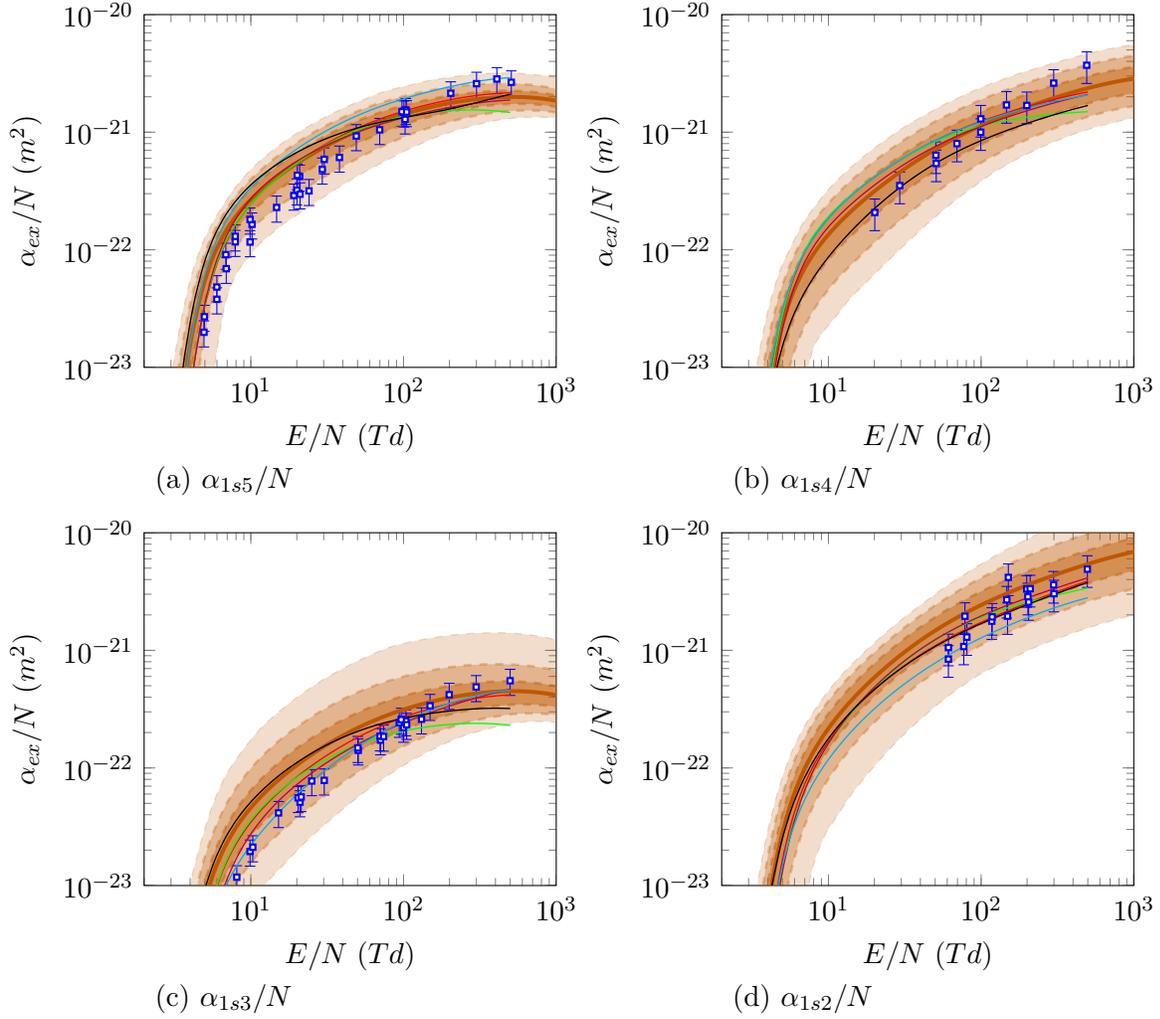}
\caption{
The cross section uncertainty propagated unto the swarm parameters,
shown with 3 credibility intervals ($68\%$, $95\%$, and $99.7\%$):
the reduced excitation coefficients to
(a) $1s_5$,
(b) $1s_4$,
(c) $1s_3$, and
(d) $1s_2$.
The measurement data are marked the same as in Figure~\ref{fig:swarm-parameter-ex}.
}
\label{fig:uncertainty-swarm-ex}
\end{figure}
Figures~\ref{fig:uncertainty-swarm}~and~\ref{fig:uncertainty-swarm-ex}
show the propagated uncertainty in the swarm parameters,
compared with the experimental data from Figures~\ref{fig:swarm-parameter}~and~\ref{fig:swarm-parameter-ex}.
As for the swarm-analysis data in Figure~\ref{fig:swarm-parameter-ex},
the evaluation of the reduced excitation coefficients in Figure~\ref{fig:uncertainty-swarm-ex}
requires the direct excitation cross sections to higher levels.
Uncertainty for these cross sections is beyond the scope of this study.
Instead, the cross sections for excitation to the $2p$-level
are taken from the BSR dataset~\cite{BSR-data}.
\par
Most of the experimental data lie within $95\%$ credibility intervals of the cross section uncertainty.
As mentioned in Section~\ref{subsec:swarm-data},
in Figure~\ref{fig:uncertainty-swarm}~(d)
the data from Specht \textit{et al.}~\cite{Specht1980} deviates from the credibility intervals,
as their measurements include contributions from the metastable population.
For the transport properties and the reduced ionization coefficients in Figure~\ref{fig:uncertainty-swarm},
the impact of the cross section uncertainties was relatively small,
and the experimental data also did not exhibit a large discrepancy with each other.
On the other hand,
the cross section uncertainties have a more significant impact on the reduced excitation coefficients, as shown Figure~\ref{fig:uncertainty-swarm-ex}.

\section{Conclusion}
\label{sec:conclusion}
A Bayesian inference technique is proposed to quantify uncertainties in plasma collision cross sections.
Cross section uncertainties of six $e$-$Ar$ collision processes are quantified with the proposed method.
Each cross section is characterized with a semi-empirical model
that captures essential features for plasma reaction and transport,
so its uncertainty is effectively approximated within a low-dimensional parameter space.
Measurement data for cross section inference are curated either from experiments or \textit{ab initio} calculations,
into two categories: direct cross section measurement and swarm-parameter measurement.
The cross section model parameters are then inferred from the direct cross section data.
The resulting distributions are forward-propagated using the \texttt{BOLSIG+} $0D$ Boltzmann equation solver
and validated against the swarm-parameter measurements.
\par
A key aspect of this process is the modeling of discrepancies between different input datasets.
Observation errors reported in direct cross section data are often notional,
not quantitatively reflecting all systematic errors underlying the experiments.
As a result, the discrepancy among datasets is much larger than the reported observation error each claims.
No detailed knowledge of these errors, upon which to base a model, is
available.  Instead, both a parametric representation and a
Gaussian-process-based model were developed, with the
Gaussian-process-based approach giving better results in the present
cases.
\par
The resulting cross section uncertainties quantified by this process
provide rich information for the plasma simulations where these calibrated cross sections are used.
For example, the impact of each collision process on the macroscopic plasma observables can be evaluated via sensitivity analysis.
Furthermore, the uncertainties of collision processes can be forward-propagated through the plasma simulations,
thereby evaluating the credibility of the predictive simulations.
\par
The uncertain cross section models calibrated for six electron-argon atom collisions
are readily applicable to practical argon plasma simulations,
just as demonstrated in Section~\ref{subsec:validation}.
Likewise, uncertainties in other plasmas can be quantified using the proposed Bayesian inference technique.
As shown throughout this study,
likelihood modeling may requre data-specific considerations reflecting the nature of cross section data.
Such extension to other plasmas will allow us to capture the impact of cross section uncertainty
in a broder range of plasma applications.
\par
In this study, the uncertainty inferred only from electron-beam measurements and \textit{ab initio} calculations
was sufficient to explain the scatter in the swarm-parameter experiments.
As mentioned in Section~\ref{subsec:swarm-data}, in principle
the proposed Bayesian inference framework can be extended to incorporate the swarm-parameter experiments,
which is a more extensive reproduction of the conventional swarm analysis.
It is worth investigating whether such incorporation would decrease the resulting uncertainties in the cross sections.
On the one hand, as the swarm parameters are determined by the entire set of collisions,
including them into the Bayesian inference may provide richer information as to the correlations between the collision processes.
Furthermore, unlike the swarm-derived datasets, the resulting full joint posterior distribution enables us
to use an individual cross section by marginalizing the posterior for a single cross section.
On the other hand, evaluating the posterior will require an additional layer of solving the Boltzmann equation model,
for which a similar challenge of handling systematic errors/model-form inadequacy is expected.
Considering these potential outcomes as a trade-off,
incorporating swarm-parameter data in the uncertainty calibration requires a careful investigation.

\section*{Acknowledgements}
This material is based upon work supported by the Department of Energy, National Nuclear Security Administration under Award Number DE-NA0003969.
This work was performed in part under the auspices of the U.S. Department of Energy by Lawrence Livermore National Laboratory under contract DE-AC52-07NA27344.
LLNL release number: LLNL-JRNL-851195.

\appendix

\section{Sensitivity study with \texttt{BOLSIG+}}\label{app:sensitivity}

\subsection{Sensitivity to the elastic collision in the large-energy range}
In Section~\ref{subsec:model-elastic},
the proposed cross section model approximates the high-energy ($>10eV$) range
with only $\cO(\epsilon^{-1})$ asymptote, neglecting smaller-scale variations observed in the measurement datasets.
\par
\begin{figure}
\centering
\input{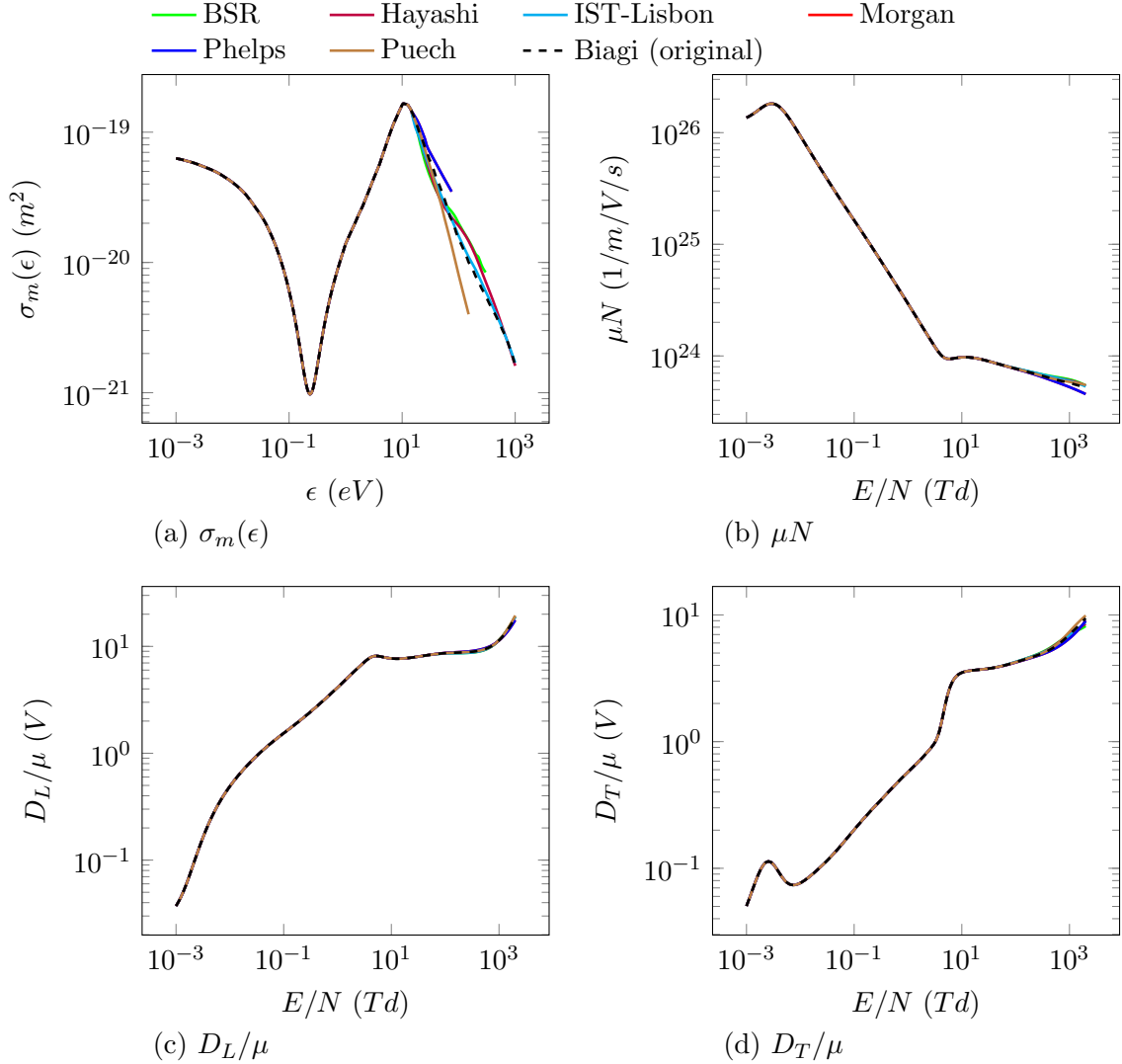}
\caption{
Sensitivity of plasma transport properties to the $\epsilon>10eV$ range of the elastic collision:
(a) the manipulated elastic cross sections for the analysis,
(b) the reduced mobility,
(c) the ratio of the longitudinal diffusivity to the mobility, and
(d) the ratio of the transverse diffusivity to the mobility computed with corresponding cross section sets.
}
\label{fig:app1}
\end{figure}
The impact of this small-scale variations on the plasma transport properties is investigated via \texttt{BOLSIG+}~\cite{Hagelaar2005}.
The Biagi dataset from the LXCat community~\cite{Biagi-data} is chosen for the baseline cross section set,
where $\epsilon>10eV$ range of the elastic momentum-transfer cross section is replaced with that from other LXCat datasets~\cite{BSR-data, Phelps-data, Puech-data, Morgan-data, IST-Lisbon-data, Hayashi-data}.
Figure~\ref{fig:app1}~(a) shows these manipulated cross section sets,
whose $\epsilon>10eV$ ranges are scattered similar to the measurement datasets in Figure~\ref{fig:elastic-model}.
Figure~\ref{fig:app1}~(b-d) show the plasma transport properties computed at the default \texttt{BOLSIG+} condition with $300K$.
Only slight variations up to $10\%$ are observed in the limited range over $E/N > 300Td$.

\subsection{Sensitivity to the small-scale features in the excitation collision}
\begin{figure}
\centering
\input{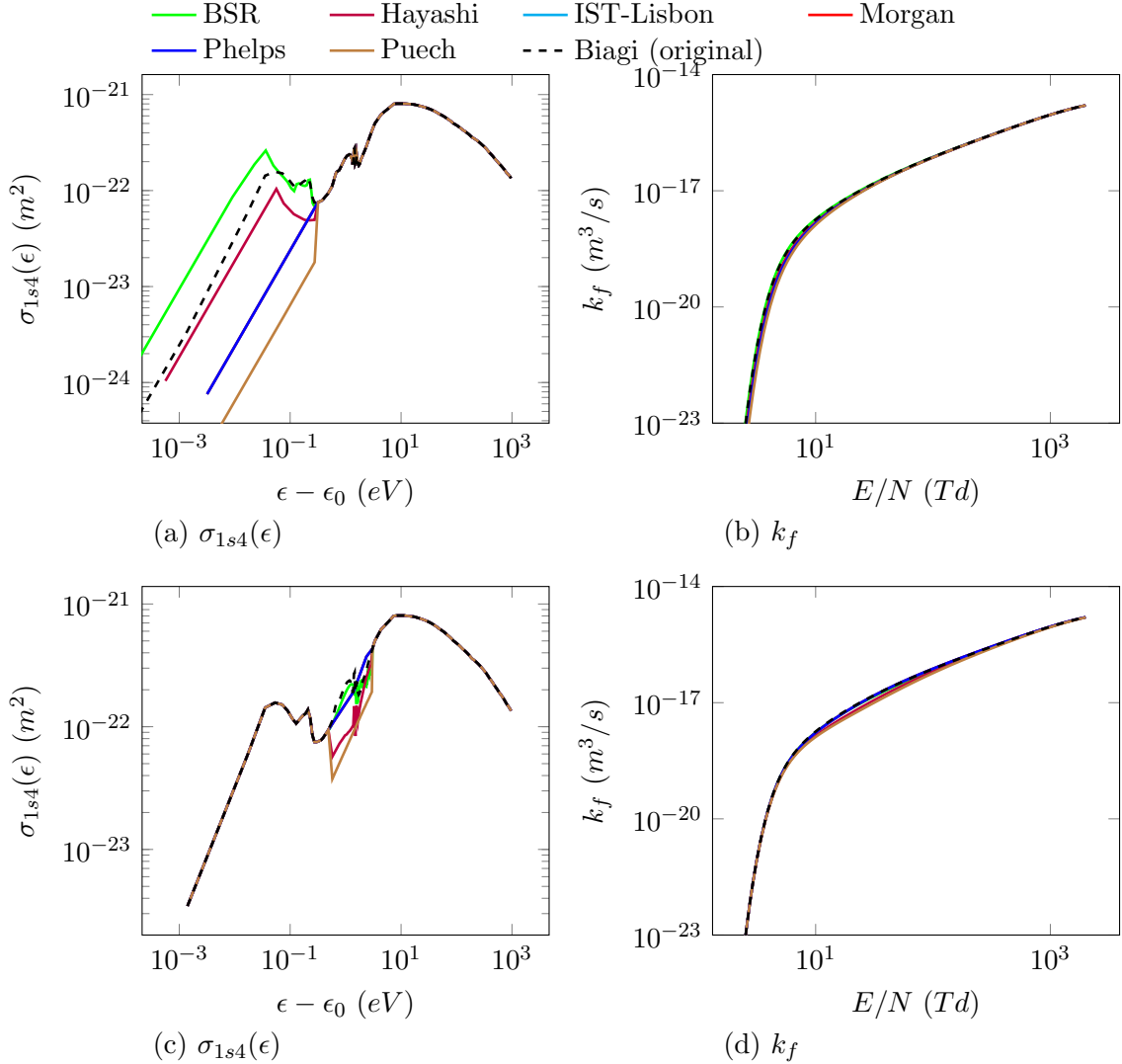}
\caption{
Sensitivity of the reaction rate constant to the small-scale features of the $1s4$ excitation collision:
(a) the $1s4$ excitation cross sections with $\epsilon-\epsilon_0<0.3eV$ range manipulated for the analysis, and
(b) the corresponding reaction rate constants;
(c) the $1s4$ excitation cross sections with $\epsilon-\epsilon_0\in[0.5, 3]eV$ range manipulated for the analysis, and
(d) the corresponding reaction rate constants;
}
\label{fig:app2}
\end{figure}
The sensitivity to the small-scale features in the excitation collision is similarly investigated.
First, the low energy $\epsilon-\epsilon_0<0.3eV$ range of the baseline Biagi dataset is replaced with that from other datasets~\cite{BSR-data, Phelps-data, Puech-data, Morgan-data, IST-Lisbon-data, Hayashi-data}.
In Figure~\ref{fig:app2}~(a), their variations shown are seemingly very large, up to two orders of magnitude.
However, in Figure~\ref{fig:app2}~(b), the resulting reaction rate constants surprisingly do not vary from each other.
\par
$\epsilon - \epsilon_0 \in [0.5, 3]eV$ range of the $1s4$ excitation cross section also exhibits some fine-scale features, which sensitivity is worth investigating.
Figure~\ref{fig:app2}~(c) shows the cross sections with this energy range manipulated.
While fine-scale features are observed in Biagi, BSR and Hayashi datasets, Phelps and Puech datasets vary smoothly in this energy range.
Figure~\ref{fig:app2}~(d) shows the corresponding reaction rate constants, which vary only slightly in the $E/N\in[10,300]Td$ range.
While Hayashi and Puech datasets seem to have lower rate constants,
this is rather due to the lower magnitude of their cross sections.
They in fact have the rate constants similar to each other, despite their difference of the fine-scale behavior.
Likewise, although Phelps dataset completely neglects fine-scale behavior and vary smoothly,
it exhibits similar rate constants with Biagi, BSR datasets, due to its similar cross section magnitude.
This sensitivity analysis suggests
that the plasma reaction rate constants are insensitive to the fine-scale features in the excitation cross sections.
\par
Similar insensitivity to the low-energy, fine-scale features is observed from other excitation cross sections as well.

\bibliography{references.bib}

\providecommand{\newblock}{}
\begin{thebibliography}{100}
\expandafter\ifx\csname url\endcsname\relax
  \def\url#1{{\tt #1}}\fi
\expandafter\ifx\csname urlprefix\endcsname\relax\def\urlprefix{URL }\fi
\providecommand{\eprint}[2][]{\url{#2}}

\bibitem{Gargioni2008}
Gargioni E and Grosswendt B 2008 {\em Reviews of Modern Physics\/} {\bf 80}
  451--480

\bibitem{Ferziger1973}
Ferziger J~H, Kaper H~G and Gross E~P 1973 {\em American Journal of Physics\/}
  {\bf 41}(4) 601--603 ISSN 0002-9505

\bibitem{Vlcek1989}
Vl{\v{c}}ek J 1989 {\em Journal of Physics D: Applied Physics\/} {\bf 22}
  632--643 ISSN 13616463

\bibitem{Bultel2002}
Bultel A, van Ootegem B, Bourdon A and Vervisch P 2002 {\em Physical Review E -
  Statistical Physics, Plasmas, Fluids, and Related Interdisciplinary Topics\/}
  {\bf 65} 16 ISSN 1063651X

\bibitem{Kapper2011}
Kapper M~G and Cambier J~L 2011 {\em Journal of Applied Physics\/} {\bf 109}
  ISSN 00218979

\bibitem{Pitchford2013}
Pitchford L, Alves L, Bartschat K, Biagi S, Bordage M, Phelps A, Ferreira C,
  Hagelaar G, Morgan W, Pancheshnyi S {\em et~al.\/} 2013 {\em Journal of
  Physics D: Applied Physics\/} {\bf 46} 334001

\bibitem{Pitchford2017}
Pitchford L~C, Alves L~L, Bartschat K, Biagi S~F, Bordage M~C, Bray I, Brion
  C~E, Brunger M~J, Campbell L, Chachereau A {\em et~al.\/} 2017 {\em Plasma
  Processes and Polymers\/} {\bf 14} 1600098

\bibitem{Oran1998}
Oran E, Oh C and Cybyk B 1998 {\em Annual Review of Fluid Mechanics\/} {\bf 30}
  403--441

\bibitem{Bird1994}
Bird G~A 1994 {\em Molecular gas dynamics and the direct simulation of gas
  flows\/}

\bibitem{Birdsall1991}
Birdsall C~K 1991 {\em IEEE Transactions on plasma science\/} {\bf 19} 65--85

\bibitem{Vahedi1993}
Vahedi V, DiPeso G, Birdsall C, Lieberman M and Rognlien T 1993 {\em Plasma
  Sources Science and Technology\/} {\bf 2} 261

\bibitem{Birdsall2004}
Birdsall C~K and Langdon A~B 2004 {\em Plasma physics via computer
  simulation\/} (CRC press)

\bibitem{Hagelaar2005}
Hagelaar G~J and Pitchford L~C 2005 {\em Plasma Sources Science and
  Technology\/} {\bf 14} 722--733 ISSN 09630252

\bibitem{petrov1997multi}
Petrov G and Winkler R 1997 {\em Journal of Physics D: Applied Physics\/} {\bf
  30} 53

\bibitem{winkler1986new}
Winkler R, Wilhelm J and Braglia G 1986 {\em Il Nuovo Cimento D\/} {\bf 7}
  641--680

\bibitem{segur1983application}
Segur P, Bordage M~C, Balaguer J~P and Yousfi M 1983 {\em Journal of
  Computational Physics\/} {\bf 50} 116--137

\bibitem{Bederson1971}
Bederson B and Kieffer L 1971 {\em Reviews of Modern Physics\/} {\bf 43} 601

\bibitem{Alves2005}
Yanguas-Gil A, Cotrino J and Alves L 2005 {\em Journal of Physics D Applied
  Physics\/} {\bf 38} 1588--1598

\bibitem{Biagi-data}
Biagi\;database retrieved on October 14, 2021
  \urlprefix\url{https://www.lxcat.net}

\bibitem{BSR-data}
BSR\;database retrieved on October 14, 2021
  \urlprefix\url{https://www.lxcat.net}

\bibitem{Hayashi-data}
Hayashi\;database retrieved on October 14, 2021
  \urlprefix\url{https://www.lxcat.net}

\bibitem{IST-Lisbon-data}
IST-Lisbon\;database retrieved on October 14, 2021
  \urlprefix\url{https://www.lxcat.net}

\bibitem{Puech-data}
Puech\;database retrieved on October 14, 2021
  \urlprefix\url{https://www.lxcat.net}

\bibitem{Filippelli1994}
A~R~Filippelli C~C~L and Anderson L~W 1994 {\em Advances in Atomic, Molecular,
  and Optical Physics\/} {\bf 33}
  \urlprefix\url{https://www.osti.gov/biblio/241114}

\bibitem{Tachibana1986}
Tachibana K 1986 {\em Physical Review A\/} {\bf 34} 1007

\bibitem{Chilton1998}
Chilton J~E, Boffard J~B, Schappe R~S and Lin C~C 1998 {\em Physical Review
  A\/} {\bf 57} 267

\bibitem{Zatsarinny2013}
Zatsarinny O and Bartschat K 2013 {\em Journal of Physics B: Atomic, Molecular
  and Optical Physics\/} {\bf 46} 112001

\bibitem{Mceachran2014}
McEachran R and Stauffer A 2014 {\em The European Physical Journal D\/} {\bf
  68} 1--8

\bibitem{Gangwar2012}
Gangwar R, Sharma L, Srivastava R and Stauffer A 2012 {\em Journal of Applied
  Physics\/} {\bf 111} 053307

\bibitem{Djuissi2022}
Djuissi E, Bultel A, Tennyson J, Schneider I and Laporta V 2022 {\em Plasma
  Sources Science and Technology\/} {\bf 31} 114012

\bibitem{Milloy1977}
Milloy H~B, Crompton R~W, Rees J~A and Robertson A~G 1977 {\em Australian
  Journal of Physics\/} {\bf 30} 61--72

\bibitem{Haddad1982}
Haddad G and O'Malley T 1982 {\em Australian Journal of Physics\/} {\bf 35}
  35--40

\bibitem{Nakamura1987}
Nakamura Y 1987 {\em Journal of Physics D: Applied Physics\/} {\bf 20} 933--938
  \urlprefix\url{https://doi.org/10.1088/0022-3727/20/7/016}

\bibitem{Biagi1989}
Biagi S~F 1989 {\em Nuclear Instruments and Methods in Physics Research Section
  A: Accelerators, Spectrometers, Detectors and Associated Equipment\/} {\bf
  283} 716--722

\bibitem{alves2018foundations}
Alves L, Bogaerts A, Guerra V and Turner M 2018 {\em Plasma Sources Science and
  Technology\/} {\bf 27} 023002

\bibitem{Stuart2010}
Stuart A~M 2010 {\em Acta numerica\/} {\bf 19} 451--559 ISSN 0962-4929

\bibitem{Smith2013}
Smith R~C 2013 {\em Uncertainty quantification: theory, implementation, and
  applications\/} vol~12 (Siam)

\bibitem{turner2015uncertainty}
Turner M~M 2015 {\em Plasma Sources Science and Technology\/} {\bf 24} 035027

\bibitem{turner2016uncertainty}
Turner M~M 2016 {\em Plasma Sources Science and Technology\/} {\bf 25} 015003

\bibitem{turner2017computer}
Turner M~M 2017 {\em Plasma Processes and Polymers\/} {\bf 14} 1600121

\bibitem{koelman2019uncertainty}
Koelman P, Yordanova D, Graef W, Mousavi S~T and van Dijk J 2019 {\em Plasma
  Sources Science and Technology\/} {\bf 28} 075009

\bibitem{berthelot2017modeling}
Berthelot A and Bogaerts A 2017 {\em Plasma Sources Science and Technology\/}
  {\bf 26} 115002

\bibitem{Kennedy2001}
Kennedy M~C and O'Hagan A 2001 {\em Journal of the Royal Statistical Society.
  Series B: Statistical Methodology\/} {\bf 63} 425--464

\bibitem{Rasmussen2006}
Rasmussen C~E and Williams C~K 2006 {\em Gaussian processes for machine
  learning\/} vol~2 (MIT press Cambridge, MA)

\bibitem{Foreman-Mackey2013}
Foreman-Mackey D, Hogg D~W, Lang D and Goodman J 2013 {\em Publications of the
  Astronomical Society of the Pacific\/} {\bf 125} 306--312 (\textit{Preprint}
  \eprint{1202.3665})

\bibitem{Bretagne1986}
Bretagne J, Callede G, Legentil M and Puech V 1986 {\em Journal of Physics D:
  Applied Physics\/} {\bf 19} 761

\bibitem{Kim1994}
Kim Y~K and Rudd M~E 1994 {\em Physical Review A\/} {\bf 50} 3954--3967

\bibitem{Srivastava1981}
Srivastava S, Tanaka H, Chutjian A and Trajmar S 1981 {\em Physical Review A\/}
  {\bf 23} 2156

\bibitem{Gibson1996}
Gibson J~C, Gulley R, Sullivan J, Buckman S, Chan V and Burrow P 1996 {\em
  Journal of Physics B: Atomic, Molecular and Optical Physics\/} {\bf 29} 3177

\bibitem{Panajotovic1997}
Panajotovic R, Filipovic D, Marinkovic B, Pejcev V, Kurepa M and Vuskovic L
  1997 {\em Journal of Physics B: Atomic, Molecular and Optical Physics\/} {\bf
  30} 5877

\bibitem{Mielewska2004}
Mielewska B, Linert I, King G~C and Zubek M 2004 {\em Physical Review A\/} {\bf
  69} 062716

\bibitem{Rapp1965}
Rapp D and Englander‐Golden P 1965 {\em The Journal of Chemical Physics\/}
  {\bf 43} 1464--1479 (\textit{Preprint}
  \eprint{https://doi.org/10.1063/1.1696957})
  \urlprefix\url{https://doi.org/10.1063/1.1696957}

\bibitem{Wetzel1987}
Wetzel R~C, Baiocchi F~A, Hayes T~R and Freund R~S 1987 {\em Physical Review
  A\/} {\bf 35} 559--577 ISSN 10502947

\bibitem{Straub1995}
Straub H, Renault P, Lindsay B, Smith K and Stebbings R 1995 {\em Physical
  Review A\/} {\bf 52} 1115

\bibitem{Chutjian1981}
Chutjian A and Cartwright D 1981 {\em Physical Review A\/} {\bf 23} 2178

\bibitem{Li1988}
Li G~P, Takayanagi T, Wakiya K, Suzuki H, Ajiro T, Yagi S, Kano S~S and Takuma
  H 1988 {\em Physical Review A\/} {\bf 38} 1240--1247 ISSN 10502947

\bibitem{Schappe1994}
Schappe R~S, Schulman M~B, Anderson L~W and Lin C~C 1994 {\em Physical Review
  A\/} {\bf 50} 444--461 ISSN 10502947

\bibitem{Filipovic2000a}
Filipovic D~M, Marinkovic B~P, Pejcev V and Vuskovic L 2000 {\em Journal of
  Physics B: Atomic and Molecular and Optical Physics\/} {\bf 33} 677

\bibitem{Filipovic2000b}
Filipovic D, Marinkovic B, Pejcev V and Vuskovic L 2000 {\em Journal of Physics
  B: Atomic, Molecular and Optical Physics\/} {\bf 33} 2081

\bibitem{Khakoo2004}
Khakoo M~A, Vandeventer P, Childers J~G, Kanik I, Fontes C~J, Bartschat K,
  Zeman V, Madison D~H, Saxena S, Srivastava R and Stauffer A~D 2004 {\em
  Journal of Physics B: Atomic, Molecular and Optical Physics\/} {\bf 37}
  247--281 ISSN 09534075

\bibitem{Zatsarinny2004}
Zatsarinny O and Bartschat K 2004 {\em Journal of Physics B: Atomic, Molecular
  and Optical Physics\/} {\bf 37} 4693

\bibitem{Zatsarinny2014}
Zatsarinny O, Wang Y and Bartschat K 2014 {\em Phys. Rev. A\/} {\bf 89}(2)
  022706 \urlprefix\url{https://link.aps.org/doi/10.1103/PhysRevA.89.022706}

\bibitem{Allan2006}
Allan M, Zatsarinny O and Bartschat K 2006 {\em Physical Review A\/} {\bf 74}
  030701

\bibitem{Khakoo2011}
Khakoo M~A, Zatsarinny O and Bartschat K 2011 {\em Journal of Physics B:
  Atomic, Molecular and Optical Physics\/} {\bf 44} 015201
  \urlprefix\url{https://doi.org/10.1088/0953-4075/44/1/015201}

\bibitem{Buckman1983}
Buckman S, Hammond P, King G and Read F 1983 {\em Journal of Physics B: Atomic
  and Molecular Physics (1968-1987)\/} {\bf 16} 4219

\bibitem{Hayashi2003}
Hayashi M 2003 {\em Institute for Fusion Science\/}

\bibitem{Yamabe1983}
Yamabe C, Buckman S~J and Phelps A~V 1983 {\em Phys. Rev. A\/} {\bf 27}(3)
  1345--1352 \urlprefix\url{https://link.aps.org/doi/10.1103/PhysRevA.27.1345}

\bibitem{Weber2003}
Weber T, Boffard J~B and Lin C~C 2003 {\em Physical Review A\/} {\bf 68} 032719

\bibitem{Drawin1967}
Drawin H~W 1967 {\em Fontenay-aux-Roses\/}
  \urlprefix\url{https://www.osti.gov/biblio/4675184}

\bibitem{Lee1973}
Lee C~M and Lu K 1973 {\em Physical Review A\/} {\bf 8} 1241

\bibitem{Bretagne1982}
Bretagne J, Godart J and Puech V 1982 {\em Journal of Physics D: Applied
  Physics\/} {\bf 15} 2205

\bibitem{Frost1964}
Frost L~S and Phelps A~V 1964 {\em Phys. Rev.\/} {\bf 136}(6A) A1538--A1545
  \urlprefix\url{https://link.aps.org/doi/10.1103/PhysRev.136.A1538}

\bibitem{Fletcher1972}
Fletcher J and Burch D 1972 {\em Journal of Physics D: Applied Physics\/} {\bf
  5} 2037

\bibitem{Schaper1969}
Schaper M and Scheibner H 1969 {\em Beitr{\"a}ge aus der Plasmaphysik\/} {\bf
  9} 45--57

\bibitem{Smith1930}
Smith P~T 1930 {\em Phys. Rev.\/} {\bf 36}(8) 1293--1302
  \urlprefix\url{https://link.aps.org/doi/10.1103/PhysRev.36.1293}

\bibitem{Fon1983}
Fon W, Berrington K, Burke P and Hibbert A 1983 {\em Journal of Physics B:
  Atomic and Molecular Physics\/} {\bf 16} 307

\bibitem{Ferreira1983}
Ferreira C and Loureiro J 1983 {\em Journal of Physics D: Applied Physics\/}
  {\bf 16} 2471

\bibitem{Townsend1922}
Townsend J and Bailey V 1922 {\em The London, Edinburgh, and Dublin
  Philosophical Magazine and Journal of Science\/} {\bf 44} 1033--1052

\bibitem{Pack1961}
Pack J and Phelps A 1961 {\em Physical Review\/} {\bf 121} 798

\bibitem{Warren1962}
Warren R~W and Parker~Jr J~H 1962 {\em Physical Review\/} {\bf 128} 2661

\bibitem{Robertson1972}
Robertson A and Rees J 1972 {\em Australian Journal of Physics\/} {\bf 25}
  637--640

\bibitem{Robertson1977}
Robertson A 1977 {\em Australian Journal of Physics\/} {\bf 30} 39--50

\bibitem{Milloy1977ratio}
Milloy H and Crompton R 1977 {\em Australian Journal of Physics\/} {\bf 30}
  51--60

\bibitem{Kucukarpaci1981}
Kucukarpaci H and Lucas J 1981 {\em Journal of Physics D: Applied Physics\/}
  {\bf 14} 2001

\bibitem{AlAmin1987}
Al-Amin S~A and Lucas J 1987 {\em Journal of Physics D: Applied Physics\/} {\bf
  20} 1590--1595 ISSN 00223727

\bibitem{Nakamura1988}
Nakamura Y and Kurachi M 1988 {\em Journal of Physics D: Applied Physics\/}
  {\bf 21} 718--723 \urlprefix\url{https://doi.org/10.1088/0022-3727/21/5/008}

\bibitem{Golden1961}
Golden D~E and Fisher L~H 1961 {\em Physical Review\/} {\bf 123} 1079--1086
  ISSN 0031899X

\bibitem{Kruithof1940}
Kruithof A 1940 {\em Physica\/} {\bf 7} 519--540

\bibitem{Specht1980}
Specht L, Lawton S and DeTemple T 1980 {\em Journal of Applied Physics\/} {\bf
  51} 166--170

\bibitem{Biagi1988}
Biagi S~F 1988 {\em Nuclear Instruments and Methods in Physics Research Section
  A: Accelerators, Spectrometers, Detectors and Associated Equipment\/} {\bf
  273} 533--535

\bibitem{Alves2014}
Alves L~L 2014 {\em Journal of Physics: Conference Series\/} {\bf 565} 012007
  \urlprefix\url{https://doi.org/10.1088/1742-6596/565/1/012007}

\bibitem{Puech1986}
Puech V and Torchin L 1986 {\em Journal of Physics D: Applied Physics\/} {\bf
  19} 2309--2323 \urlprefix\url{https://doi.org/10.1088/0022-3727/19/12/011}

\bibitem{Wiese1969}
Wiese W~L, Smith M~W and Miles B 1969 Atomic transition probabilities. volume
  2. sodium through calcium Tech. rep. NATIONAL STANDARD REFERENCE DATA SYSTEM

\bibitem{Morgan-data}
Morgan\;database retrieved on October 14, 2021
  \urlprefix\url{https://www.lxcat.net}

\bibitem{Tachibana1981}
Tachibana K and Phelps A~V 1981 {\em The Journal of Chemical Physics\/} {\bf
  75} 3315--3320 (\textit{Preprint} \eprint{https://doi.org/10.1063/1.442483})
  \urlprefix\url{https://doi.org/10.1063/1.442483}

\bibitem{Phelps-data}
Phelps\;database retrieved on October 14, 2021
  \urlprefix\url{https://www.lxcat.net}

\bibitem{carbone2021data}
Carbone E, Graef W, Hagelaar G, Boer D, Hopkins M~M, Stephens J~C, Yee B~T,
  Pancheshnyi S, Van~Dijk J and Pitchford L 2021 {\em Atoms\/} {\bf 9} 16

\bibitem{Crompton1994}
Crompton R 1994 {\em Advances in atomic, molecular, and optical physics\/} {\bf
  33}

\bibitem{OMalley1963}
O'Malley T~F 1963 {\em Physical Review\/} {\bf 130} 1020--1029 ISSN 0031899X

\bibitem{Kramida2022}
Kramida A, {Yu~Ralchenko}, Reader J and {and NIST ASD Team} 2022 {NIST Atomic
  Spectra Database (ver. 5.10), [Online]. Available:
  {\tt{https://physics.nist.gov/asd}} [2022, November 7]. National Institute of
  Standards and Technology, Gaithersburg, MD.}

\end{thebibliography}

\end{document}